\documentclass[12pt]{JHEP3}
\pdfoutput=1
\usepackage{graphicx}
\usepackage{epsfig}
\usepackage{subfig}
\usepackage{amsmath}
\usepackage{multirow}
\usepackage{listings}
\usepackage{float}

\def\ss{{\bigl.^3\hspace{-1mm}S^{[1]}_1}}
\def\sps{{\bigl.^1\hspace{-1mm}S^{[8]}_0}}
\def\spso{{\bigl.^1\hspace{-1mm}S^{[1,8]}_0}}
\def\spo{{\bigl.^1\hspace{-1mm}S^{[1]}_0}}
\def\so{{\bigl.^3\hspace{-1mm}S^{[8]}_1}}
\def\pjs{{\bigl.^3\hspace{-1mm}P^{[1]}_J}}
\def\pjso{{\bigl.^3\hspace{-1mm}P^{[1,8]}_J}}
\def\opos{{\bigl.^1\hspace{-1mm}P^{[1]}_1}}
\def\opoo{{\bigl.^1\hspace{-1mm}P^{[8]}_1}}
\def\oposo{{\bigl.^1\hspace{-1mm}P^{[1,8]}_1}}
\def\tpzs{{\bigl.^3\hspace{-1mm}P^{[1]}_0}}
\def\tpos{{\bigl.^3\hspace{-1mm}P^{[1]}_1}}
\def\tpts{{\bigl.^3\hspace{-1mm}P^{[1]}_2}}
\def\pj{{\bigl.^3\hspace{-1mm}P^{[8]}_J}}
\def\p0{{\bigl.^3\hspace{-1mm}P^{[8]}_0}}
\def\sso{\bigl.^3\hspace{-1mm}S^{[1,8]}_1}

\def\bqa{\begin{eqnarray}}
\def\eqa{\end{eqnarray}}
\def\bc{\begin{center}}
\def\bc{\end{center}}
\def\cCode#1{\begin{lstlisting}[mathescape,basicstyle=\small
\ttfamily,frame=leftline,aboveskip=4mm,belowskip=4mm,xleftmargin=20pt,framexleftmargin=10pt,
numbers=none,framerule=2pt,abovecaptionskip=0.0mm,belowcaptionskip=3.5mm #1]}
\def\chCode#1{\begin{lstlisting}[mathescape=true,basicstyle=\small
\ttfamily]}
\newcommand{\figuredir}{./figures}

\normalbaselines

\title{Boosting perturbative QCD stability in quarkonium production}

\author{Hua-Sheng Shao\\
Laboratoire de Physique Th\'eorique et Hautes Energies (LPTHE), UMR 7589, Sorbonne Universit\'e et CNRS, 4 place Jussieu, 75252 Paris Cedex 05, France\\
E-mail: \email{huasheng.shao@lpthe.jussieu.fr}}

\abstract{The aim of this paper is to introduce a general way to stabilize the perturbative QCD computations of heavy quarkonium production in the boosted or high-momentum transferring region with tree-level generators only. Such an approach is possible by properly taking into account the power-enhanced perturbative contributions in a soft and collinear safe manner without requiring any complete higher-order computations. The complicated NLO results for inclusive quarkonium hadroproduction can be well reproduced within our approach based on a tree-level generator {\sc\small HELAC-Onia}. We have applied it to estimate the last missing leading-twist contribution from the spin-triplet color-singlet S-wave production at $\mathcal{O}(\alpha_s^5)$, which is a NNLO term in the $\alpha_s$ expansion for the quarkonium $P_T$ spectrum. We conclude that the missing NNLO contribution will not change the order of the magnitude of the short-distance coefficient. Such an approach is also quite appealing as it foresees broad applications in quarkonium-associated production processes, which are mostly absent of complete higher-order computations and fragmentation functions.}

\keywords{QCD, NRQCD, Quarkonium}

\begin{document}

\section{Introduction and motivations\label{sec:intro}}

As a class of the simplest hadrons, heavy quarkonium is usually viewed as the ``hydrogen atom" in the strong-interaction theory QCD. While the knowledge of the nonperturbative aspect in QCD is still quite limited, heavy quarkonium provides a unique opportunity to probe the quark confinement in QCD by exploring the physics at the scale around the nonperturbative and perturbative boundary. The intrinsic scales of the heavy quark mass $m_Q$ and their binding energy $m_Qv^2$  lie in the perturbative and nonperturbative regimes respectively, where $v$ is the relative velocity between the heavy quark pair in the rest frame of the quarkonium. Due to the smallness of the relative velocity $v$ (e.g. $v^2\simeq 0.3$ and $v^2\simeq 0.1$ for the charmonium and bottomonium),  the relativistic QCD can be reorganized via the operator product expansion in the power counting of $v$. The effective theory was dubbed as non-relativistic QCD (NRQCD)~\cite{Bodwin:1994jh}. The reformulation of QCD provides a factorization conjecture for calculating the rates of the quarkonium production and decay. In the case of the quarkonium ${\cal H}$ production, the (differential) cross section at leading-order (LO) in the QCD strong coupling constant $\alpha_s$ can be schematically written as
\begin{eqnarray}
d\sigma({\cal H})&=&\sum_{n}{d\hat{\sigma}(n)\langle\mathcal{O}^{\cal H}(n)\rangle},\label{eq:xsfac}
\end{eqnarray}
where $n$ represents a Fock state, $d\hat{\sigma}(n)$ is a perturbatively calculable short-distance coefficient (SDC) with the heavy quark pair in the Fock state $n$ and $\langle\mathcal{O}^{\cal H}(n)\rangle$ is the vacuum expectation number of an operator $\mathcal{O}^{\cal H}(n)$. If the factorization formula Eq.(\ref{eq:xsfac}) holds, the nonperturbative long-distance matrix element (LDME) $\langle\mathcal{O}^{\cal H}(n)\rangle$ is independent of quarkonium production process as well as the production environment. The universal LDMEs, which are analogous to the parton-distribution functions (PDFs) in the perturbative QCD factorization, are to be determined from a subset of the experimental data and to predict all of the rest experimental measurements. They have the probability explanations at LO, while LDMEs depend on the renormalization scheme and they are not physical objects.

The prediction power of Eq.(\ref{eq:xsfac}) heavily relies on the perturbative convergences of $v^2$ and $\alpha_s$ in $d\sigma({\cal H})$. The leading power counting of various Fock states up to $\mathcal{O}(v^7)$ for S-wave and P-wave quarkonia is listed in Table.~\ref{tab:powcnt} according to the NRQCD velocity scaling rule~\cite{Bodwin:1994jh}. The convergence in $v^2$ can be improved by including the relativistic corrections. However, the prices to pay are that one has to introduce more nonperturbative LDMEs that can not be determined from the first principle, and the good relations like heavy-quark spin symmetry holding at LO in $v$ will be violated too.

\begin{table}[ht!]
\begin{center}
\begin{tabular}{{|c|}*{4}{c|}}\hline
Power counting & $\eta_Q$ & $\psi,\Upsilon$ & $h_Q$ & $\chi_{QJ}$\\\hline
$v^3$ & $\spo$ & $\ss$ & $-$ & $-$\\
$v^5$ & $-$ & $-$ & $\opos,\sps$ & $\pjs,\so$\\
$v^7$ & $\sps,\so,\opoo$ & $\sps,\so,\pj$ & $-$ & $-$
\\\hline
\end{tabular}
\end{center}
\caption{The leading power counting of various Fock states contributing to various quarkonium within NRQCD velocity scaling rule~\cite{Bodwin:1994jh}.\label{tab:powcnt}}
\end{table}

The most subtle part is the $\alpha_s$ stability in the SDCs $d\hat{\sigma}(n)$, which is the main point to be discussed in this paper. For the high-transverse-momentum ($P_T$) quarkonium production at a high-energy hadron collider, it was found that $\ss$ receives a giant K factor from QCD corrections to its SDC a decade ago~\cite{Campbell:2007ws}, which was understood by the fact that due to the quantum number conservation, there is a factor $\alpha_s \frac{P_T^2}{4m_Q^2}$ enhancement at $\mathcal{O}(\alpha_s^4)$ (next-to-leading order, NLO) compared to $\mathcal{O}(\alpha_s^3)$ (LO). This enhancement spoils the perturbative convergence in $\alpha_s$, shedding light on another possible enhancement from $\mathcal{O}(\alpha_s^5)$ (next-to-next-to-leading order, NNLO) corrections, while the accomplishment of the full NNLO calculation is even lacking today. The sole reason is the partonic cross sections $\frac{d\hat{\sigma}}{dP_T^2}$, before convoluting PDFs, are asymptotically scaling as $\left(\frac{2m_Q}{P_T}\right)^4\frac{1}{P_T^4}$ (next-to-next-to-leading power, NNLP) , $\left(\frac{2m_Q}{P_T}\right)^2\frac{1}{P_T^4}$ (next-to-leading power, NLP) and $\frac{1}{P_T^4}$ (leading power, LP) at LO, NLO, NNLO respectively.~\footnote{Rigorously speaking, the associated production of $\ss$ with the same flavoured heavy quark pair contributes $\mathcal{O}(\alpha_s^4)$ and is LP in $P_T$. We guide the readers to the discussion on this part in section~\ref{sec:charmfrag}.} Therefore, even with a full NNLO calculation at $\mathcal{O}(\alpha_s^5)$, the accuracy for the LP part of $\ss$ hadroproduction is still at LO level, while the NLP piece is indeed NLO accurate. A NLO accuracy of the LP contribution can only be achieved with a next-to-NNLO calculation in $\alpha_s$ for the SDC. The situation is slightly better though still similar for the other Fock states listed in Table~\ref{tab:powcnt}. Like $\spso,\pjso,\oposo$, the NLP (LP) parts of SDCs appear firstly at LO (NLO) in $\alpha_s$. On the other hand, because of the same quantum number as the gluon, $\so$ has the leading $P_T$ behaviour as the jet, which means the LP channel is already opened at LO $\mathcal{O}(\alpha_s^3)$. In Table~\ref{tab:accuracy}, we have collected the first $\alpha_s$ powers in order to achieve the LO and NLO QCD accuracies for various Fock states at both LP and NLP in $P_T$.

\begin{table}[ht!]
\begin{center}
\begin{tabular}{{|c|}*{7}{c|}}\hline
 \multirow{2}{*}{Accuracy} & \multicolumn{2}{c|}{$\ss$} & \multicolumn{2}{c|}{$\so$} & \multicolumn{2}{c|}{$\spso,\pjso,\oposo$}\\\cline{2-7}
 & {\rm LP} & {\rm NLP} & {\rm LP} & {\rm NLP} & {\rm LP} & {\rm NLP}\\\hline
{\rm LO} & $\alpha_s^5$ & $\alpha_s^4$ & $\alpha_s^3$ & $\alpha_s^3$ & $\alpha_s^4$ & $\alpha_s^3$\\\hline
{\rm NLO} & $\alpha_s^6$ & $\alpha_s^5$ & $\alpha_s^4$ & $\alpha_s^4$ & $\alpha_s^5$ & $\alpha_s^4$\\\hline
\end{tabular}
\end{center}
\caption{The first $\alpha_s$ orders needed in the SDCs for both LP and NLP in $P_T$ of various Fock states in their hadroproduction in order to achieve the LO and NLO QCD accuracies.\label{tab:accuracy}}
\end{table}

Following this observation, the complete NLO result for $\ss$ production is possible to be reproduced by the tree-level matrix element alone at $\mathcal{O}(\alpha_s^4)$ after introducing an {\it ad hoc} infrared cutoff. A first attempt was given in Ref.~\cite{Artoisenet:2008fc} to introduce an invariant-mass cut on any final-final and initial-final massless parton pairs, which was called NLO$^\star$. It can successfully reproduce the high-$P_T$ NLO calculation for $\ss$ production.~\footnote{Besides the single $\ss$ production, NLO$^\star$ cut was also applied to $\ss+\ss$ hadroproduction in Ref~\cite{Lansberg:2013qka}. NLO$^\star$ calculation is able to well reproduce the complete NLO result~\cite{Sun:2014gca} in the double charmonium/bottomonium production. Its good performance may rely on the fact that like the single $\ss$ production, the LO SDC of $\ss+\ss$ is also NNLP in $P_T$ in the large transverse momentum region.} The same infrared cut can be imposed in the phase-space integration of the $\mathcal{O}(\alpha_s^5)$ tree-level matrix element. Another giant K factor was observed compared to the NLO calculation at high $P_T$, which may question on the extractions of color-octet LDMEs based on NLO calculations~\cite{Butenschoen:2010rq,Ma:2010yw,Chao:2012iv,Gong:2012ug,Bodwin:2014gia}. In contrast, a suspicion in Ref.~\cite{Ma:2010jj} on the size of $\mathcal{O}(\alpha_s^5)$ was given from their $P_T$ scaling reanalysis of the NNLO$^\star$ curves. Instead of the $P_T$ power enhancement, the observed giant K factor $\frac{d\sigma^{\rm NNLO^\star}}{d\sigma^{\rm NLO}}$ is mainly attributed to the introduction of the infrared cutoff. Therefore, a reliable estimate of the size of $\mathcal{O}(\alpha_s^5)$ is still missing. It is necessary to clarify the situation before drawing a solid conclusion.

The aim of this paper is to introduce an infrared-safe method to cure the problematic giant K factors appearing in the SDC calculations in particular for high-$P_T$ quarkonium production without performing complete higher-order calculations.~\footnote{In the processes of elementary particle production, a few proposals to cure the giant K factors, which are mainly from logarithmic terms in perturbative calculations, are present~\cite{Caravaglios:1998yr,Mangano:2001xp,Catani:2001cc,Krauss:2002up,Lonnblad:2001iq,
Lavesson:2005xu,Lonnblad:2011xx,Hamilton:2010wh,Hoche:2010kg,Lavesson:2008ah,Lonnblad:2012ix,
Gehrmann:2012yg,Hoeche:2012yf,Frederix:2012ps,Lonnblad:2012ng,Hamilton:2013fea,Rubin:2010xp}. Unfortunately, none of them is straightforwardly applicable to the power-enhanced contributions in quarkonium production.} In contrast to the NLO$^\star$ calculations, the new method will not introduce the logarithmic dependence from the infrared cutoff. The estimate of the missing higher orders is to use the conventional renormalization and factorization scale variations. It is complemented with the fragmentation function approach, which requires the analytical calculations of different single- and double-parton fragmentation functions for single and multiple quarkonium production. Another nontrivial task to use the fragmentation function approach is to solve the corresponding coupled evolution equations. It has been shown in Ref.~\cite{Ma:2014svb} that the fragmentation function approach without scale evolution can reproduce the spin-summed NLO cross sections of $\ss,\so,\sps,\pj$ at high $P_T$, which shows the necessity of taking into account both the single-parton (at LP) and the double-parton (at NLP) fragmentation contributions. The factorization theorem for the single-inclusive quarkonium production cross sections in terms of single- and double-parton fragmentation functions was first proven in Ref.~\cite{Kang:2011mg} under the assumption of perturbative QCD factorization. 

There are also other appealing reasons to introduce such a method. First of all, it can be used to stabilize the higher-order QCD corrections in quarkonium associated production processes, where most of them are absent of complete NLO calculations. The possible cancellations between S-wave and P-wave are guaranteed in our approach. For instance, in the double $J/\psi$ at the LHC, it requires a NNLO calculation to have the full cancellations between S-wave and P-wave Fock states. As we will see later in this paper, the good reproduction of the NLO results both in the spin-summed and spin-dependent cross sections for single quarkonium production at high $P_T$ can serve as a fast way to the future phenomenology studies. In practice, the phenomenology from a complicated calculation scales as an inverse power of the computation time.

The outline of the remaining context is following. After introducing the remainders of P-wave counterterms in section ~\ref{sec:pCTs}, we will show that one can reproduce the NLO results for most of the Fock states (except $\so$) with fairly simple cuts based on tree-level matrix elements in section \ref{sec:simpleNLO}. These simple cuts are not sufficient to remove large logarithms introduced by the phase space cut parameters. Hence, a general infrared-safe method is introduced to obtain the giant K factors for all the Fock states relevant for $J/\psi$ and $\chi_{cJ}$ production in section \ref{sec:generalSTOP}. Finally, we draw our conclusions in section \ref{sec:summary}. An instruction on how to use {\sc\small HELAC-Onia}~\cite{Shao:2012iz,Shao:2015vga} to perform the calculations done in this paper is given in appendix \ref{app:helaconia}. The appendix \ref{app:moreplots} contains supplemental figures.

\section{Remainders of P-wave counterterms\label{sec:pCTs}}

It is well-known that the remaining infrared divergences in the SDC computations for the productions and decays of P-wave Fock states should be cancelled by the P-wave counterterms arising from the renormalization group running of S-wave LDMEs beyond LO in $\alpha_s$, which is analogous to the remaining collinear divergences absorbed by the PDF counterterms in a peturbative QCD calculation. The renormalization of NRQCD operators links the S-wave LDMEs with the P-wave LDMEs as shown in Eq.(150) of Ref.~\cite{Petrelli:1997ge}. Such counterterms, after cancelling infrared divergences with the real and virtual matrix elements, will leave finite remainders proportional to the S-wave  SDCs and P-wave LDMEs. The introduction of the P-wave counterterms is crucial especially in the case that the S-wave SDCs are much larger than the P-wave SDCs. In particular, the negative P-wave SDCs for heavy quarkonium hadroproduction at high $P_T$ could be attributed to these negative remainders. We have implemented the following finite remainders of P-wave counterterms:
\begin{eqnarray}
d\sigma^{\mathcal{C}}(\pj)&=&d\hat{\sigma}^{\rm Born}(\ss)\times\left(\frac{4}{3}\frac{\alpha_s}{\pi}\frac{\log{\frac{m_Q^2}{4\mu_{\Lambda}^2}}}{m_Q^2}\right)\times \langle \mathcal{O}(\pj) \rangle\nonumber\\
&&+d\hat{\sigma}^{\rm Born}(\so)\times\left(\frac{5}{9}\frac{\alpha_s}{\pi}\frac{\log{\frac{m_Q^2}{4\mu_{\Lambda}^2}}}{m_Q^2}\right)\times \langle \mathcal{O}(\pj) \rangle,\nonumber\\
d\sigma^{\mathcal{C}}(\pjs)&=&d\hat{\sigma}^{\rm Born}(\so)\times\left(\frac{8}{27}\frac{\alpha_s}{\pi}\frac{\log{\frac{m_Q^2}{4\mu_{\Lambda}^2}}}{m_Q^2}\right)\times \langle \mathcal{O}(\pjs) \rangle,\nonumber\\
d\sigma^{\mathcal{C}}(\opoo)&=&d\hat{\sigma}^{\rm Born}(\spo)\times\left(\frac{4}{3}\frac{\alpha_s}{\pi}\frac{\log{\frac{m_Q^2}{4\mu_{\Lambda}^2}}}{m_Q^2}\right)\times \langle \mathcal{O}(\opoo) \rangle\nonumber\\
&&+d\hat{\sigma}^{\rm Born}(\sps)\times\left(\frac{5}{9}\frac{\alpha_s}{\pi}\frac{\log{\frac{m_Q^2}{4\mu_{\Lambda}^2}}}{m_Q^2}\right)\times \langle \mathcal{O}(\opoo),\nonumber\\
d\sigma^{\mathcal{C}}(\opos)&=&d\hat{\sigma}^{\rm Born}(\sps)\times\left(\frac{8}{27}\frac{\alpha_s}{\pi}\frac{\log{\frac{m_Q^2}{4\mu_{\Lambda}^2}}}{m_Q^2}\right)\times \langle \mathcal{O}(\opos) \rangle,
\end{eqnarray} 
where $m_Q$ is the mass of the heavy quark and $\mu_{\Lambda}$ is the NRQCD scale. In the following, we will set $\mu_{\Lambda}=m_Q$ as usually done in the complete NLO calculations. These remainders have already been implemented in the {\sc\small HELAC-Onia}~\cite{Shao:2012iz,Shao:2015vga}. They are necessary ingredients to reproduce the NLO results, which we will show in the following two sections.

\section{A first step towards NLO\label{sec:simpleNLO}}

From the discussion in the section~\ref{sec:intro}, it is known that large NLO QCD corrections to the $J/\psi$ production at a high-energy hadron collider are mainly due to the emergence of new $P_T$ power-enhanced fragmentation contributions. Hence, all S- and P-wave Fock states except $\so$ receive giant K factors from NLO QCD calculations. 

We first introduce the following basic phase space cuts in order to take into account the hard radiations without using virtual amplitudes. In real part at $\mathcal{O}(\alpha_s^4)$, exact $2$  light-flavoured jets~\footnote{We mean ``light-flavoured jet" here as a cluster of gluon, up, down, strange (anti-)quarks. Similarly, the light-flavoured partons are defined as gluon, up, down, strange (anti-)quarks.} satisfying $P_T(j)>P_{T}^{\rm min}$ and $|y(j)|<y^{\rm max}$ are required, which is denoted as $d\sigma^{\mathcal{R}_0}$. The phase space integrations of Born $d\sigma^{\mathcal{B}}$ ($\mathcal{O}(\alpha_s^3)$) and the remainders of the NRQCD P-wave counterterms $d\sigma^{\mathcal{C}}$ ($\mathcal{O}(\alpha_s^4)$) are infrared safe with $P_T(\rm{onium})$ larger than a given positive value $P_T^{\rm min}(\rm{onium})$. We call the summed results of  $d\sigma^{\mathcal{B}}+d\sigma^{\mathcal{R}_0}+d\sigma^{\mathcal{C}}$ as approximated NLO (aNLO).

In the following, we take $P_T^{\rm min}(\rm{onium})=5$ GeV, and light-flavoured jets are clustered with anti-$k_T$ algorithm~\cite{Cacciari:2008gp} using radius $R=0.5$ and $|y(j)|<5,P_T(j)>P_{T}^{\rm min}$ by {\sc\small FastJet}~\cite{Cacciari:2011ma}. We will vary $P_{T}^{\rm min}$ from 3 GeV to 6 GeV as a way to estimate the infrared-cut dependence. We have shown  the spin-summed double differential distributions for the $c\bar{c}$ Fock state $\ss$ in Fig.~\ref{Fig:aNLOvsNLO} with the $13$ TeV proton-proton collisions, while the distributions for the 5 Fock states $\sps,\pj,\tpzs,\tpos,\tpts$ are displayed in Fig.~\ref{Fig:aNLOvsNLO2} as our supplemental material. The complete NLO curves (denoting as NLO) from Refs.~\cite{Ma:2010vd,Ma:2010yw} are also shown in order to have a comparison. The red-hatched bands represent the infrared cut variations $P_{T}^{\rm min}\in\ [3, 6]$ GeV, and the grey bands are the uncertainty from the independent variations of renormalization and factorization scales $\mu_R,\mu_F$ around the central value $\mu_0=\sqrt{P_{T}^2({\rm{onium}})+4m_c^2}$ by a factor of 2. It is interesting to notice that the scale uncertainty in general captures the missing virtual and soft/collinear pieces. The agreements between NLO and aNLO are improved as $P_{T}({\rm{onium}})$ increases. A similar behaviour can be observed for the spin-dependent differential cross sections shown in Fig.~\ref{Fig:aNLOvsNLOSpin} for $\ss$ and in Figs.~\ref{Fig:aNLOvsNLOSpin2}, \ref{Fig:aNLOvsNLOSpin3} for $\pj,\tpos,\tpts$ Fock states, where the NLO curves are from Refs.~\cite{Chao:2012iv,Shao:2014fca}. The spin-density matrix elements of the scalars $\sps,\tpzs$ are trivial.~\footnote{The spin-density matrix elements $\frac{d\sigma_{J_zJ_z}}{dP_T}$ shown in this paper are defined in the usual helicity frame.} We have utilized  CTEQ6M PDF~\cite{Pumplin:2002vw} to be consistent with the NLO results. For the reproducible purpose, the values of LDMEs for the distributions of the Fock states are listed in Table~\ref{tab:LDMEs}.

\begin{figure}[ht!]
\vspace{-1cm}
\centering
\includegraphics[width=.99\textwidth,draft=false]{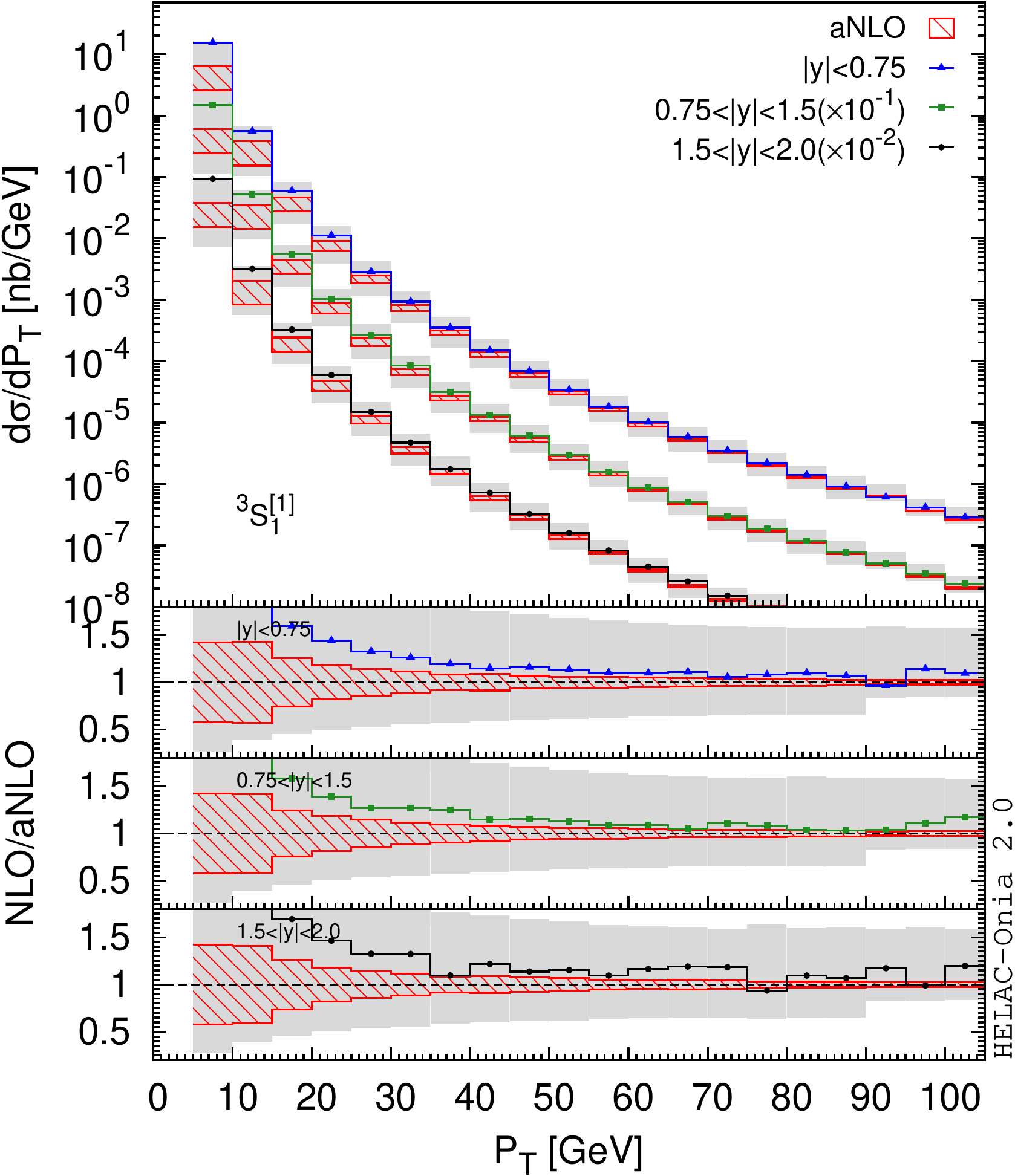}
\caption{Comparison of spin-summed differential cross sections for the Fock state $\ss$ between our aNLO calculations and the complete NLO calculations.\label{Fig:aNLOvsNLO}}
\end{figure}

\begin{figure}[ht!]
\centering
\includegraphics[width=.45\textwidth,draft=false]{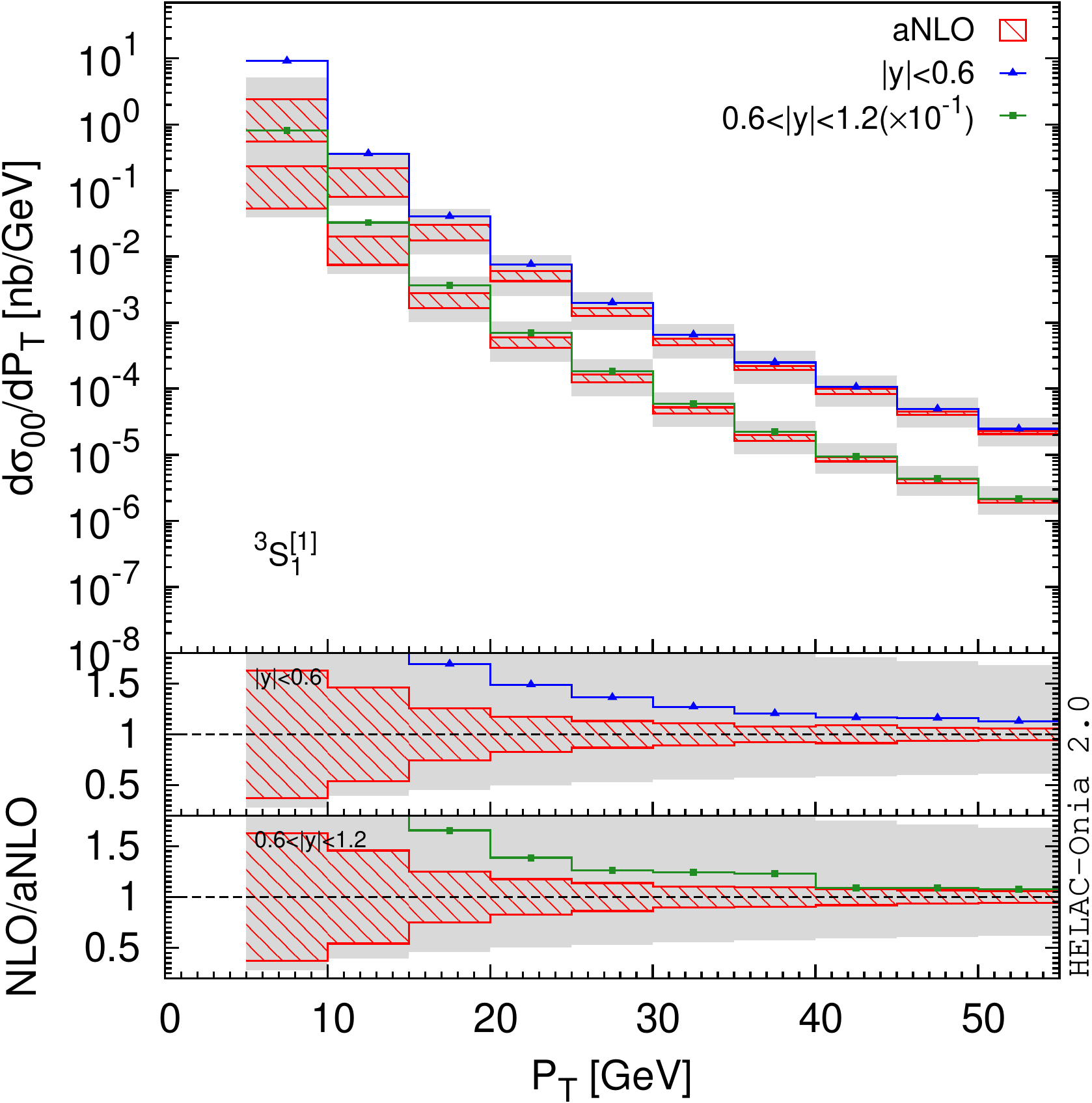}
\includegraphics[width=.45\textwidth,draft=false]{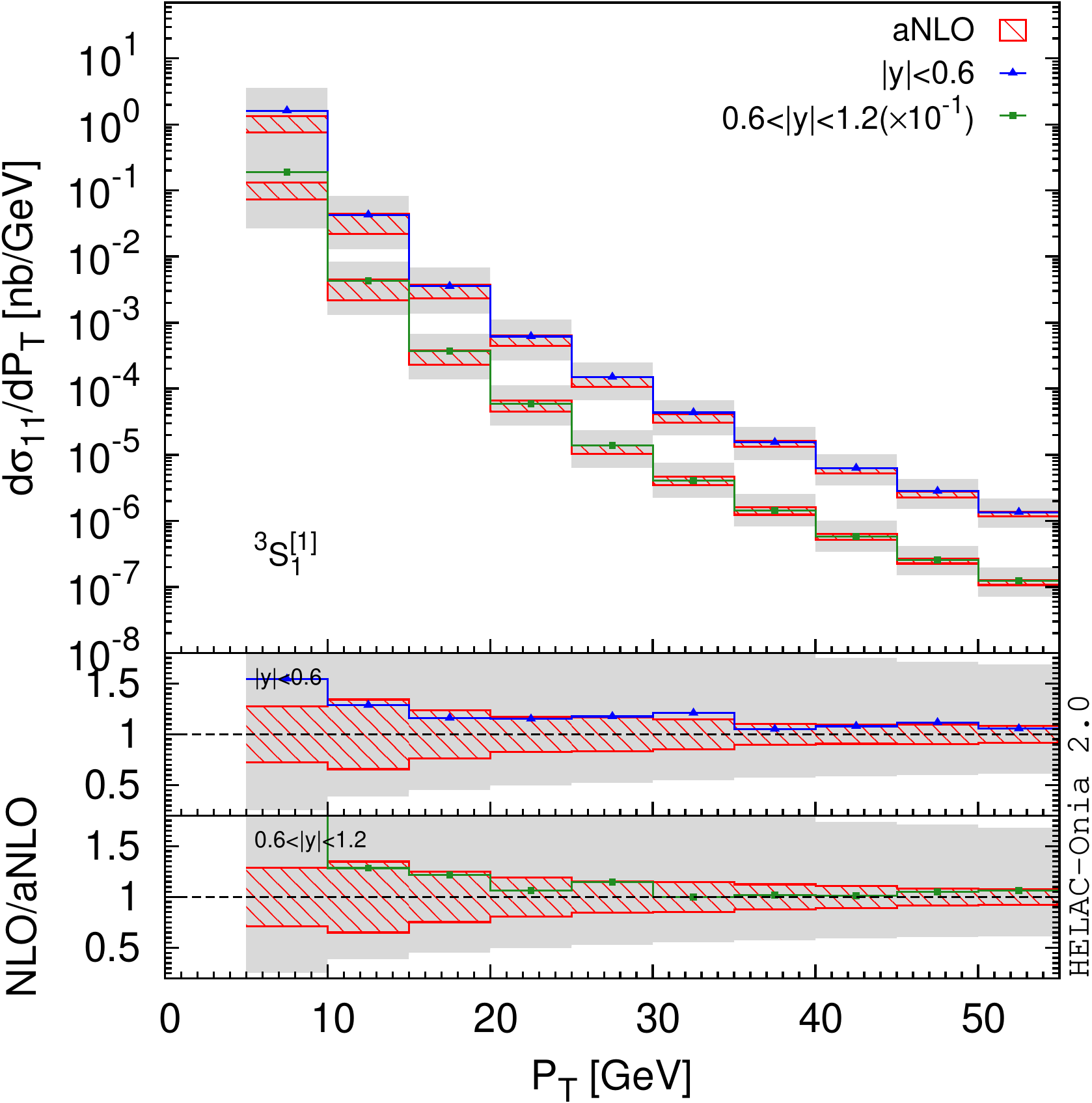}
\caption{Comparisons of spin-dependent differential cross sections for the Fock state $\ss$ between our aNLO calculations and the complete NLO calculations.\label{Fig:aNLOvsNLOSpin}}
\end{figure}

\begin{table}[ht!]
\begin{center}
\begin{tabular}{{|c|}*{5}{c|}}\hline
$\langle\mathcal{O}(\ss)\rangle$ & $\langle\mathcal{O}(\so)\rangle$ & $\langle\mathcal{O}(\sps)\rangle$ & $\langle\mathcal{O}(\pj)\rangle/(2J+1)$ & $\langle\mathcal{O}(\pjs)\rangle/(2J+1)$\\\hline
$1.16$ GeV$^3$ & $9.03\cdot 10^{-3}$ GeV$^3$ & $1.46\cdot 10^{-2}$ GeV$^3$ & $3.43\cdot 10^{-2}$ GeV$^5$ & $0.107$ GeV$^5$
\\\hline 
\end{tabular}
\end{center}
\caption{The values of LDMEs used in the differential distributions of various Fock states.\label{tab:LDMEs}}
\end{table}

Because the LP in $P_T$ for $\so$ already exists at Born $d\sigma^{\mathcal{B}}$ (i.e. $\mathcal{O}(\alpha_s^3)$) from the gluon fragmentation, it is expected that the scale uncertainty at LO would already give a reliable estimate of the missing NLO QCD corrections, which is indeed observed from the left-panel of Fig.~\ref{Fig:aNLO3S18vsNLO}. In such a case, a request of $2$ light-flavoured jets in the computation of $d\sigma^{\mathcal{R}_0}$ is insufficient to obtain an infrared-safe differential cross section. From the right-panel of Fig.~\ref{Fig:aNLO3S18vsNLO}, the aNLO $P_T$ spectra are too hard compared to the complete NLO ones. The reason is because of the large logarithms arising from the very asymmetric dijet system $P_T(j_1)\gg P_T(j_2)$. Such a configuration is suppressed in other Fock states, because the leading fragmentation topologies require at least one light-flavoured parton along with the quarkonium direction at high $P_T$. The weights of the asymmetric dijet events will be enhanced due to the unphysical logarithm $\log{\frac{P_T(j_1)}{P_T(j_2)}}$ in the aNLO calculations of $\so$, which should be in principle cancelled by the virtual contributions because of the unitarity. Therefore, one must introduce a more general infrared-safe method to avoid these large logarithms, and at meantime one should maintain the hard radiations from the real contributions. 

\begin{figure}[hbt!]
\centering
\includegraphics[width=.45\textwidth,draft=false]{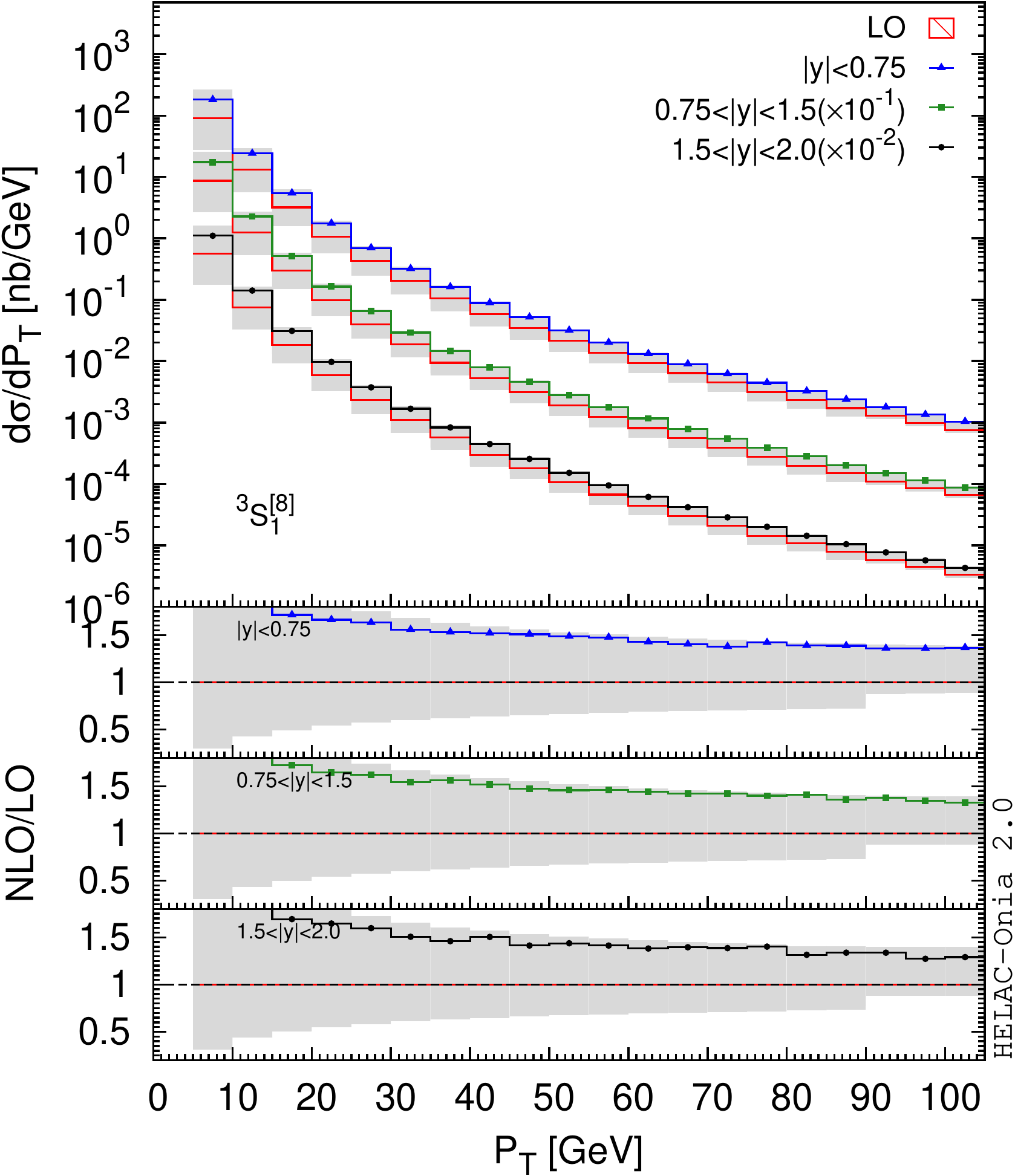}
\includegraphics[width=.45\textwidth,draft=false]{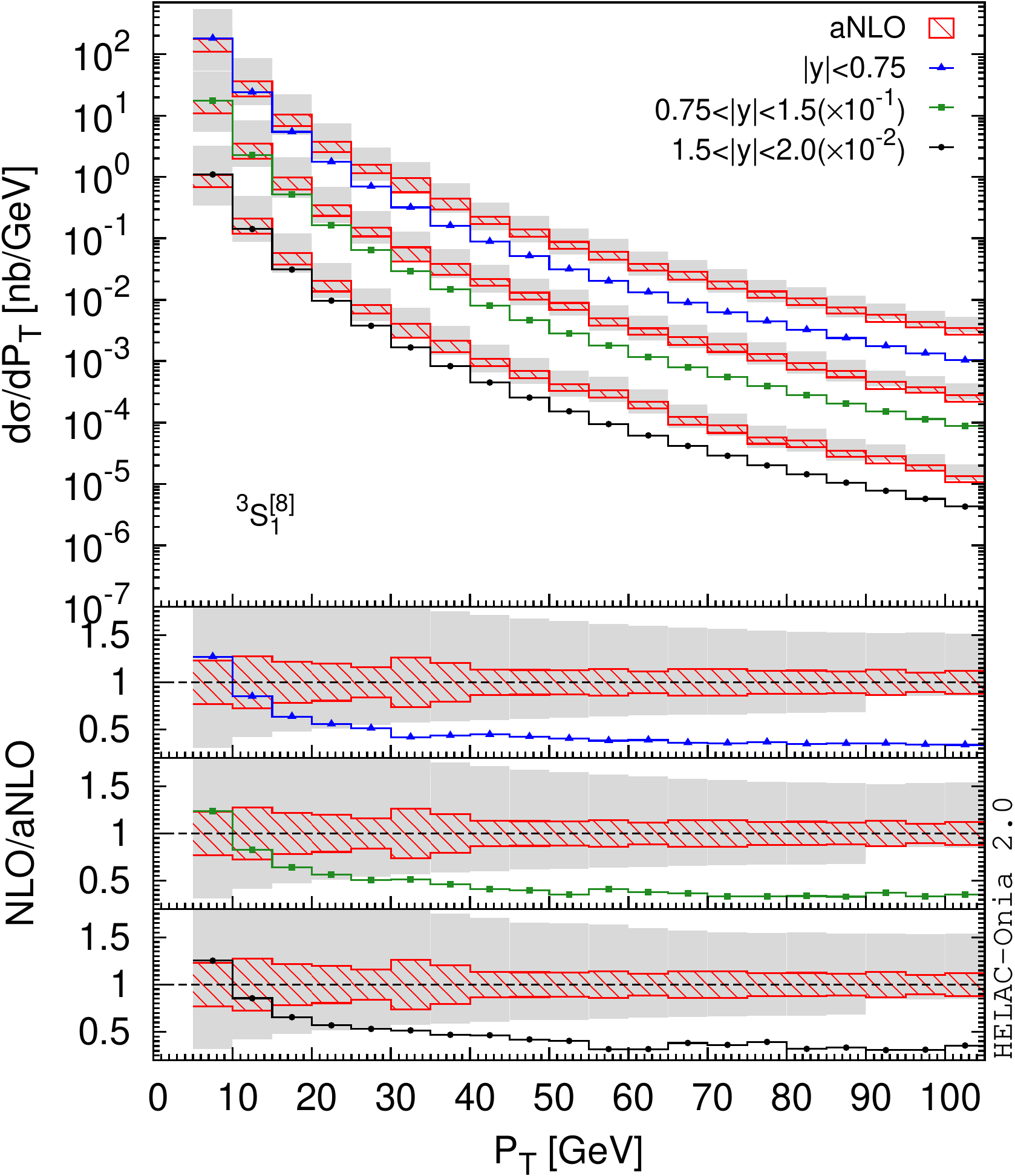}
\caption{Comparisons of spin-summed differential cross sections for the Fock state $\so$ between LO (left), aNLO (right) calculations and the complete NLO calculations.\label{Fig:aNLO3S18vsNLO}}
\end{figure}

\section{A general infrared-safe method\label{sec:generalSTOP}}

\subsection{Infrared-safe cutoffs\label{sec:stopcuts}}

\begin{figure}[hbt!]
\centering
\includegraphics[width=.90\textwidth,draft=false]{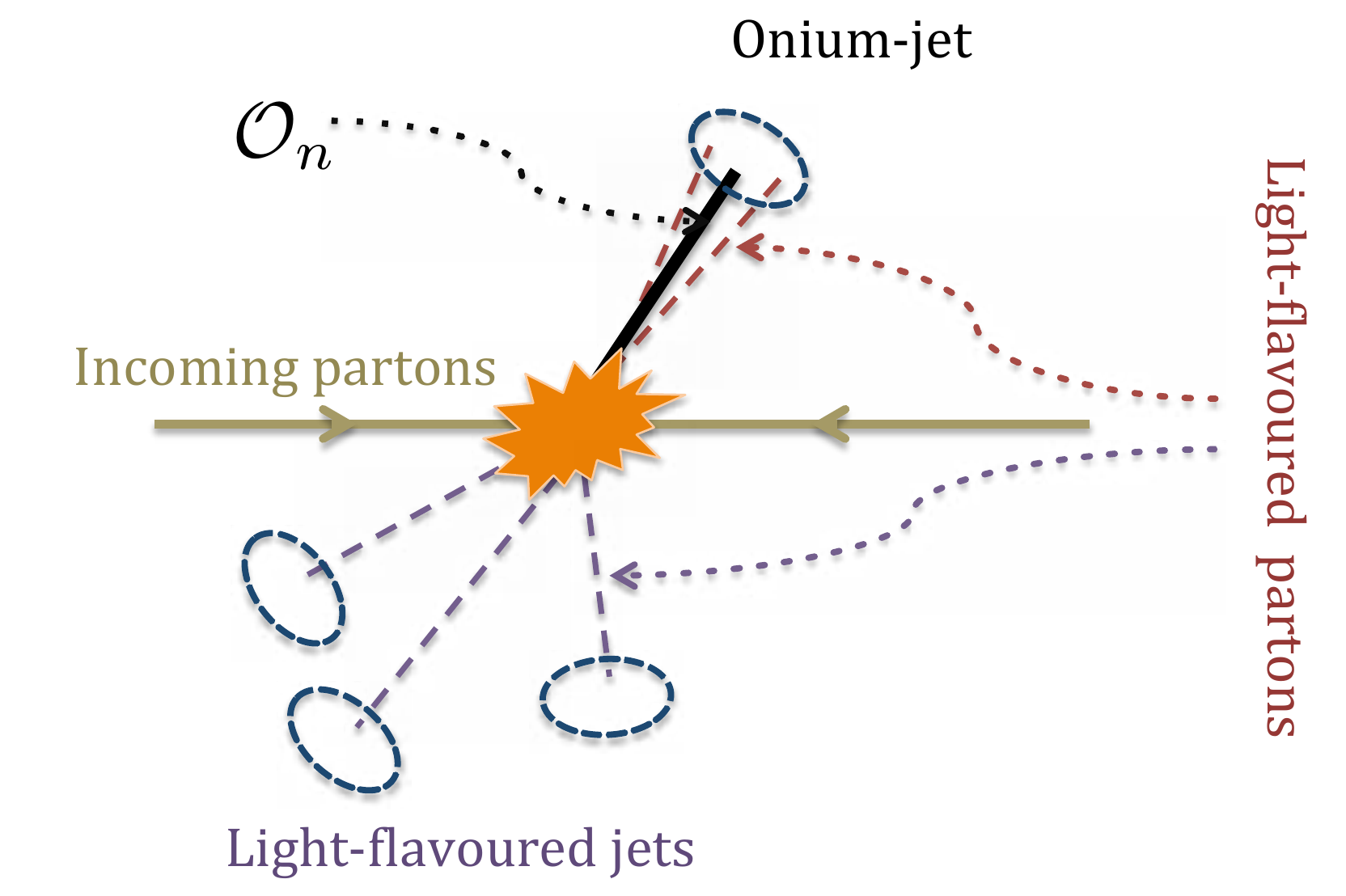}
\caption{Schematic depiction of inclusive quarkonium ${\cal O}_n$ production.\label{Fig:pp2oniumX}}
\end{figure}

\begin{figure}[hbt!]
\centering
\includegraphics[width=.90\textwidth,draft=false]{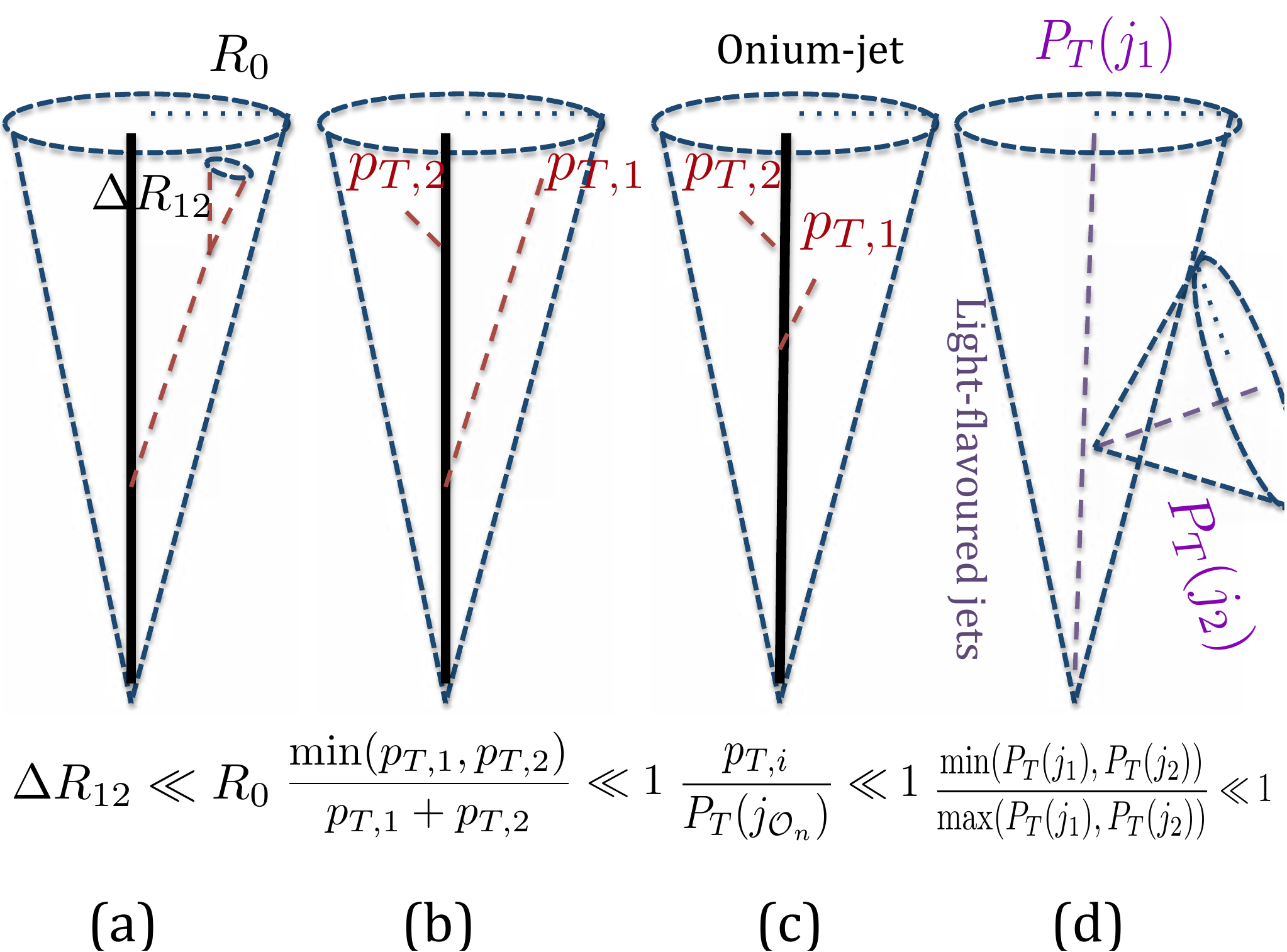}
\caption{Infrared unsafe configurations to be considered in inclusive quarkonium production, where the first 3 subfigures are for the onium jet $j_{{\cal O}_n}$ and the last one is for the light-flavoured jets.\label{Fig:infraredconfigs}}
\end{figure}

Let us assume a LO fragmentation process for a given Fock state ${\cal O}_n$ is accompanying with $k$ final massless partons:
\begin{eqnarray}
{\cal F}: p_0 \rightarrow P_{{\cal O}_n}+p_1+\cdots+p_k. \label{eq:fragprocess}
\end{eqnarray}
For a given observable, one needs to consider ${\cal O}_n$ plus $i$ recoiling partons. For example, in the case of the transverse-momentum distribution for a single quarkonium production (schematically depicted in Fig.~\ref{Fig:pp2oniumX}), the quarkonium at least recoils against one parton at the lowest order (bar the zero transverse momentum bin). The soft- and collinear-safe calculations can be achieved based on pure tree-level matrix elements via the following conditions:
\begin{enumerate}
\item The number of jets is larger than $i+1$ with the transverse momentum of jet $P_T(j)>P_{T}^{\rm min}$ and the rapidity $|y(j)|<y^{\rm max}$. ${\cal O}_n$ is also taken into account in the jet-clustering procedure. One should make sure that there is exactly one jet containing ${\cal O}_n$ passing the above $P_T$ and rapidity cuts. Such a jet is called an {\it onium-jet} here.
\item In the onium-jet, there are at least $k$ light-flavoured partons to fulfill the above fragmentation process. Let us say there are $m$ light-flavoured partons inside the onium-jet with $m\ge k$.~\footnote{At LO one should have $m=k$ since the configuration of $i+k-m<i$ recoiling partons is zero by definition for the given observable when $m>k$.}
\item If $m \ge 2$, each pair of parton $1$ and $2$ inside the onium-jet should pass the following soft drop condition~\cite{Larkoski:2014wba}
\begin{eqnarray}
\frac{{\rm min}(p_{T,1},p_{T,2})}{p_{T,1}+p_{T,2}} > z_{\rm cut}\left(\frac{\Delta R_{12}}{R_0}\right)^\beta~\label{eq:softdropcut}
\end{eqnarray}
where $p_{T,i}$ is the transverse momentum of parton $i$ and $\Delta R_{12}=\sqrt{\Delta \phi_{12}^2+\Delta y_{12}^2}$. The above cut already excludes the soft singularity as long as $z_{\rm cut}>0$, while the requirement of the collinear safety is guaranteed by choosing $\beta<0$. $R_0$ is the original jet radius, which is an order one number.
\end{enumerate}
The condition Eq.(\ref{eq:softdropcut}) in item 3 is chosen to kill the infrared unsafe configurations (a) and (b) given in Fig.~\ref{Fig:infraredconfigs}. Either when partons $1$ and $2$ are close to be collinear $\Delta R_{12}\ll R_0$ or if one parton is soft $p_{T,2}\ll p_{T,1}$, Eq.(\ref{eq:softdropcut}) cannot be fulfilled when $z_{\rm cut}>0,\beta<0$. In practice, the absolute value of $\beta$ is at order one and $z_{\rm cut}$ is at the order of $v^2$.

If one goes to extra $o$ radiations (i.e. ${\cal O}_n$ plus $i+k+o$ final light-flavoured QCD partons), one should impose the following additional cuts:
\begin{enumerate}
\item[4.]There are $i+k+o-m$ partons outside the onium-jet. Each parton should form a single jet within $P_T(j)>P_{T}^{\rm min}$ and $|y(j)|<y^{\rm max}$ to avoid the collinear divergences. In order to get rid of large logarithms from infrared cuts arising from the soft large-angle radiations illustrated in the case (d) in Fig.~\ref{Fig:infraredconfigs}, when $i+k+o-m\ge 2$, it is necessary to impose an asymmetric cut on these parton jets
\begin{eqnarray}
\frac{{\rm min}\left(P_T(j_1),\cdots, P_T(j_{i+k+o-m})\right)}{{\rm max}\left(P_T(j_1),\cdots, P_T(j_{i+k+o-m})\right)}> z_{\rm cut, a}.
\end{eqnarray}
The value of $z_{\rm cut,a}$ should be a positive number smaller than 1 but not close to 0. It is important to vary its value in order to assess this cut dependence.
\item[5.]If $k=0$~\footnote{For example, ${\cal O}_n=Q\bar{Q}({}^3S_1^{[8]})$ has $g\rightarrow Q\bar{Q}({}^3S_1^{[8]})$ fragmentation at LO.} and $m>0$, each parton $1$ in the onium jet $j_{{\cal O}_n}$ should pass the soft cut
\begin{eqnarray}
z_1> z_{\rm cut,s}
\end{eqnarray}
where $z_1$ can be the energy fraction $\frac{E_1}{E(j_{{\cal O}_n})}$, the transverse mass fraction $\frac{p_{T,1}}{\sqrt{P_T^2(j_{{\cal O}_n})+M^2(j_{{\cal O}_n})}}$, the transverse momentum fraction $\frac{p_{T,1}}{P_T(j_{{\cal O}_n})}$ or other similar fractions corresponding to $z_1\rightarrow 0$ when the parton $1$ is soft. This condition is needed in order to kill the case (c) in Fig.~\ref{Fig:infraredconfigs}, where all light-flavoured partons in the onium-jet $j_{{\cal O}_n}$ can be soft and the condition Eq.(\ref{eq:softdropcut}) is still satisfied. Similar to the value of $z_{\rm cut}$ in Eq.(\ref{eq:softdropcut}), the proper value of $z_{\rm cut,s}$ should be $\mathcal{O}(v^2)$ as the effect of the soft radiations should be absorbed into the long-distance part of the quarkonium. 
\end{enumerate}

In fact, the combination of items 1-5 introduces a general infrared-safe method for any Fock state production if $k=0$ is assumed at the beginning.~\footnote{When $k=0$, the cut in item 2 will not be applied.} In other words, we do not need to pay a special attention to which kind of fragmentation process $\mathcal{F}$ is allowed for a given Fock state. We call such cuts as STOP cuts, where ``STOP" is an acronym of ``STabilize quarkOnium Production".

In the case of the $P_T$ spectrum of a quarkonium $\mathcal{O}_n$ production at a hadron collider, $i$ is equal to $1$ and the LO process is $\mathcal{O}_n$ plus one parton. For a real emission process $\mathcal{O}_n$ plus $o+1$ partons with $o>0$, we should impose the cuts listed in items 1-5 with $k=0$, where the condition in item 2 is fulfilled automatically. Same as the previous section, we will denote the Born contribution at $\mathcal{O}(\alpha_s^3)$ as $d\sigma^{\mathcal{B}}$ and the remainders of the P-wave counterterms at $\mathcal{O}(\alpha_s^4)$ as $d\sigma^{\mathcal{C}}$. $d\sigma^{\mathcal{R}_{\rm STOP}}$ ($d\sigma^{\mathcal{R}^2_{\rm STOP}}$) stands for the contribution from $\mathcal{O}_n$ plus two (three) partons within the STOP cuts.

\subsection{Reproducing NLO results\label{sec:NLOStop}}

In this section, we will present the results up to NLO QCD corrections (i.e. $\mathcal{O}(\alpha_s^4)$). In order to differentiate our partial NLO calculations with the complete NLO results, we will denote our partial NLO calculations by imposing STOP cuts as ``nLO", i.e. $d\sigma^{\rm nLO}\equiv d\sigma^{\mathcal{B}}+d\sigma^{\mathcal{C}}+d\sigma^{\mathcal{R}_{\rm STOP}}$. In the following, we will illustrate that the complete NLO results can be reproduced with the tree-level generators under the following setup of the STOP cuts:
\begin{eqnarray}
&&P_T(j)>P_T^{\rm min}, P_T^{\rm min}\in\ [3, 6]~{\rm GeV}, |y(j)|<5.0,\nonumber\\
&&z_{\rm cut}=0.1, \beta=-1, R_0=1.0,\nonumber\\
&&z_{\rm cut,a} \in\ [0.1,0.7], z_{\rm cut,s}=\frac{0.1}{m},
\end{eqnarray}
where $m$ is the number of light-flavoured partons inside the onium jet. Jets are reconstructed with the anti-$k_T$ clustering algorithm using {\sc\small FastJet}. Since there is no infrared divergence in the Born after imposing $P_T({\rm onium})>0$ cut, the STOP cuts will not be applied to the Born and Born-like counterterm events.

\subsubsection{Reproducing $\so$}

After imposing the STOP cuts on $\so$, we can reproduce the complete NLO curves within the theoretical uncertainties. They are shown in Fig.~\ref{Fig:nLO3S18vsNLO} and Fig.~\ref{Fig:nLO3S18vsNLOSpin} for the spin-summed and spin-dependent differential cross sections respectively. In the left panel of Fig.~\ref{Fig:nLO3S18vsNLO} and the upper panels of Fig.~\ref{Fig:nLO3S18vsNLOSpin}, we estimate infrared cutoff dependence (the red-hatched bands) via the combined variations of $P_T^{\rm min}\in\ [3, 6]$ GeV and $z_{\rm cut,a} \in\ [0.1,0.7]$. The grey-shadowed bands represent the scale uncertainties. Opposed to the aNLO results in the right panel of Fig.~\ref{Fig:aNLO3S18vsNLO}, it indeed shows that the STOP cuts improve the perturbative calculations, and the transverse-momentum dependence in $\frac{d\sigma^{\rm nLO}}{dP_T}$ is the same as the NLO distributions $\frac{d\sigma^{\rm NLO}}{dP_T}$. It demonstrates that the large logarithmic dependence from the simple cuts in section \ref{sec:simpleNLO} disappears after imposing the STOP cuts. The STOP-cut dependence (the red-hatched bands) is not reduced by increasing the $P_T$ of the quarkonium. It is expected since the LP contribution is already present at LO. The variations of the STOP cut variables only alter the fractions of hard radiations in the real matrix elements, which are not logarithmically enhanced. In fact, a careful tuning of STOP cut parameters can reproduce the NLO results at high precision. In the right panel of Fig.~\ref{Fig:nLO3S18vsNLO} and the lower panels of Fig.~\ref{Fig:nLO3S18vsNLOSpin}, we have calculated the $\so$ differential distributions after using $z_{\rm cut,a}=0.6$ and $z_{\rm cut,s}=\frac{0.2}{m}$. The comparisons to the full NLO calculations imply that the $P_T$ spectra of $\so$ in different rapidity intervals can be precisely reproduced as long as $P_T(\so)>10$ GeV.

\begin{figure}[ht!]
\centering
\includegraphics[width=.45\textwidth,draft=false]{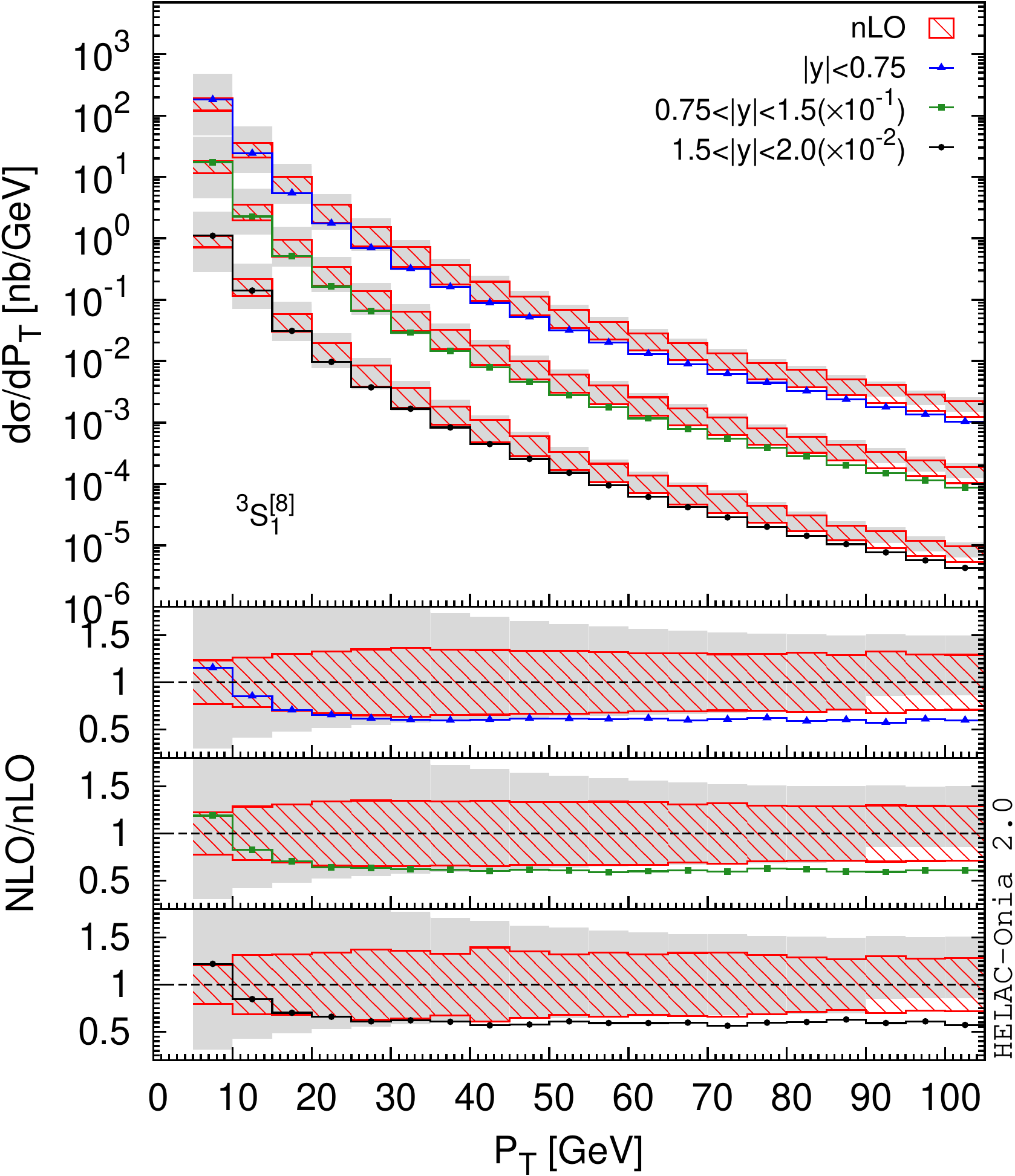}
\includegraphics[width=.45\textwidth,draft=false]{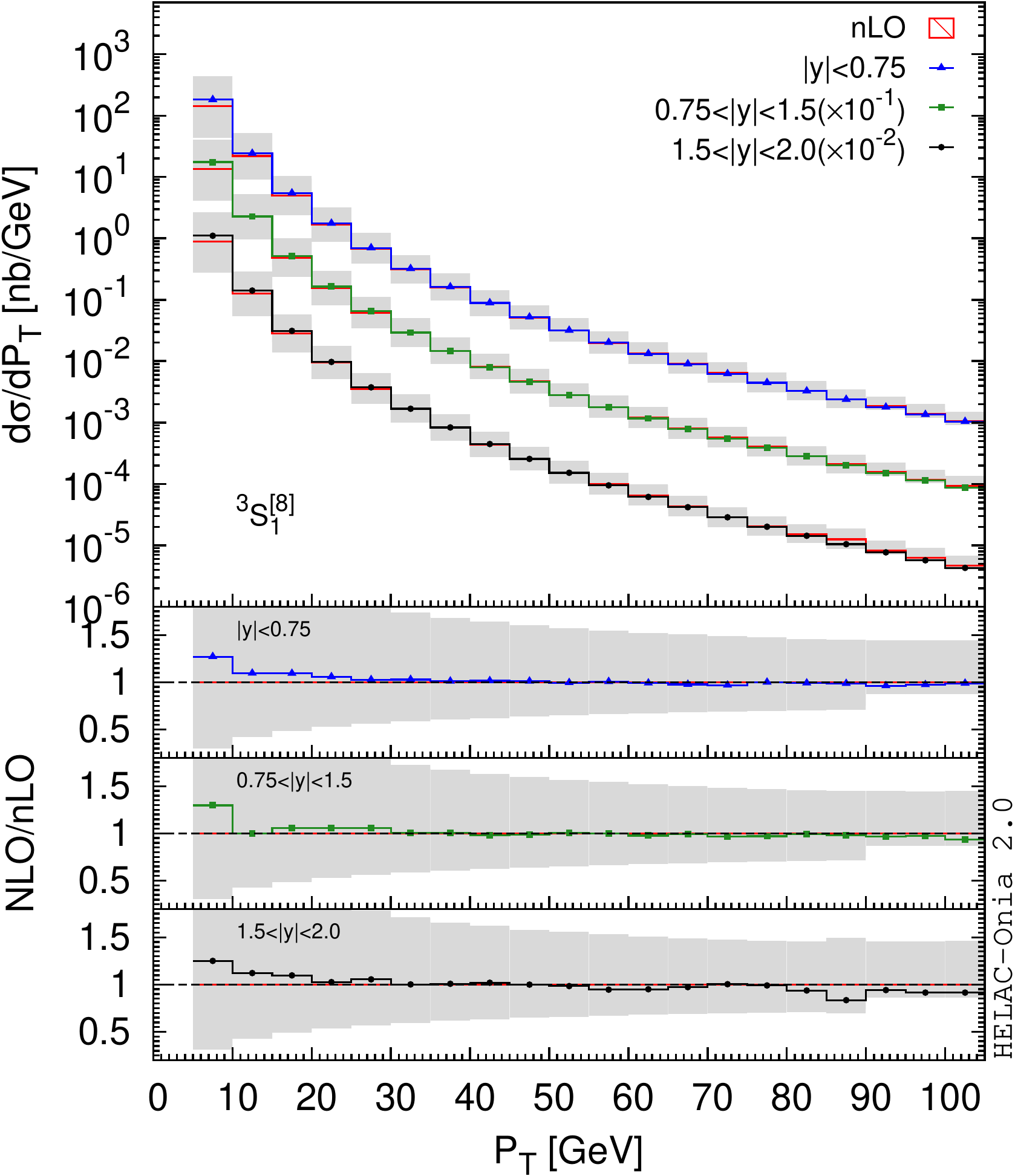}
\caption{Comparisons of spin-summed differential cross sections for the Fock state $\so$ between nLO (left), tunned nLO with $z_{\rm cut,a}=0.6, z_{\rm cut,s}=\frac{0.2}{m}$ (right) calculations and the complete NLO calculations.\label{Fig:nLO3S18vsNLO}}
\end{figure}

\begin{figure}[ht!]
\centering
\includegraphics[width=.45\textwidth,draft=false]{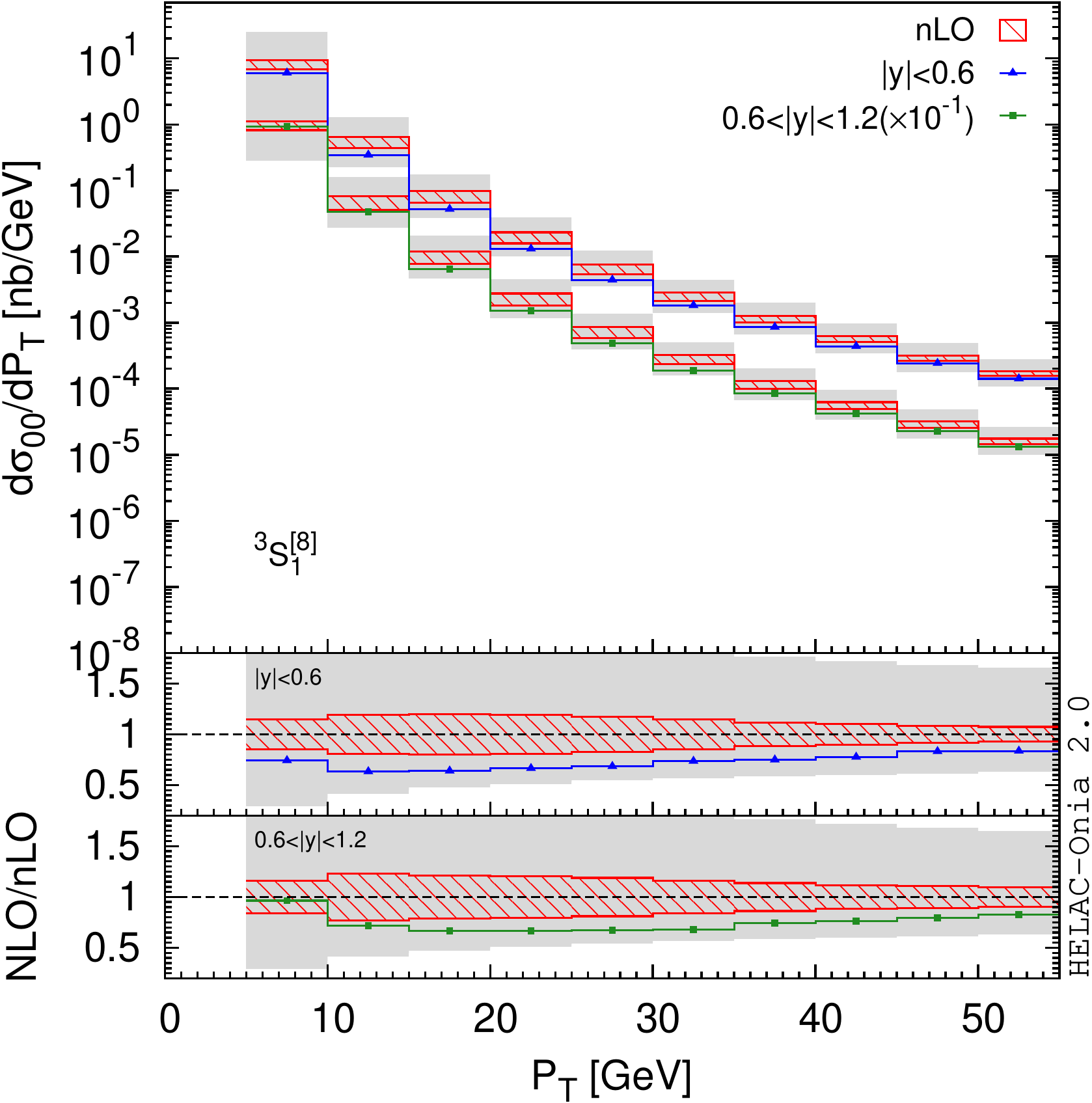}
\includegraphics[width=.45\textwidth,draft=false]{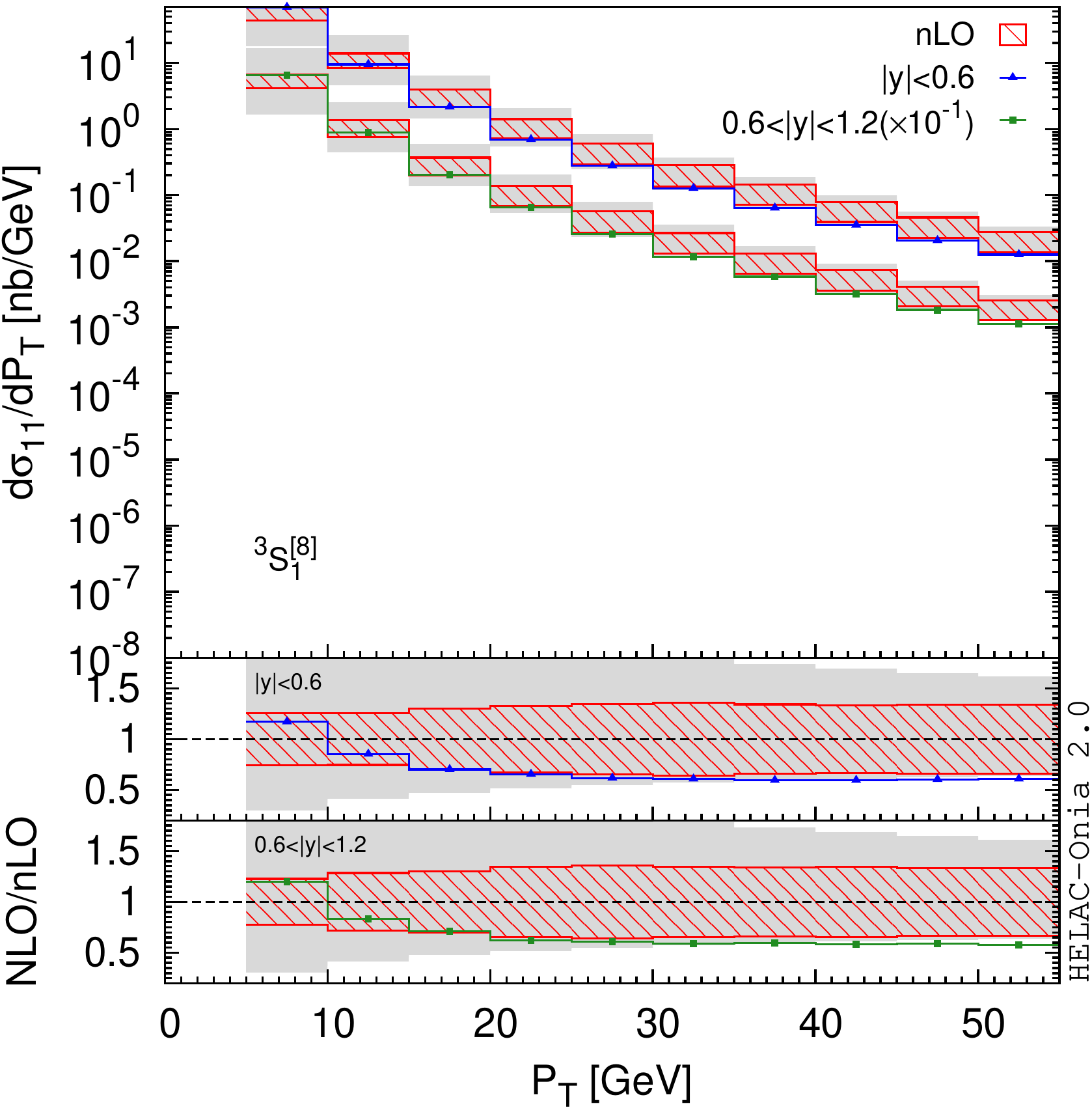}\\
\includegraphics[width=.45\textwidth,draft=false]{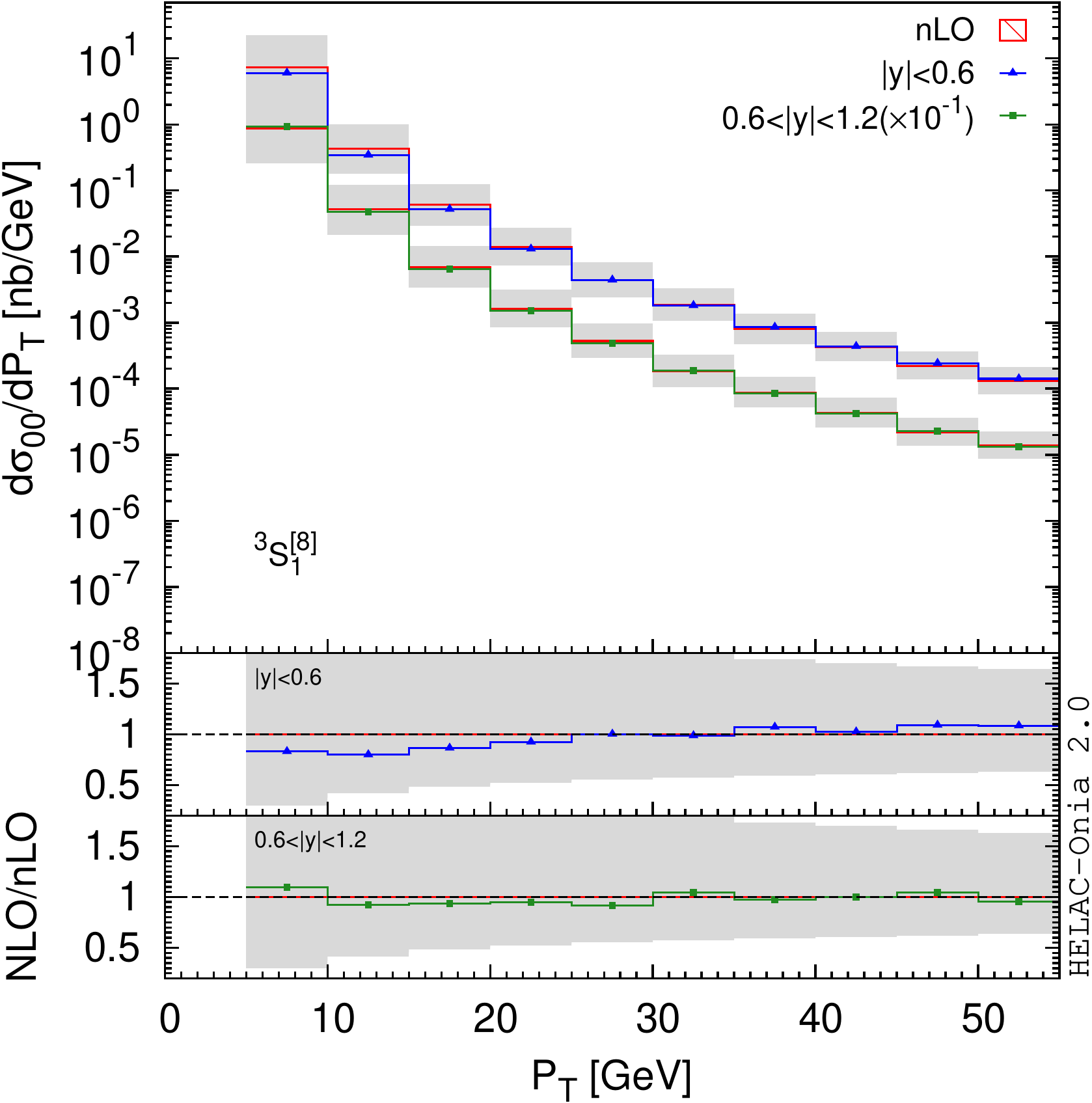}
\includegraphics[width=.45\textwidth,draft=false]{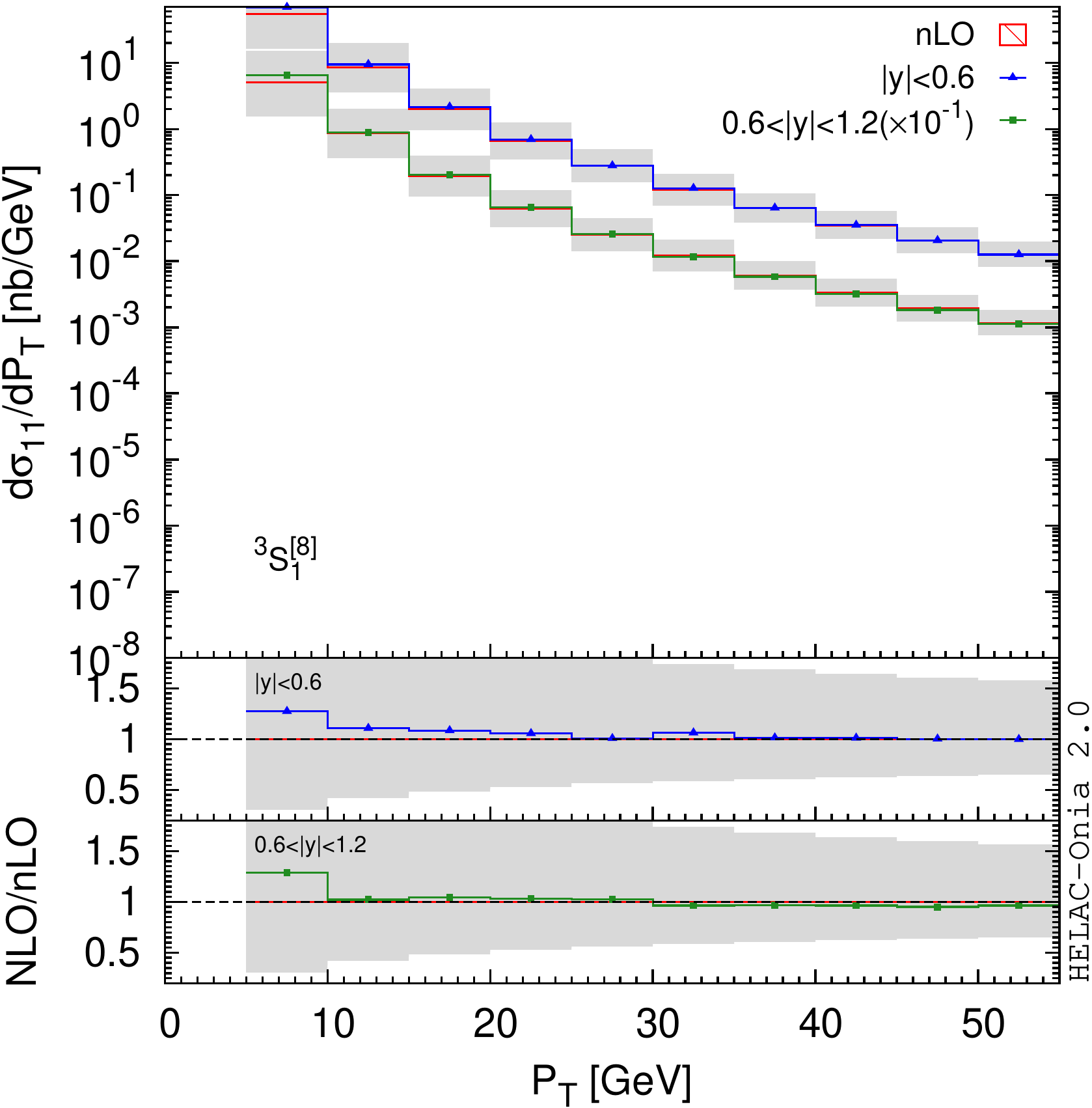}
\caption{Comparisons of spin-dependent differential cross sections for the Fock state $\so$ between nLO (up), tunned nLO with $z_{\rm cut,a}=0.6, z_{\rm cut,s}=\frac{0.2}{m}$ (down) calculations and the complete NLO calculations.\label{Fig:nLO3S18vsNLOSpin}}
\end{figure}

\subsubsection{Reproducing other Fock states}
 
We are now in the position to check the calculations for the other Fock states with the STOP cuts. Like the case of the simple cuts in section~\ref{sec:simpleNLO}, the general infrared-safe STOP cuts can reproduce the complete NLO results within theoretical uncertainties well. These Fock states do not show LP behaviour at LO. The comparisons of nLO calculations to NLO calculations for spin-summed and spin-dependent differential cross sections in the states $\ss,\sps$ are displayed in Fig.~\ref{Fig:nLOvsNLO} and Fig.~\ref{Fig:nLOvsNLOSpin} respectively, while we put the supplemental plots Figs.~\ref{Fig:nLOvsNLO2},~\ref{Fig:nLOvsNLOSpin2},~\ref{Fig:nLOvsNLOSpin3} for the other Fock states in the appendix \ref{app:moreplots}. With the scale variations shown by the grey bands, nLO results in general are able to successfully reproduce the NLO calculations in both cases. The only exception is the differential cross section of $\sps$ (see the right panel of Fig.~\ref{Fig:nLOvsNLO}) at very large $P_T$, i.e. $P_T>90$ GeV. Such a discrepancy in $\sps$ can be better understood from the LO fragmentation function $g\rightarrow \sps+g$~\cite{Braaten:1996rp}, which has the functional form 
\begin{eqnarray}
D_{g\rightarrow \sps}(z_{\rm onium})&\propto&3z_{\rm onium}-2z_{\rm onium}^2+2(1-z_{\rm onium})\log{(1-z_{\rm onium})},
\end{eqnarray}
where $z_{\rm onium}$ is the momentum fraction of $\sps$. The function peaks at $z_{\rm onium}=1$. A finite value of $z_{\rm cut,s}$ in the STOP cuts will remove a non-negligible fraction of radiations in the LP contributions. In fact, we have explicitly checked that if we set $z_{\rm cut,s}=\frac{10^{-2}}{m}$ instead of $z_{\rm cut,s}=\frac{0.1}{m}$, the agreement between nLO and NLO results are significantly improved at large $P_T$, which can be found in Fig.~\ref{Fig:nLO1S08vsNLO}. In the spirit of the NRQCD factorization, the soft gluons from the heavy quark pair with the momentum fraction smaller than $v^2$ should be absorbed into the LDMEs as well as their energy evolutions, where $v^2$ is around $0.3$ for the charmonium. Therefore, without taking into account the relativistic corrections, the resolution of NRQCD in describing the heavy quarkonium production should be not better than $v^2$. Hence, it is not straightforward to judge which is a better choice between the two different values $z_{\rm cut,s}=\frac{0.1}{m}$ and $z_{\rm cut,s}=\frac{10^{-2}}{m}$. In fact, we believe $z_{\rm cut,s}=\frac{0.1}{m}$ is a compromising choice in order to avoid spoiling the perturbative convergence in the fixed-order calculations by a large logarithm $\log{z_{\rm cut,s}}$.

We have compared the recent CMS measurement~\cite{Sirunyan:2017qdw} to our nLO calculations (with and without STOP cut tuning on $\so$) for $\psi(2S)$ production at 13 TeV LHC in Fig.~\ref{Fig:nLOvsCMS}, where the nonperturbative LDMEs are taken from Eqs.(2.17) and (2.18) in Ref.~\cite{Shao:2014yta}. A factor $10^{-1}$ has been multiplied to the nLO results with tuned $\so$ in order to improve the visibility between the two theoretical bands. Without surprising, the CMS data agree very well with our nLO calculations, because nLO does a similarly good job as NLO.

\begin{figure}[ht!]
\vspace{-1cm}
\centering
\includegraphics[width=.45\textwidth,draft=false]{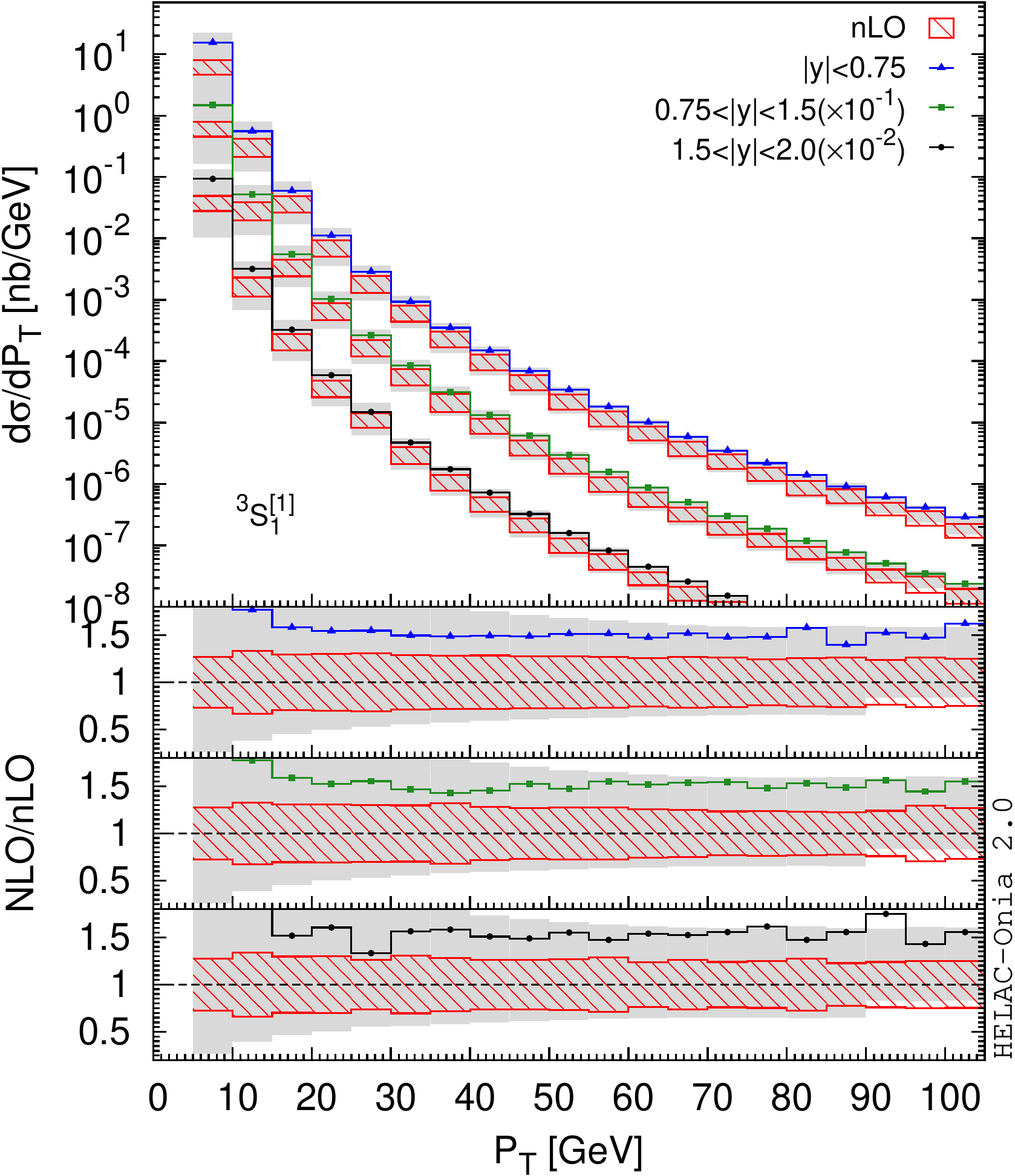}
\includegraphics[width=.45\textwidth,draft=false]{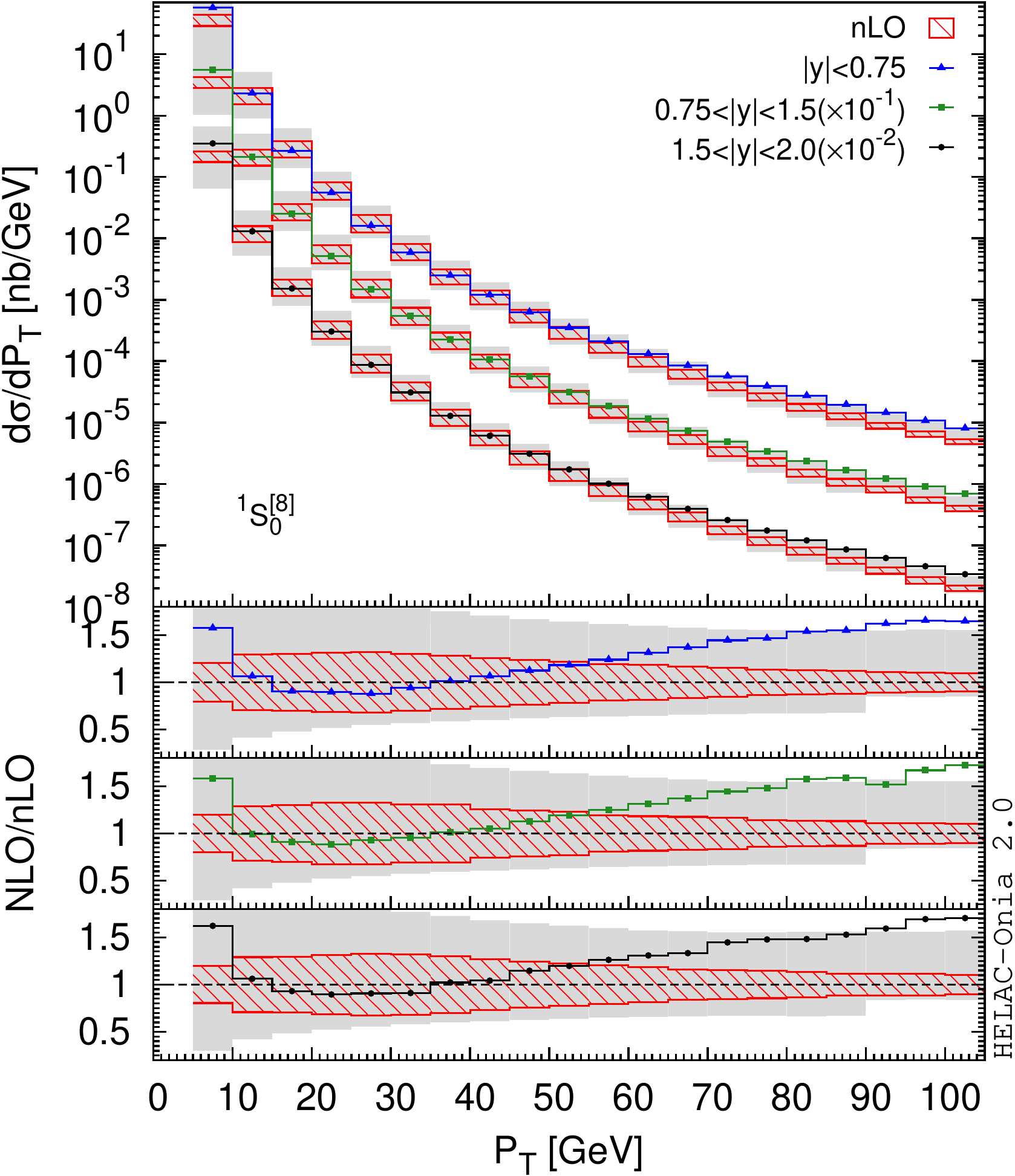}
\caption{Comparisons of spin-summed differential cross sections for the Fock states $\ss,\sps$ between our nLO calculations and the complete NLO calculations.\label{Fig:nLOvsNLO}}
\end{figure}

\begin{figure}[ht!]
\centering
\includegraphics[width=.45\textwidth,draft=false]{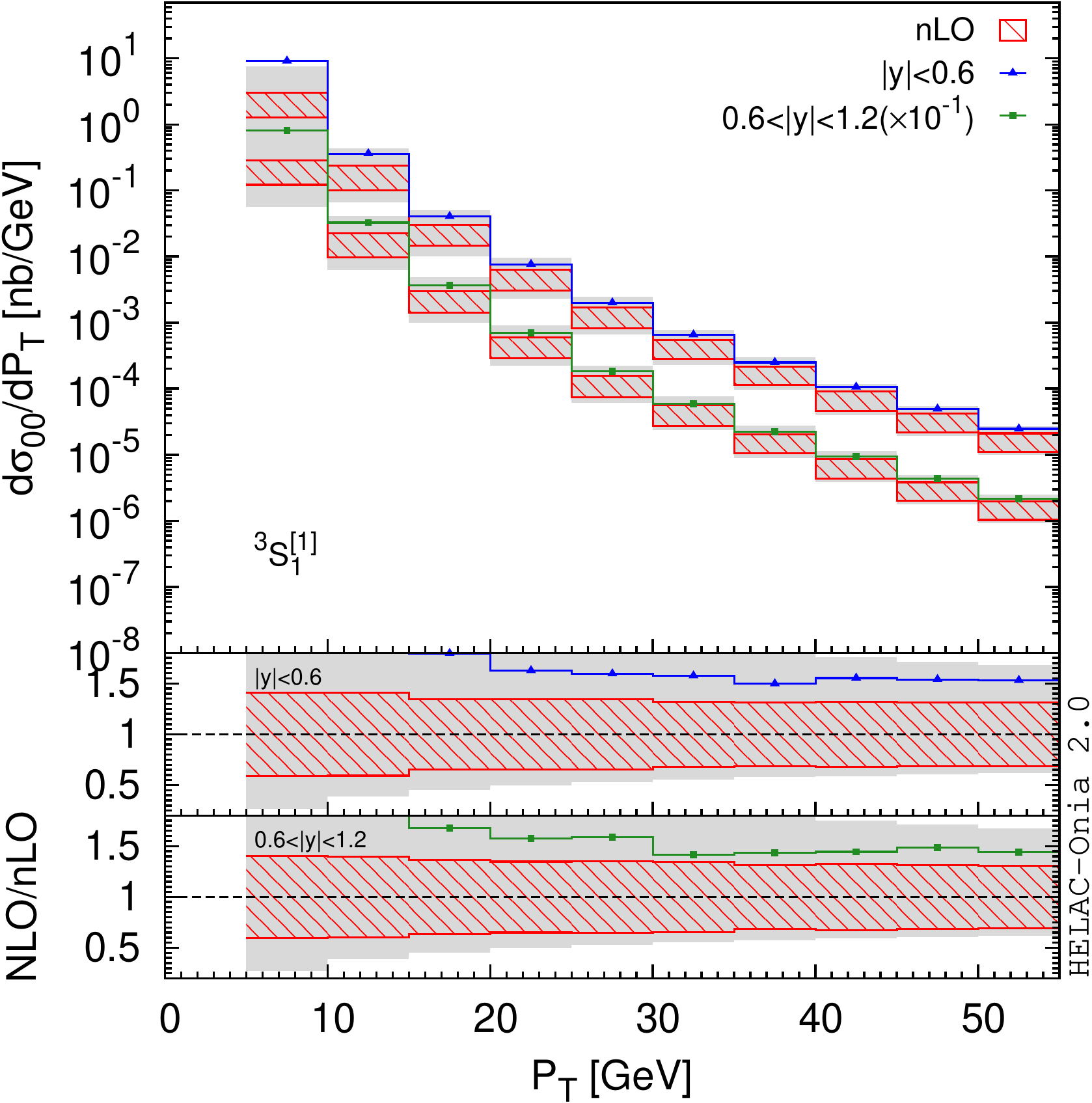}
\includegraphics[width=.45\textwidth,draft=false]{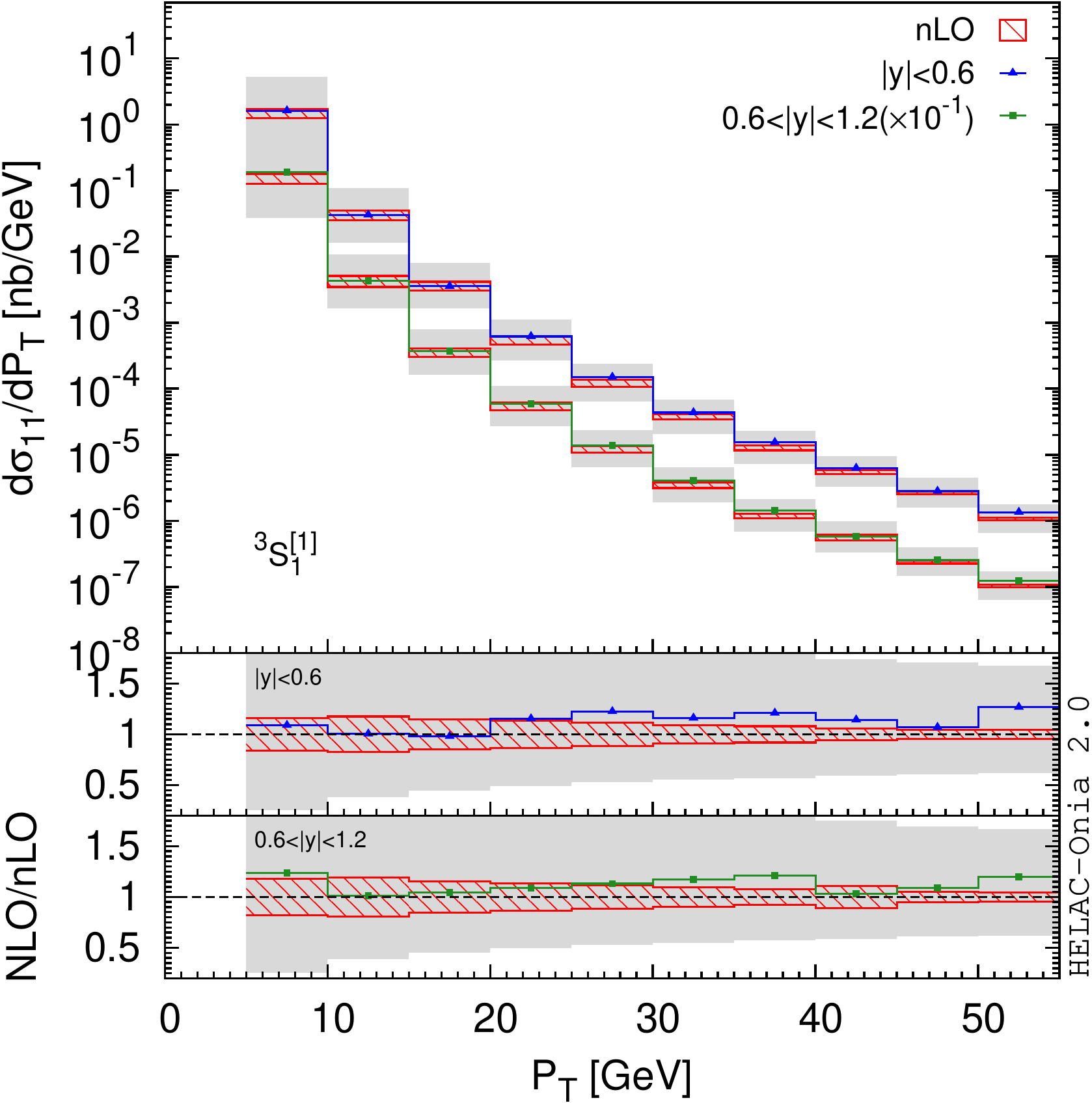}
\caption{Comparisons of spin-dependent differential cross sections for the Fock state $\ss$ between our nLO calculations and the complete NLO calculations.\label{Fig:nLOvsNLOSpin}}
\end{figure}

\begin{figure}[ht!]
\centering
\includegraphics[width=.99\textwidth,draft=false]{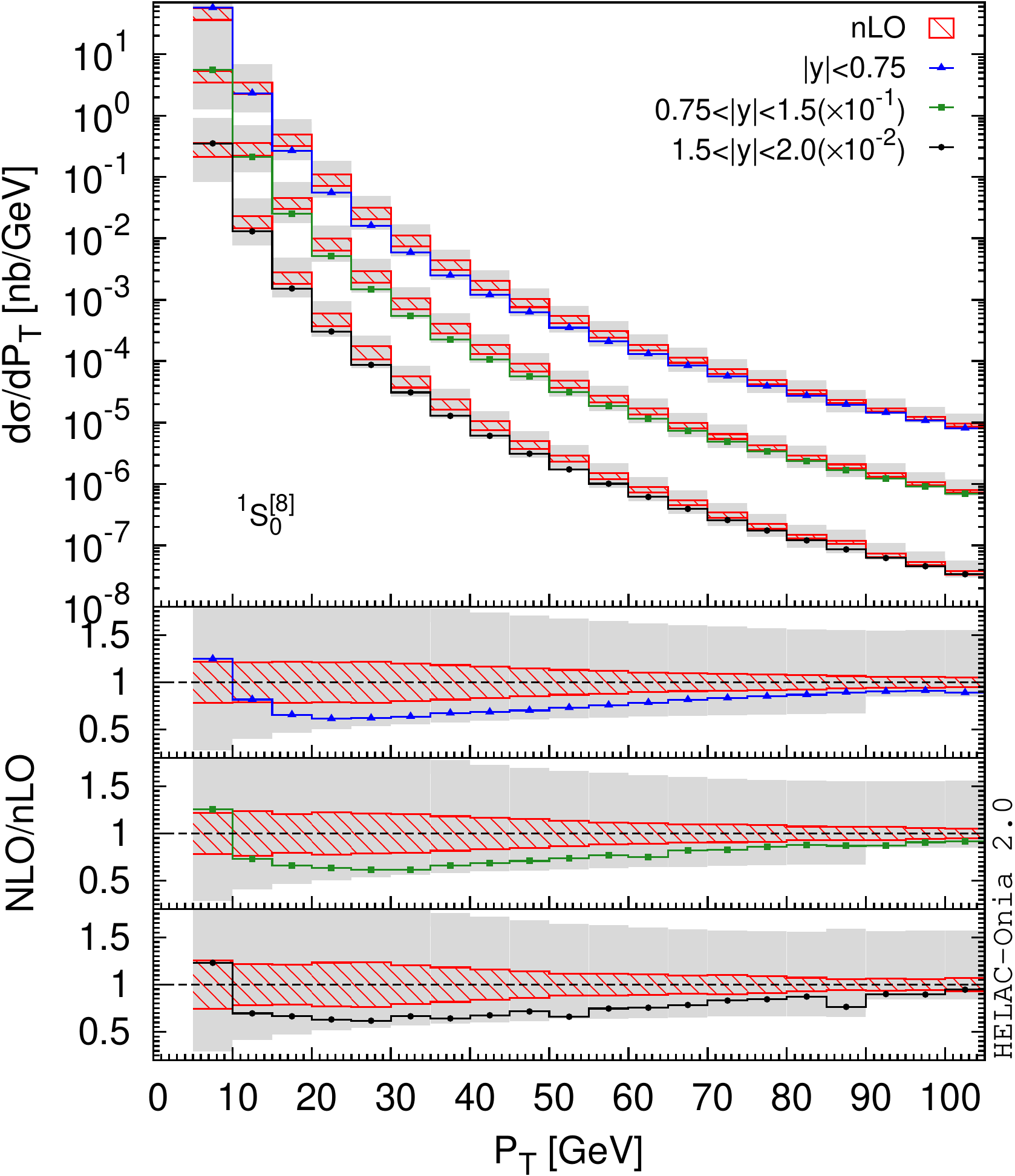}
\caption{Comparisons of the differential cross sections for $\sps$ between our nLO calculations with $z_{\rm cut,s}=\frac{10^{-2}}{m}$ and the complete NLO calculations.\label{Fig:nLO1S08vsNLO}}
\end{figure}

\begin{figure}[ht!]
\centering
\includegraphics[width=.99\textwidth,draft=false]{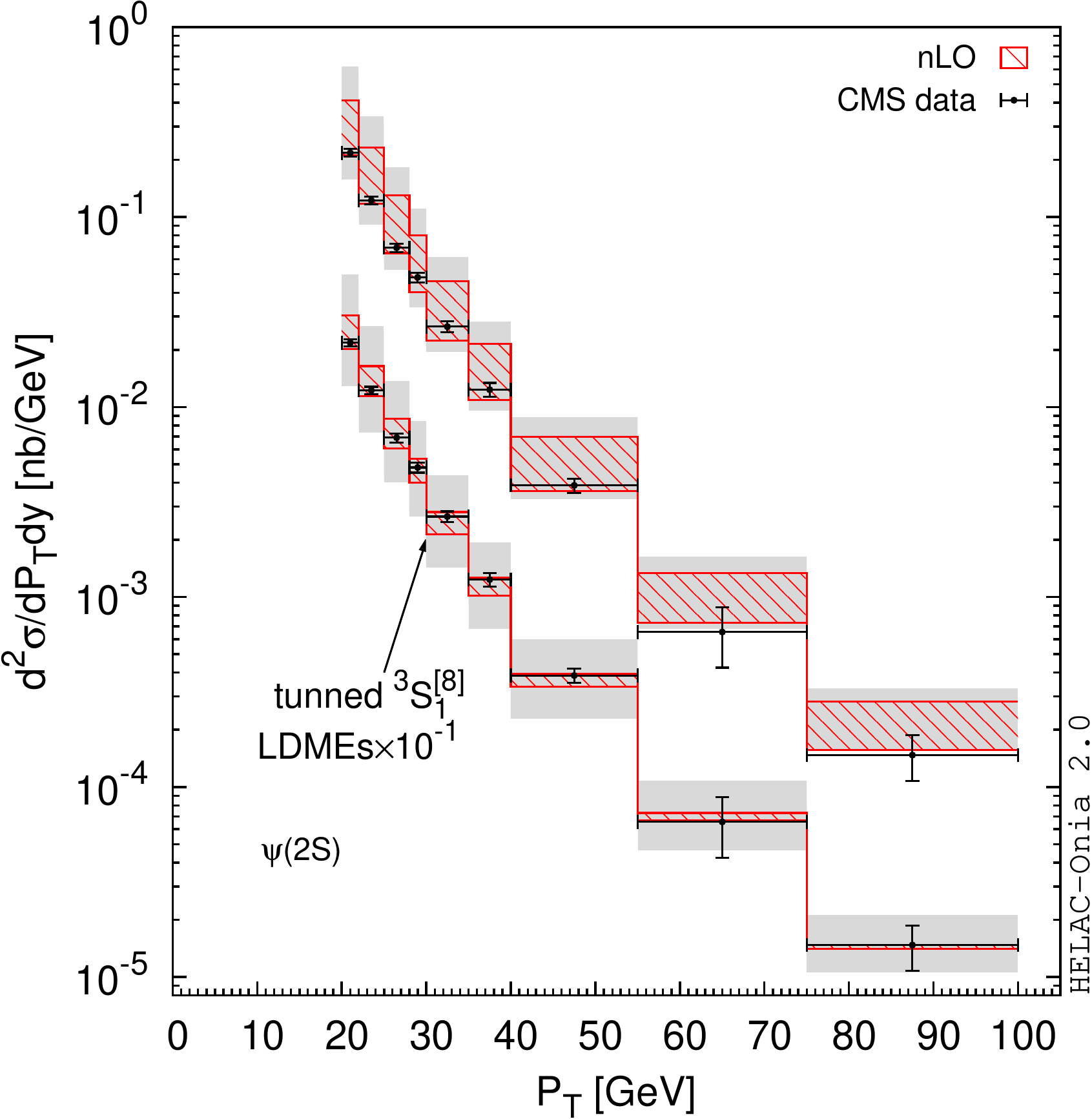}
\caption{Comparisons of the nLO $\psi(2S)$ differential cross sections $\frac{d^2\sigma}{dP_Tdy}$ in $|y|<0.6$ with the CMS measurement~\cite{Sirunyan:2017qdw}.\label{Fig:nLOvsCMS}}
\end{figure}

\subsection{Going beyond NLO}

It is usually believed that the color-octet states for $J/\psi$ hadroproduction will not receive giant K factors beyond NLO as the LP topologies in $P_T$ appear at NLO. On the other hand, the color-singlet Fock state $\ss$, which is LO in $v^2$ expansion, contains the LP single-gluon fragmentation contributions starting from NNLO in $\alpha_s$ (i.e. $\mathcal{O}(\alpha_s^5)$). A giant K factor for $\ss$ from NLO to NNLO might be possible in $J/\psi$ production, though the NLO calculation shows that the $\ss$ contribution to $J/\psi$ hadroproduction seems to be negligible compared to the color-octet contributions. If it is true, the extractions of color-octet NRQCD LDMEs solely based on NLO calculations will be questionable. This is one of the reasons why the importance of color-octet contributions in $J/\psi$ hadroproduction is still under debate. Although the accomplishment of NNLO calculations for $\ss$ is still beyond state-of-the art, it was indeed suggested in Ref.~\cite{Artoisenet:2008fc} that the partial calculation shows a giant K factor $\frac{d\sigma^{\rm NNLO^\star}}{d\sigma^{\rm NLO}}$. Later on, it was pointed out in Ref.~\cite{Ma:2010jj} that the giant K factor observed in Ref.~\cite{Artoisenet:2008fc} is in fact due to the logarithmic enhancement induced by the infrared cutoff. Such a logarithm is expected to be absent in a full NNLO calculation because of the infrared safety. 

We have the opportunity to clarify the situation with our infrared-safe STOP cut method. With the same setup done in section \ref{sec:NLOStop}, we have performed the calculations for $\ss$ plus three light-flavored jets production at $\mathcal{O}(\alpha_s^5)$ at the $13$ TeV. The spin-summed $P_T$ differential distributions are shown in Fig.~\ref{Fig:nnLOvsNLO}, where we have used ``nnLO" and ``nNLO" for the $\mathcal{O}(\alpha_s^n),n\leq 4$ parts being nLO and NLO cross sections respectively. In other words, we have used $d\sigma^{\rm nnLO}\equiv d\sigma^{\rm nLO}+d\sigma^{\mathcal{R}^2_{\rm STOP}}$ and $d\sigma^{\rm nNLO}\equiv d\sigma^{\rm NLO}+d\sigma^{\mathcal{R}^2_{\rm STOP}}$. In the nNLO results, no theoretical uncertainties are taking into account from the NLO piece $d\sigma^{\rm NLO}$. In contrast to the finding made in Ref.~\cite{Artoisenet:2008fc}, we do not observe any giant K factor up to $P_T\simeq 100$ GeV. In fact, the $P_T$ spectra of nnLO and nNLO are not harder than NLO ones. Such an observation can be explained if the coefficient of the LP $P_T$ part arising from the single-gluon fragmentation is much smaller than the coefficient of the NLP $P_T$ part and/or if the average momentum fraction of $\ss$ taking from the original gluon is significantly smaller than $1$. The calculation based on the gluon fragmentation function shows a similar behaviour, and the normalization of $\ss$ is significantly smaller than the color-octet contributions~\cite{Brambilla:2004wf}. In our calculation, the K factor $\frac{d\sigma^{\rm nnLO}}{d\sigma^{\rm NLO}}$ is ranging from 1 to 3 depending on the infrared cutoff choices. A similar conclusion can be drawn for the spin-dependent differential distributions from Fig.~\ref{Fig:nnLOvsNLOSpin}. We believe a complete NLO calculations of $\ss$ plus two jets will help to reduce the remaining large infrared cutoff as well as the renormalization/factorization scale dependence. 

\begin{figure}[ht!]
\centering
\includegraphics[width=.45\textwidth,draft=false]{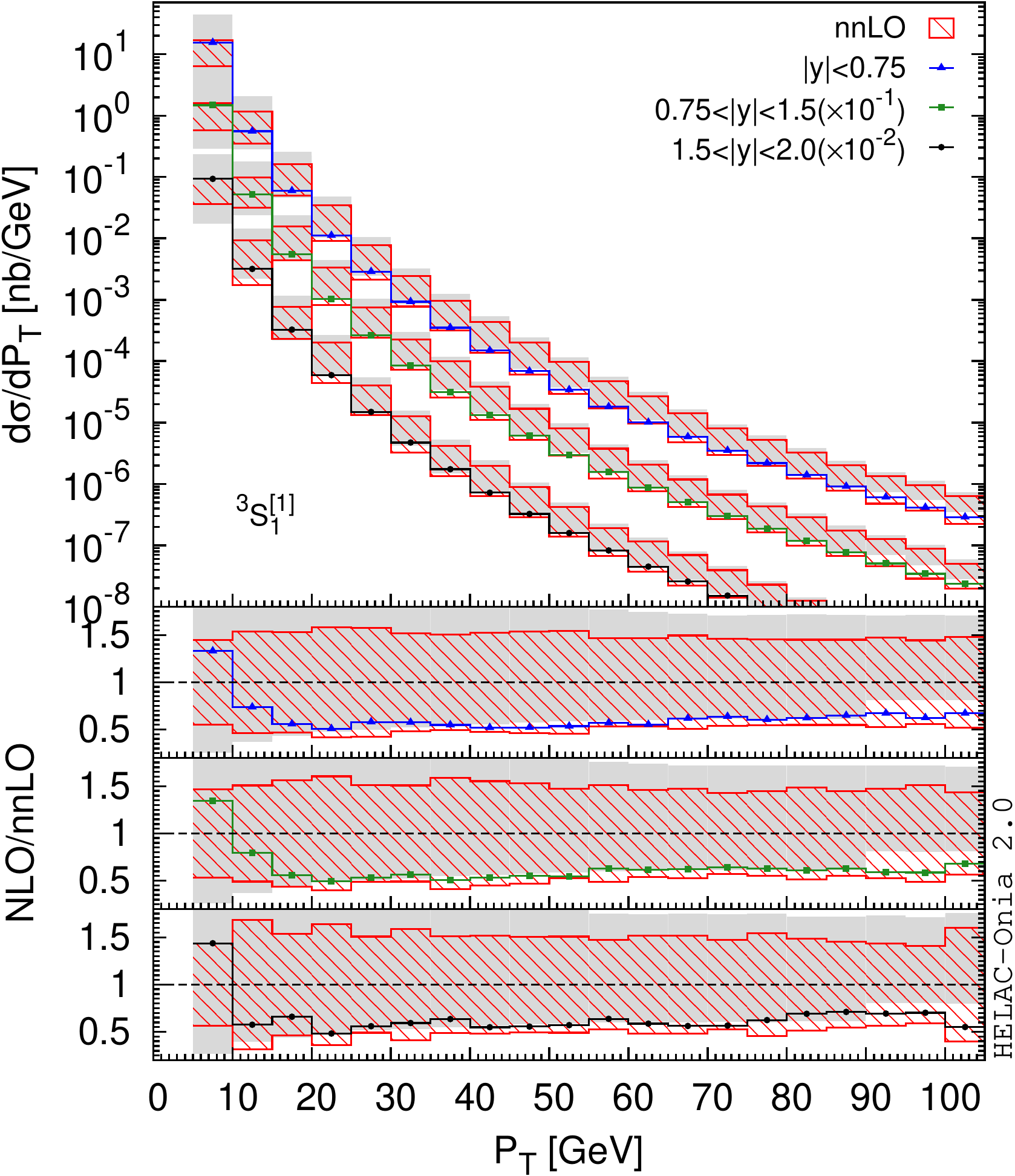}
\includegraphics[width=.45\textwidth,draft=false]{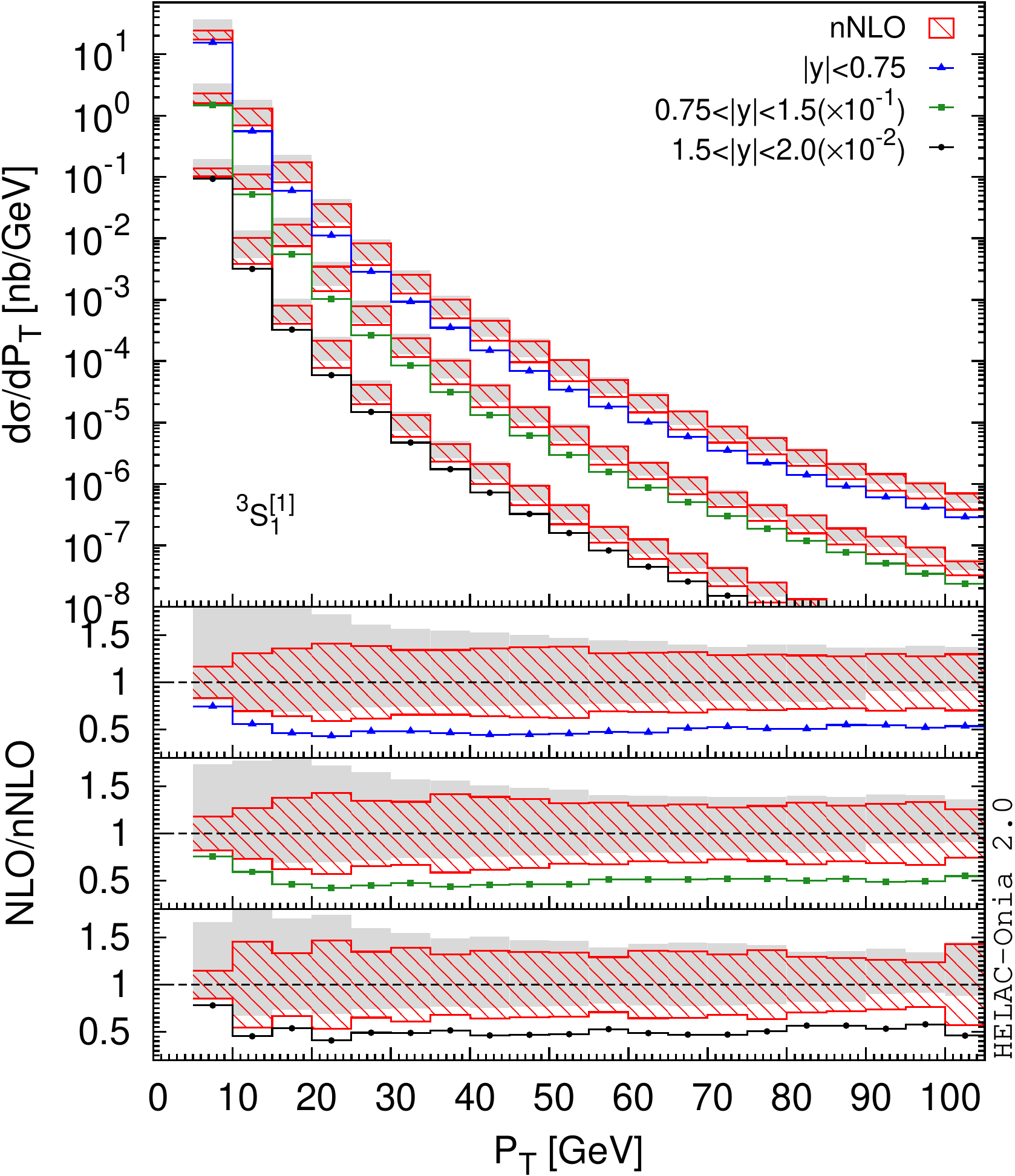}
\caption{Comparisons of spin-summed differential cross sections for the Fock state $\ss$ between nnLO (left), nNLO (right) calculations and the complete NLO calculations.\label{Fig:nnLOvsNLO}}
\end{figure}

\begin{figure}[ht!]
\centering
\includegraphics[width=.45\textwidth,draft=false]{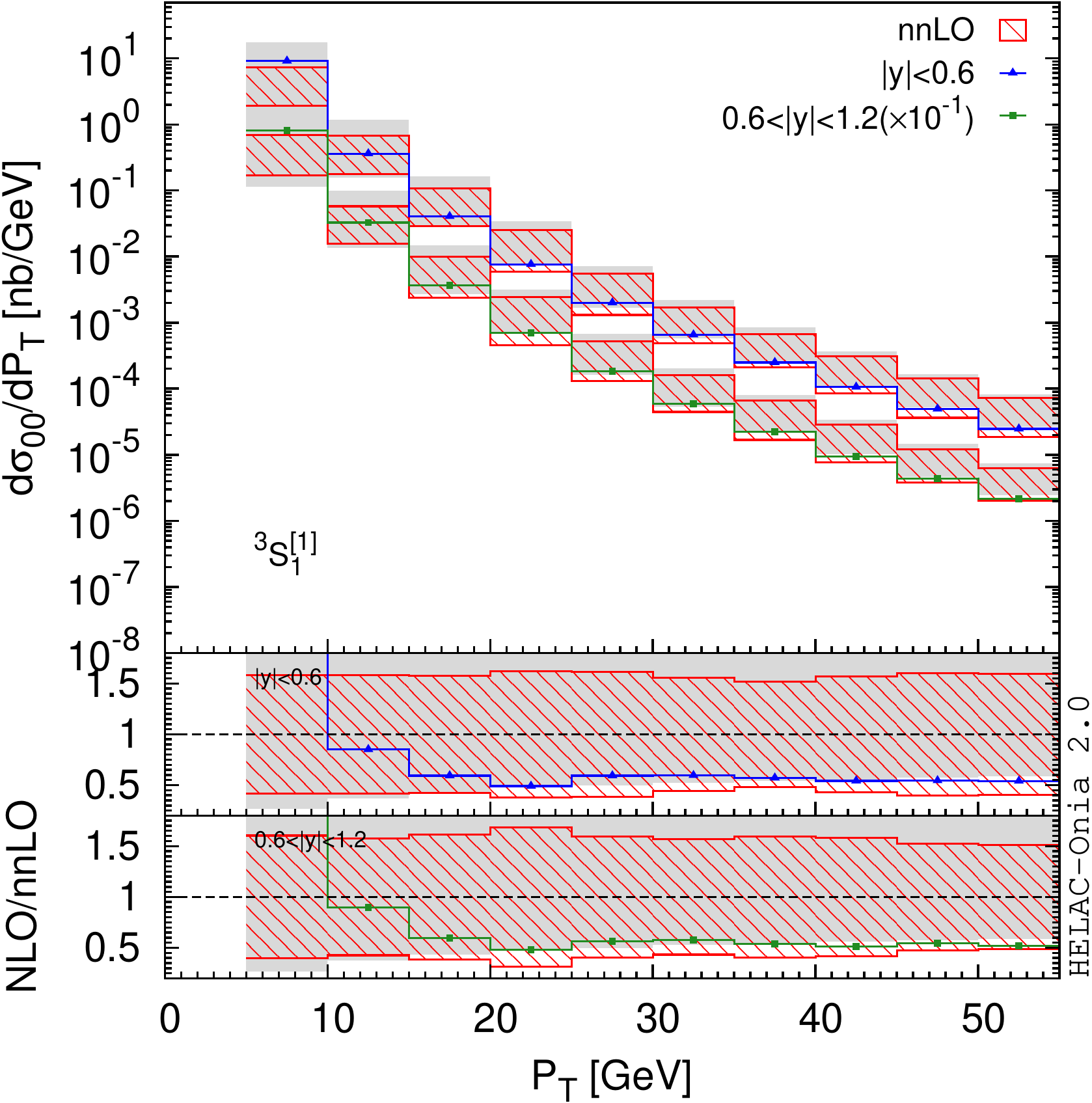}
\includegraphics[width=.45\textwidth,draft=false]{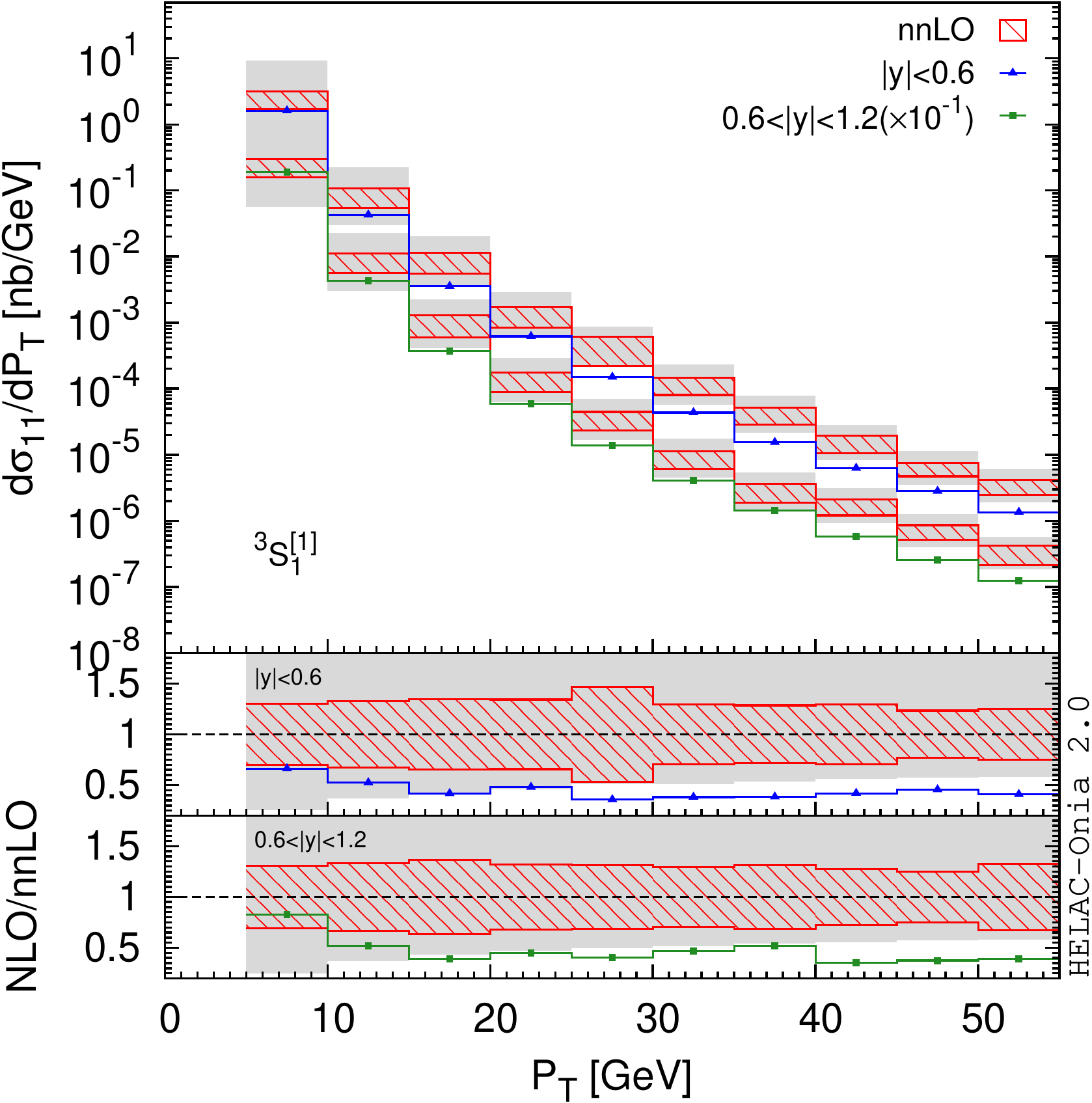}
\caption{Comparisons of spin-dependent differential cross sections for the Fock state $\ss$ between nnLO calculations and the complete NLO calculations.\label{Fig:nnLOvsNLOSpin}}
\end{figure}

\subsection{Reassessing the charm fragmentation\label{sec:charmfrag}}

So far, we have only considered the light-flavoured jet(s) accompanying with the quarkonium, which is usually thought to be dominant because the gluons are more often produced than the heavy quarks at high-energy hadron colliders. However, since the LP $P_T$ contribution from the charm quark fragmentation appears at $\mathcal{O}(\alpha_s^4)$, one should not overlook the associated production processes of a quarkonium plus a heavy quark pair. They were first studied in Ref.~\cite{Artoisenet:2007xi} for $\ss$, in Ref.~\cite{Artoisenet:2009zwa} for $\so,\sps,\pj$ and in Ref.~\cite{Li:2011yc} for $\tpzs,\tpos,\tpts$. To the best of our knowledge, the existing calculations only focus on the spin-summed differential cross sections, while we will also present the spin-dependent results in this section. In fact, one has to examine the relevance of these contributions if large cancellations between various Fock states happen. 

In Fig.~\ref{Fig:nLOvsccx3S11} and Fig.~\ref{Fig:nLOvsccx3S11Spin}, we compared the tree-level $\ss+c\bar{c}$ (tagged as ``$c\bar{c}$") production with the nnLO calculations of $\ss$ plus light-flavoured partons.  The $\ss+c\bar{c}$ contribution has a harder $P_T$ spectrum than the nnLO contribution. The former one exceeds the latter one above $P_T\simeq 55$ GeV in the spin-summed case, while such a kind of crossover happens earlier for the spin transverse component $\frac{d\sigma_{11}}{dP_T}$ around $P_T\simeq 20$ GeV.

On the other hand, the charm quark associated contributions are orders of magnitude smaller than the light-flavoured jet contributions for $\so$ productions as clearly shown in Fig.~\ref{Fig:nLOvsccx4CO} and Fig.~\ref{Fig:nLOvsccxSpin4CO} for the spin-summed and spin-dependent distributions. The similar conclusion can be drawn for the other Fock states $\sps,\pj,\tpzs,\tpos,\tpts$ as shown in Figs.~\ref{Fig:nLOvsccx4CO2},~\ref{Fig:nLOvsccxSpin4CO2},~\ref{Fig:nLOvsccxSpin4CO3} in the appendix \ref{app:moreplots}.

\begin{figure}[ht!]
\vspace{-1cm}
\centering
\includegraphics[width=.99\textwidth,draft=false]{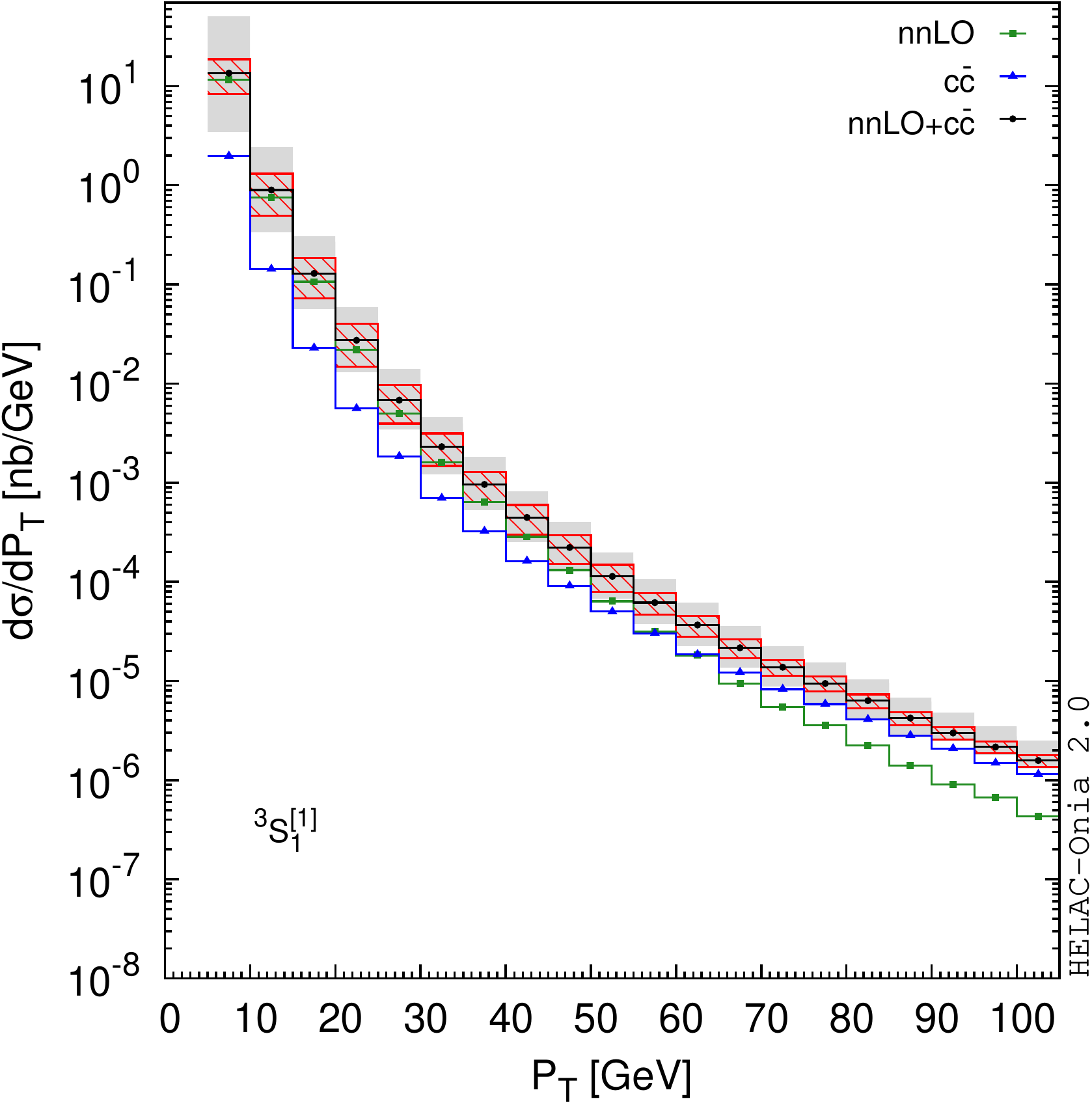}
\caption{Comparisons of spin-summed differential cross section $\frac{d\sigma}{dP_T}$ for the Fock state $\ss$ between our nnLO calculations and the LO charmonium plus charm quark pair calculations.\label{Fig:nLOvsccx3S11}}
\end{figure}

\begin{figure}[ht!]
\vspace{-1cm}
\centering
\includegraphics[width=.45\textwidth,draft=false]{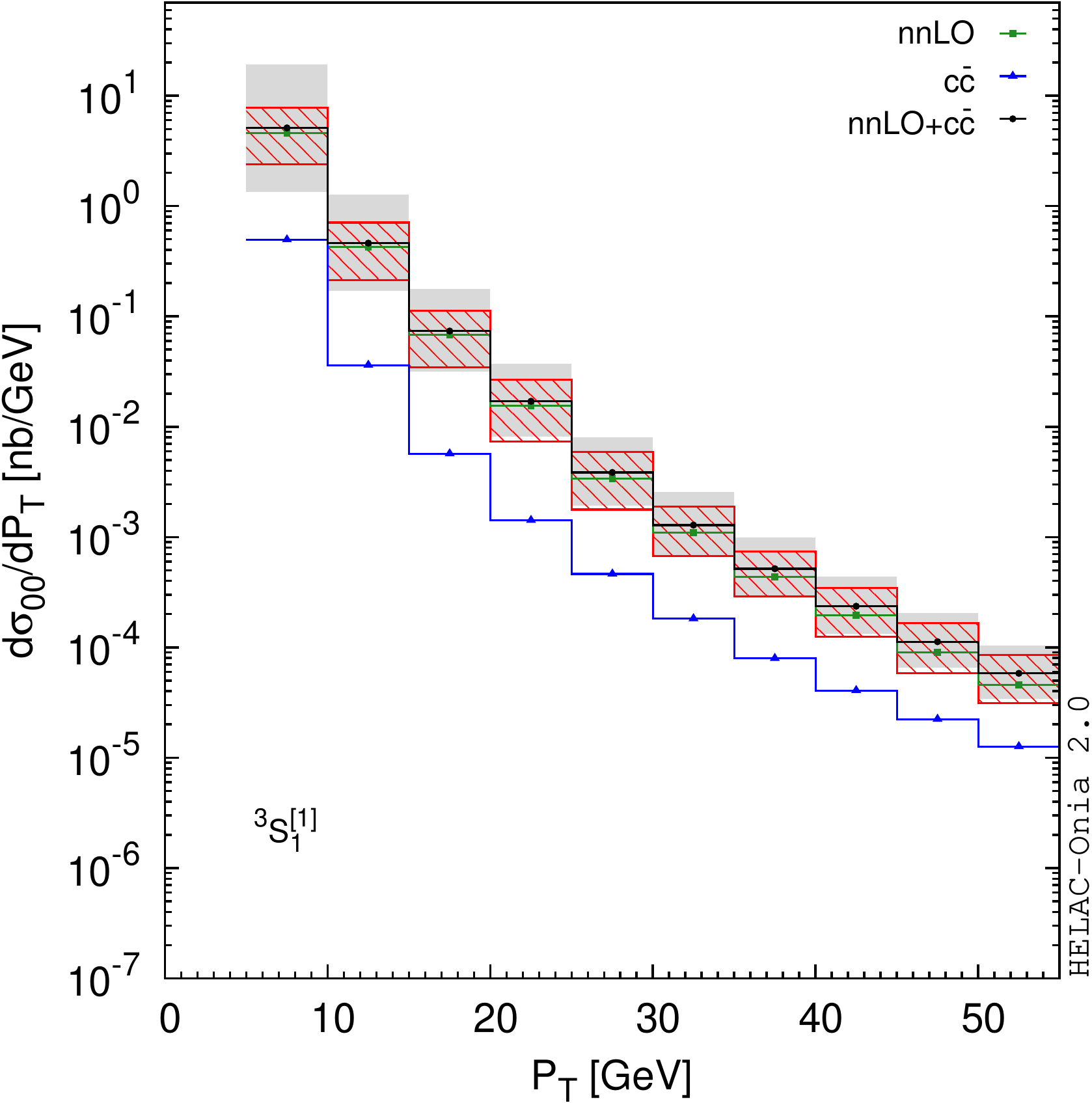}
\includegraphics[width=.45\textwidth,draft=false]{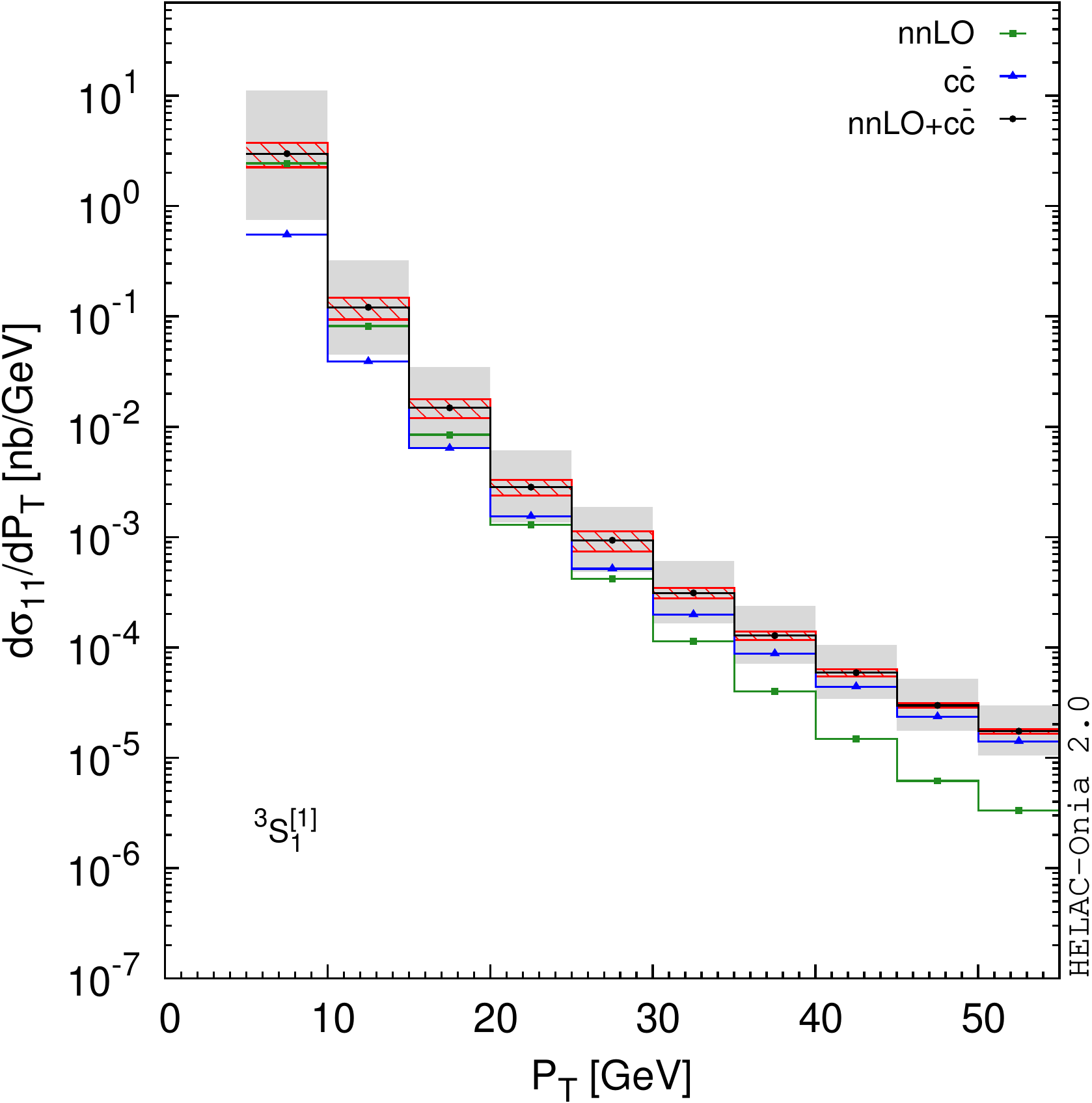}
\caption{Comparisons of spin-dependent differential cross section $\frac{d\sigma}{dP_T}$ for the Fock state $\ss$ between our nnLO calculations and the LO charmonium plus charm quark pair calculations.\label{Fig:nLOvsccx3S11Spin}}
\end{figure}

\begin{figure}[ht!]
\vspace{-1cm}
\centering
\includegraphics[width=.99\textwidth,draft=false]{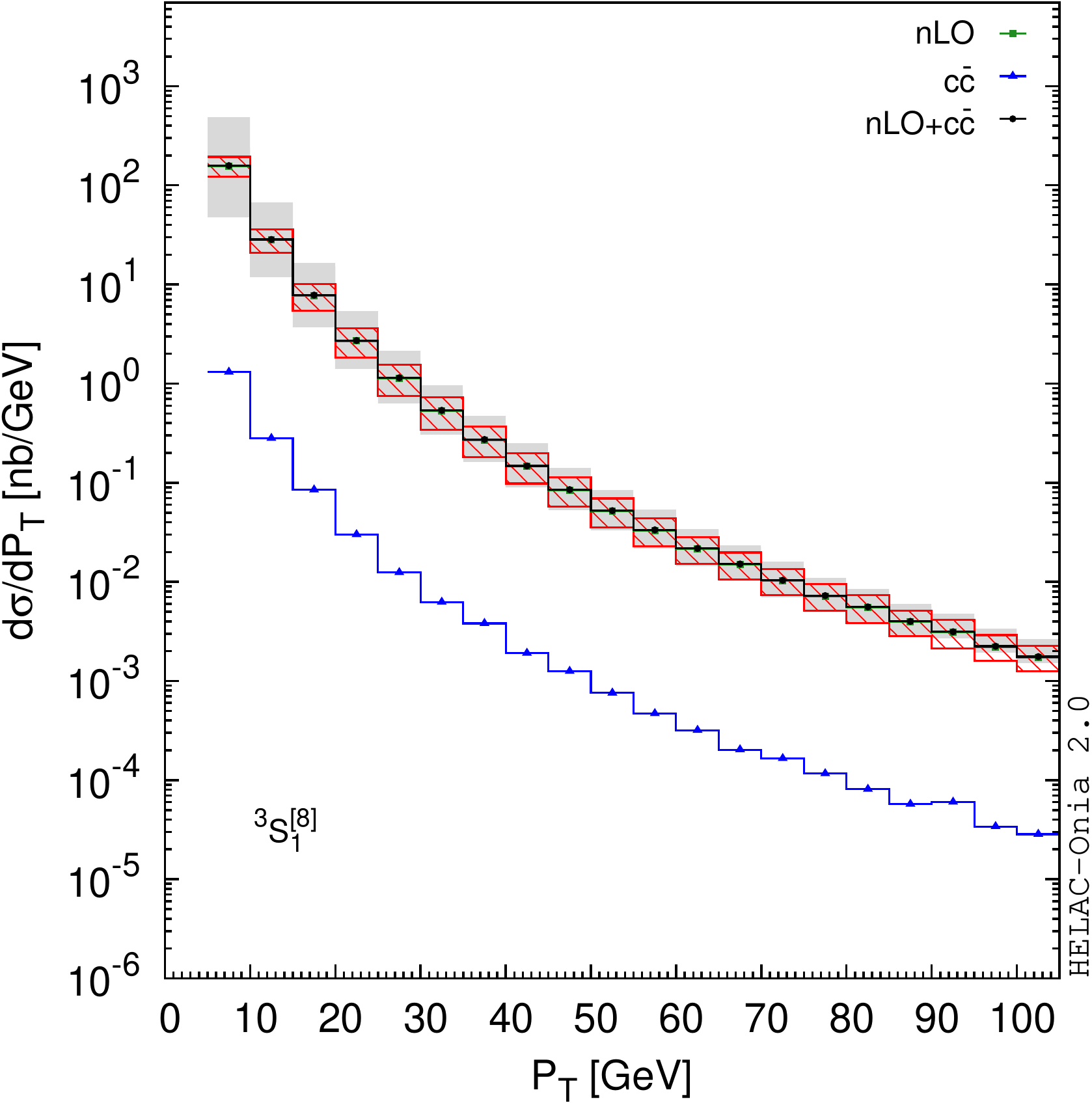}
\caption{Comparisons of spin-summed differential cross sections $\frac{d\sigma}{dP_T}$ for the Fock state $\so$ between our nLO calculations and the LO charmonium plus charm quark pair calculations.\label{Fig:nLOvsccx4CO}}
\end{figure}

\begin{figure}[ht!]
\centering
\includegraphics[width=.45\textwidth,draft=false]{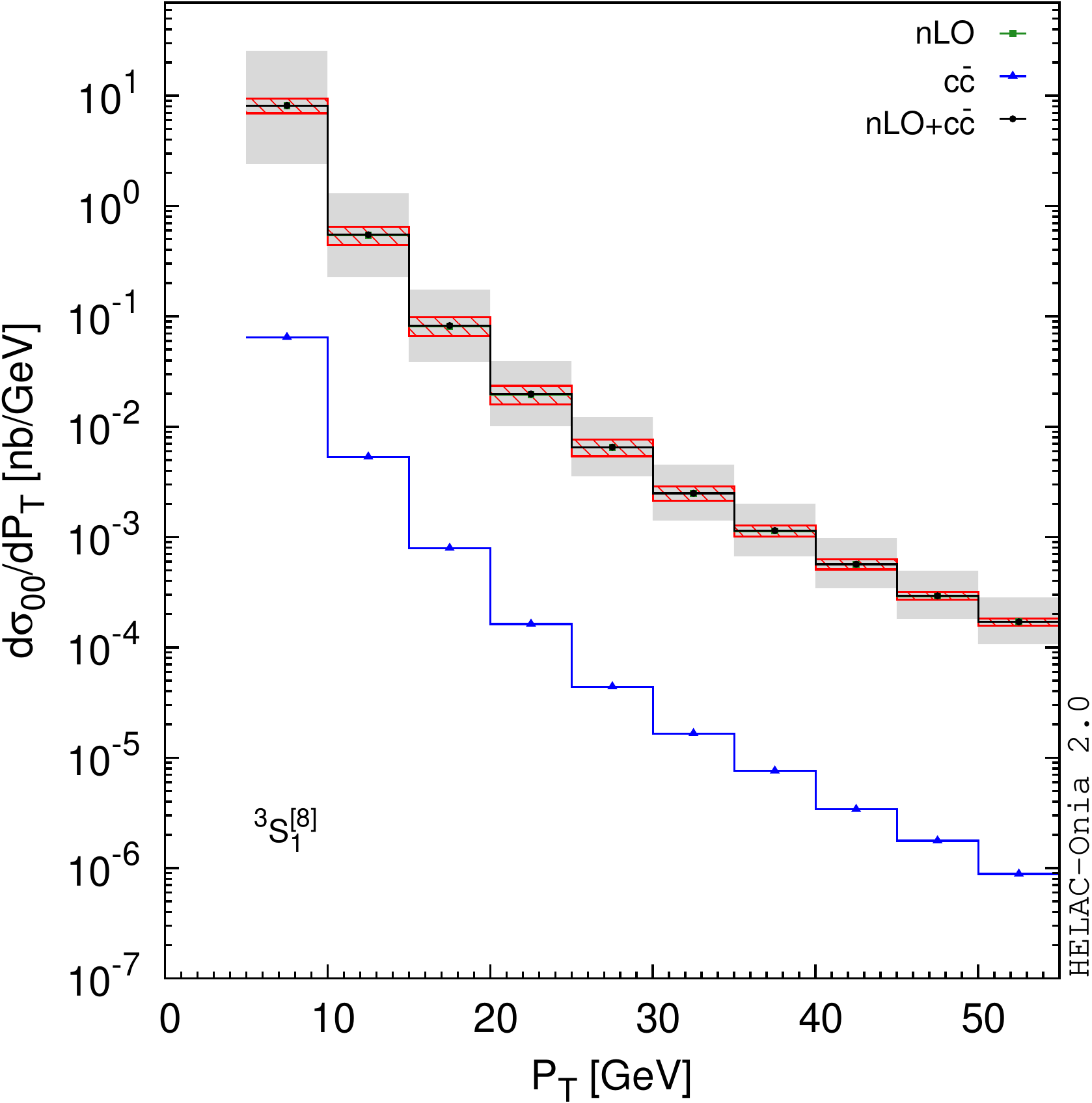}
\includegraphics[width=.45\textwidth,draft=false]{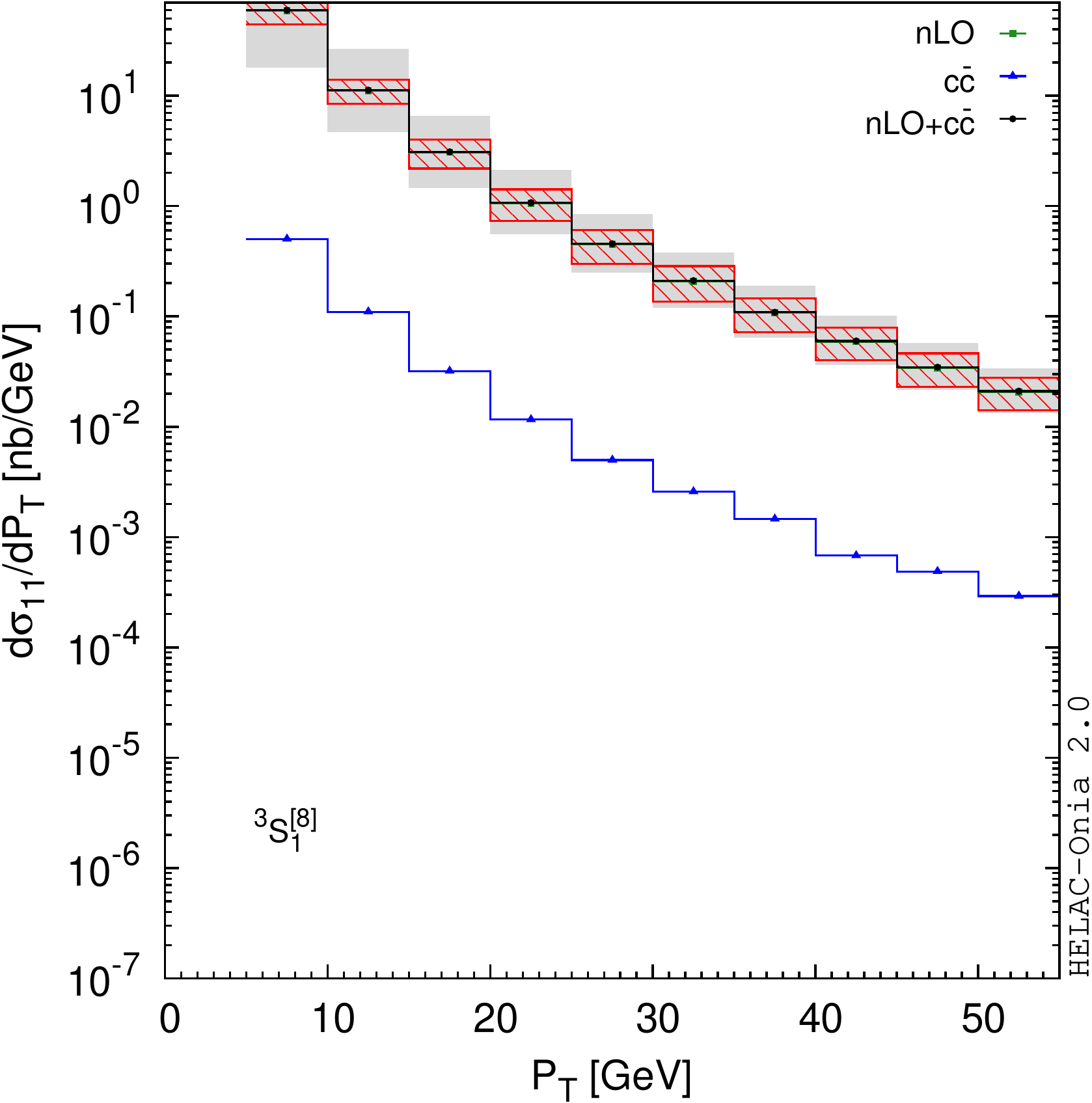}
\caption{Comparisons of spin-dependent differential cross sections $\frac{d\sigma}{dP_T}$ for the Fock state $\so$ between our nLO calculations and the LO charmonium plus charm quark pair calculations.\label{Fig:nLOvsccxSpin4CO}}
\end{figure}

\section{Summary and outlooks\label{sec:summary}}

After implementing the remainders of P-wave counterterms in section~\ref{sec:pCTs}, we have introduced a general infrared-safe method to estimate the giant K factors in quarkonium production in high $P_T$ region. As a proof of concept, we have validated our approach with the existing complete NLO QCD calculations of the Fock states $\sso,\sps,\pjso$ in both spin-summed and spin-dependent cases. They are relevant for $J/\psi$ and $\chi_{cJ}$ hadroproduction up to $\mathcal{O}(v^7)$. Our approach only requires the tree-level amplitudes provided by {\sc\small HELAC-Onia}. To the best of our knowledge, it is the first time to be able to reproduce the complete NLO spin-dependent results with tree-level amplitudes only. These spin-dependent results can be used to predict the polarization observables. We are also firstly able to obtain the spin-summed NLO results for $\pjs$ without performing complete NLO calculations. With our new approach, we have estimated the partial NNLO contributions at $\mathcal{O}(\alpha_s^5)$ for $\ss$ production. It is believed to be at LP in $P_T$ scaling starting at this order, and is the last missing piece for the heavy quarkonium $P_T$ spectrum up to $\mathcal{O}(v^7)$. In contrast to the NNLO$^\star$ calculations based on the simple invariant-mass cuts~\cite{Artoisenet:2008fc}, we do not observe the similar orders of magnitude enhancement compared to the NLO calculations, while an enhancement factor of 1 to 3 is still possible up to $P_T\simeq 100$ GeV depending on the infrared cutoff choices. We believe the complete NLO calculations of $\ss$ plus 2 jets will reduce this uncertainty. Finally, we have also calculated the charmonium plus a charm quark pair production, where the spin-dependent differential cross sections presented here are new. Their contributions to the inclusive $P_T$ distributions of charmonium are only relevant in the $\ss$ channel.

Our approach stabilizes the QCD corrections in the heavy quarkonium production rate calculations at high $P_T$. It is quite appealing not only because it provides a fast way to perform the phenomenology studies of inclusive quarkonium production but also it can be used to improve the predictions in the associated quarkonium production processes. Together with the controlled perturbative SDCs, it is feasible to study various nonperturbative effects in the heavy quarkonium hadroproduction in an acceptable amount of computation time. Last but not least, with a similar method, we believe that we are able to promote the accuracy of both LP and NLP pieces to NLO level simultaneously with the full one-loop calculations.

\begin{acknowledgments}
I thank Jean-Philippe Lansberg for useful discussions. 
This work is supported by the ILP Labex (ANR-11-IDEX-0004-02, ANR-10-LABX-63). The computations in this paper were performed with the help of the computing facilities at IPN Orsay.
\end{acknowledgments}

\appendix

\section{Calculations with {\sc\small HELAC-Onia}\label{app:helaconia}}


In this section, we will give an instruction on how to use {\sc\small HELAC-Onia} to perform nLO and nnLO calculations. The implementations are available from version 2.3.6 and onwards, which can be downloaded from http://hshao.web.cern.ch/hshao/helaconia.html. An example of a {\sc\small Fortran} analysis file {\tt plot\_pp\_psinjets\_spin2.f90} is given in the subdirectory {\tt analysis/user} in order to get the differential distributions in this paper. The common calculation setup is given by the following entries in {\tt user.inp}:\\

\begin{minipage}{1.1\textwidth}
{\tt
$\#$ basic setup for running\\
colpar 1  $\#$ colliding particles: 1=pp, 2=ppbar, 3=e+e-\\
energy\_beam1 6500d0  $\#$ beam 1 energy (GeV)\\
energy\_beam2 6500d0  $\#$ beam 2 energy (GeV)\\
alphasrun 1 $\#$ 0=alpha QCD not running, 1=alpha QCD running\\
useMCFMrun F $\#$ alphas running with MCFM (if False run it with original one)\\
qcd 2 $\#$ 0=only electroweak, 1=electroweak and QCD, 2=only QCD,3=only QED, 4=QCD and QED\\
cmass 1.5d0 $\#$ charm quark mass\\
unwevt F $\#$ unweighting on/off\\
reweight\_Scale T $\#$ reweight to get scale dependence (only when alphasrun=T)\\
hwu\_output T $\#$ hwu output file (T) or not (F)\\
ranhel 4 $\#$ doing Monte Carlo over helicities\\
pdf  1000 $\#$ 10000=cteq6m\\
Scale 1  $\#$ central value of the renormalization/factorization scale = Sqrt(m1**2+pt1**2)\\
\\$\#$ basic kenematic cuts\\
minptq 0d0 $\#$ minimum gluon/light-quark pt\\
minptc 0d0 $\#$ minimum charm quark pt\\
minptconia 5d0 $\#$ minimum charmonium pt\\
maxrapq 30d0    $\#$ maximum gluon/light-quark pesudorapidity\\
maxrapc  30d0   $\#$ maximum charm quark pesudorapidity\\
maxrapconia 30d0   $\#$ maximum charmonium pesudorapidity\\
maxyrapconia 4.5d0  $\#$ maximum charmonium rapidity\\
mindrqq 0d0   $\#$ minimum delta R (quark/gluon-quark/gluon)\\
\\$\#$ technical details on the numerical integration\\
gener 0 $\#$ onte Carlo generator: 0 PHEGAS 1 RAMBO 2 DURHAM 3 VEGAS -1 From PS.input\\
nmc  10000000 $\#$ maximal number of weighted events\\
nopt 1000000\\
nopt\_step 1000000\\
noptlim 10000000\\
nlimit 1 $\#$ The lower limit of the number of channels\\
grid\_nchmax 3000 $\#$ maximum number of channels for griding\\
}
\end{minipage}

\begin{minipage}{1.1\textwidth}
{\tt
$\#$ Long Distance Matrix Elements For Charmonium\\
$\#$ Long Distance Matrix Element <O(3S1[1])>=|R(0)|**2/4/Pi\\
$\#$ in JHEP 02 (2008) 102  <O(3S1[1])>=(2J+1)*2Nc*|R(0)|**2/4/Pi,\\
$\#$ i.e. LDME****1=<O(2S+1)LJ[1]>/2Nc/(2J+1)\\
$\#$ For p-wave <O(3P0[1])>=<O(3P1[1])>=<O(3P2[1])>=3*|R'(0)|**2/4/Pi\\
$\#$ in JHEP 02 (2008) 102 LDME****8=<O((2S+1)LJ[8])>/(Nc**2-1)/(2J+1)\\
LDMEcc3S11 0.064444444444d0 $\#$ LDME for 3S1[1] charmonium\\
LDMEcc3S18 0.00037621791666666665d0 $\#$ LDME for 3S1[8] charmonium\\
LDMEcc1S01 0.064444444444d0 $\#$ LDME for 1S0[1] charmonium\\
LDMEcc1S08 0.001825d0 $\#$ LDME for 1S0[8] charmonium\\
LDMEcc3P08 0.00428365d0 $\#$ LDME for 3P0[8] charmonium\\
LDMEcc3P18 0.00428365d0 $\#$ LDME for 3P1[8] charmonium\\
LDMEcc3P28 0.00428365d0 $\#$ LDME for 3P2[8] charmonium\\
LDMEcc3P01 0.017904931097838226d0 $\#$ LDME for 3P0[1] charmonium\\
LDMEcc3P11 0.017904931097838226d0 $\#$ LDME for 3P1[1] charmonium\\
LDMEcc3P21 0.017904931097838226d0 $\#$ LDME for 3P2[1] charmonium\\
}
\end{minipage}

\subsection{Born and counterterms}

The Born $d\sigma^{\mathcal{B}}$ at $\mathcal{O}(\alpha_s^3)$ for S-wave Fock states can be achieved via the following commands:
\cCode{}
HO> define ppsi = cc~(1s08) cc~(3s11) cc~(3s18)
HO> decay ppsi > m+ m- @ 1.0
HO> generate p g > ppsi j
HO> generate g p > ppsi j
HO> launch
\end{lstlisting}
where we have always excluded the quark-quark initial states due to their very small parton luminosity from PDFs. No extra kinematical cuts are needed for the Born-like events.

The color-octet P-wave Fock states $\pj$ can be grouped together as we will always use the relation from heavy-quark spin symmetry $\langle \mathcal{O}(\pj) \rangle=(2J+1)\langle \mathcal{O}(\p0) \rangle$. The contributions from the Born and the remainders of the counterterms $d\sigma^{\mathcal{B}}+d\sigma^{\mathcal{C}}$ can be included via:
\cCode{}
HO> set generate_CT = T
HO> decay cc~(3p08) > m+ m- @ 1.0
HO> generate p g > cc~(3p08) j
HO> generate g p > cc~(3p08) j
HO> launch
\end{lstlisting}
The command sets {\tt generate\_CT} to be {\tt T} in order to get the contributions from counterterms $d\sigma^{\mathcal{C}}$.

The color-singlet P-wave Fock states $\pjs$ will be calculated separately since they contribute to $\chi_{cJ},J=0,1,2$ respectively. The commands are:
\cCode{}
HO> set exp3pjQ = T
HO> set generate_CT = T
HO> define pchic = cc~(3p01) cc~(3p11) cc~(3p21)
HO> decay pchic > cc~(3s11) a @ 1.0
HO> decay cc~(3s11) > m+ m- @ 1.0
HO> generate p g > pchic j
HO> generate g p > pchic j
HO> launch
\end{lstlisting}
We set {\tt exp3pjQ}$=${\tt T} in order to get $\pjs,J=0,1,2$ individually. At the meantime, $\pjs$ are cascaded decaying to $\ss+\gamma\rightarrow \mu^+\mu^-+\gamma$. The counterterms should be taken into account by setting {\tt generate\_CT}$=${\tt T}.

\subsection{Real terms\label{app:real}}

One should apply the STOP cuts to the real terms at $\mathcal{O}(\alpha_s^n),n\geq 4$. It requires us to implement the following additional entries in {\tt user.inp}:

\begin{minipage}{1.1\textwidth}
{\tt
$\#$ STOP cuts\\
use\_stop\_cut T $\#$ whether use stop cuts (following cuts will be ignored if it is F)\\
stop\_minptjet 3d0  $\#$ minimum jet pt cut (include onium in the jet clustering)\\
stop\_maxrapjet 5d0  $\#$ max jet rapidity, negative no such a cut\\
stop\_zcut  0.1d0  $\#$ zcut in the soft drop\\
stop\_beta  -1d0   $\#$ beta in the soft drop (negative to make it collinear safe)\\
stop\_jet\_alg -1   $\#$ 1: kt; 0: C/A; -1: anti-kt\\
stop\_jet\_radius 1.0d0  $\#$ jet radius R\\
stop\_jet\_dyn\_radius -1d0 $\#$ if it is > 0, it will use dynamical jet radius R=max[stop\_jet\_radius,stop\_jet\_dyn\_radius*M\_{onium}/P\_{T,onium}]\\
stop\_min\_n\_jet 2  $\#$ min number of jet (should be n hard jet + 1 onium jet), negative no such a cut\\
stop\_max\_n\_jet -1 $\#$ max number of jet, negative no such a cut\\
stop\_n\_frag\_gluon 0 $\#$ minimal number of final gluon in the LO fragmentation process\\
stop\_n\_frag\_quark 0 $\#$ minimal number of final bare quark in the LO fragmentation process\\
stop\_zsoftcut 0.1d0 $\#$ z\_{s,cut} for the soft cut applied to (stop\_n\_frag\_gluon+stop\_n\_frag\_quark)==0\\
                                  $\#$ It will be divided by number of partons inside the onium jet\\
stop\_zasymcut 0.1d0 $\#$ asymmetric cut for the light-flavoured jets if the number of light jets are >= 2
}
\end{minipage}

The {\sc\small HELAC-Onia} commands to calculate the $\mathcal{O}(\alpha_s^4)$ real terms $d\sigma^{\mathcal{R}_{\rm STOP}}$ are
\cCode{}
HO> set exp3pjQ = F
HO> define ppsi = cc~(1s08) cc~(3s11) cc~(3s18) cc~(3p08)
HO> decay ppsi > m+ m- @ 1.0
HO> generate p g > ppsi j j
HO> generate g p > ppsi j j
HO> launch
\end{lstlisting}
for the S-wave and color-octet P-wave Fock states, while for $\pjs$ one should type
\cCode{}
HO> set exp3pjQ = T
HO> define pchic = cc~(3p01) cc~(3p11) cc~(3p21)
HO> decay pchic > cc~(3s11) a @ 1.0
HO> decay cc~(3s11) > m+ m- @ 1.0
HO> generate p g > pchic j j
HO> generate g p > pchic j j
HO> launch
\end{lstlisting}

With the same STOP cuts, the generation of weighted events at $\mathcal{O}(\alpha_s^5)$ $d\sigma^{\mathcal{R}^2_{\rm STOP}}$ for $\ss$ can be achieved by the following commands:
\cCode{}
HO> decay cc~(3s11) > m+ m- @ 1.0
HO> generate p g > cc~(3s11) j j j
HO> generate g p > cc~(3s11) j j j
HO> launch
\end{lstlisting}

\section{Supplemental plots\label{app:moreplots}}

We provided the supplemental plots in this appendix for the sake of completeness. The comparisons of spin-summed differential cross sections for the 5 Fock states $\sps,\pj,\tpzs,\tpos,\tpts$ between aNLO and NLO are shown in Fig.~\ref{Fig:aNLOvsNLO2}, while the spin-dependent ones for $\pj,\tpos,\tpts$ can be found in Fig.~\ref{Fig:aNLOvsNLOSpin2} and Fig.~\ref{Fig:aNLOvsNLOSpin3}. The nLO versus NLO plots for $\pj,\tpzs,\tpos,\tpts$ are available in Fig.~\ref{Fig:nLOvsNLO2} (spin-summed ones) and in Figs.~\ref{Fig:nLOvsNLOSpin2},\ref{Fig:nLOvsNLOSpin3} (spin-dependent ones). The contributions from ${\cal O}_n+c\bar{c}$ with ${\cal O}_n=\sps,\pj,\tpzs,\tpos,\tpts$ are shown in Figs.~\ref{Fig:nLOvsccx4CO2},~\ref{Fig:nLOvsccxSpin4CO2},~\ref{Fig:nLOvsccxSpin4CO3}.

\begin{figure}[ht!]
\vspace{-1cm}
\centering
\includegraphics[width=.45\textwidth,draft=false]{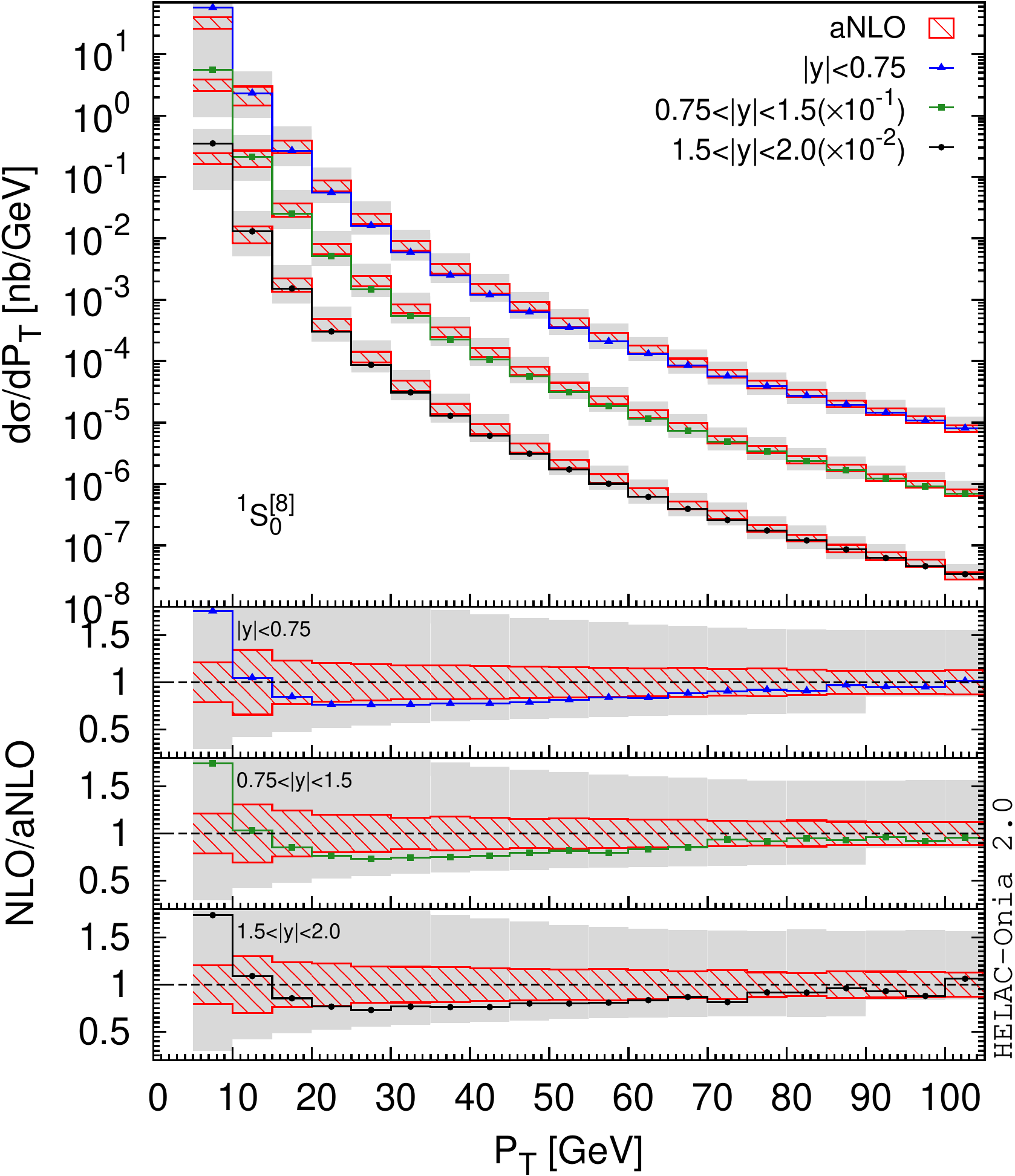}
\includegraphics[width=.45\textwidth,draft=false]{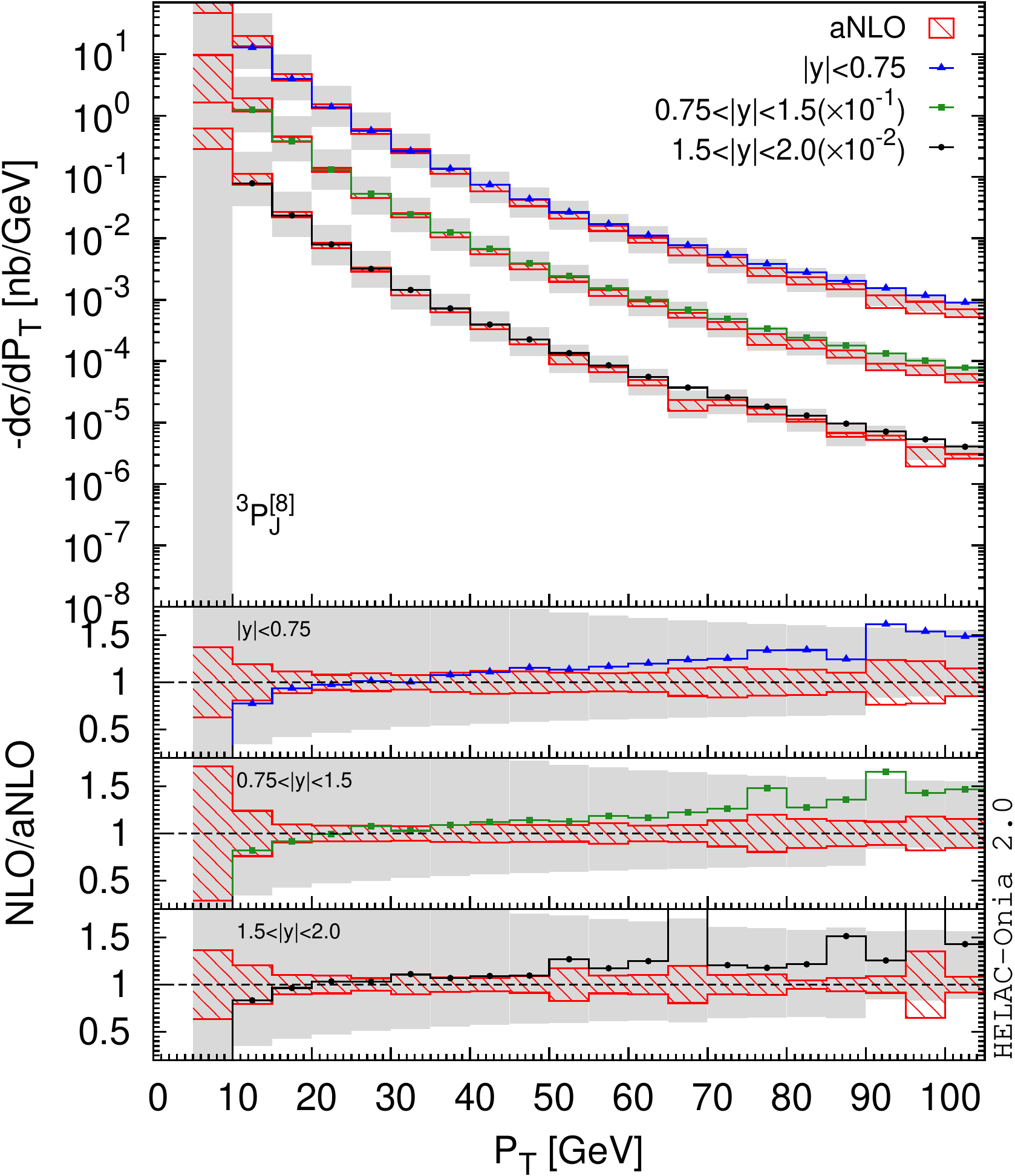}\\
\includegraphics[width=.45\textwidth,draft=false]{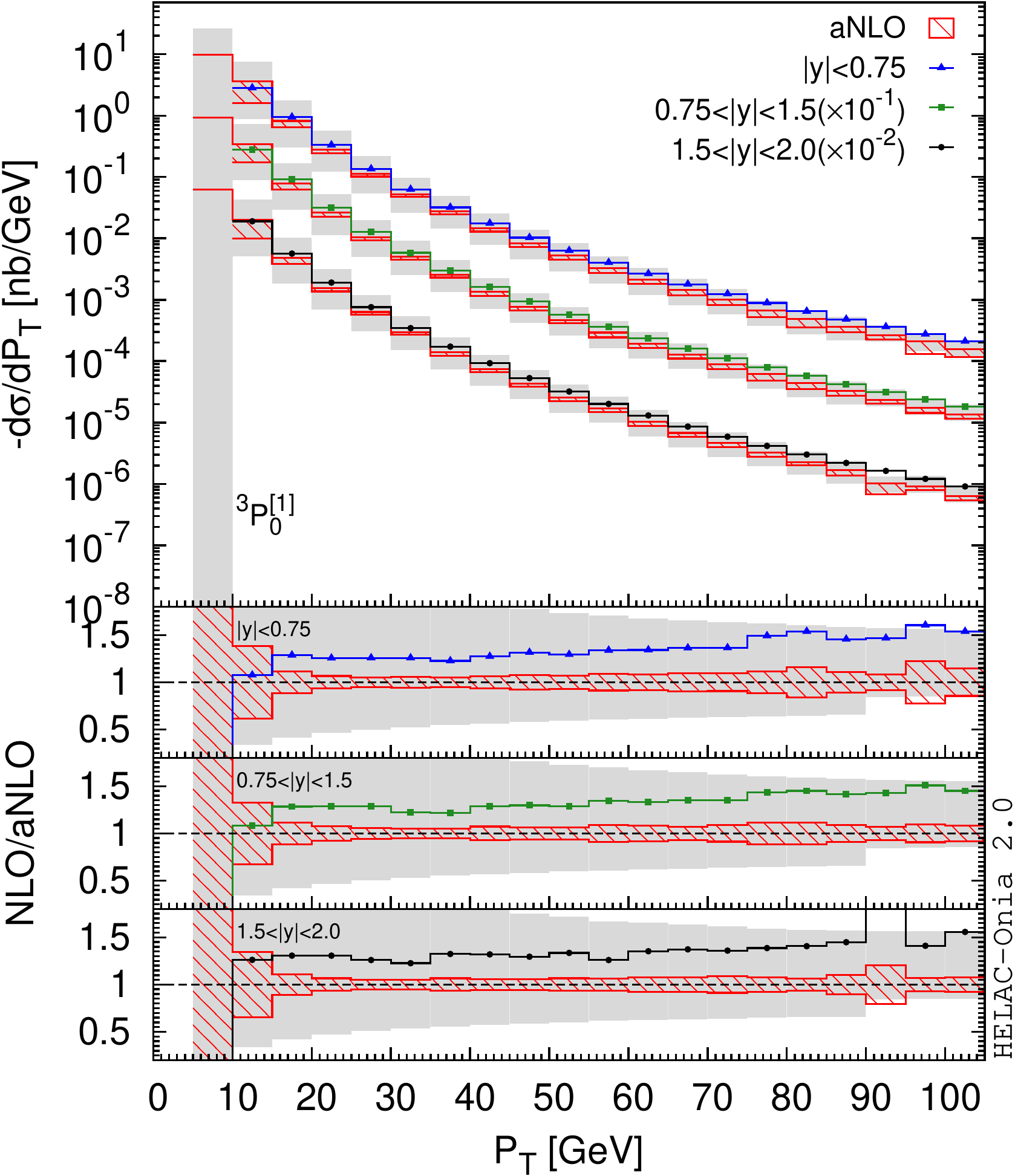}
\includegraphics[width=.45\textwidth,draft=false]{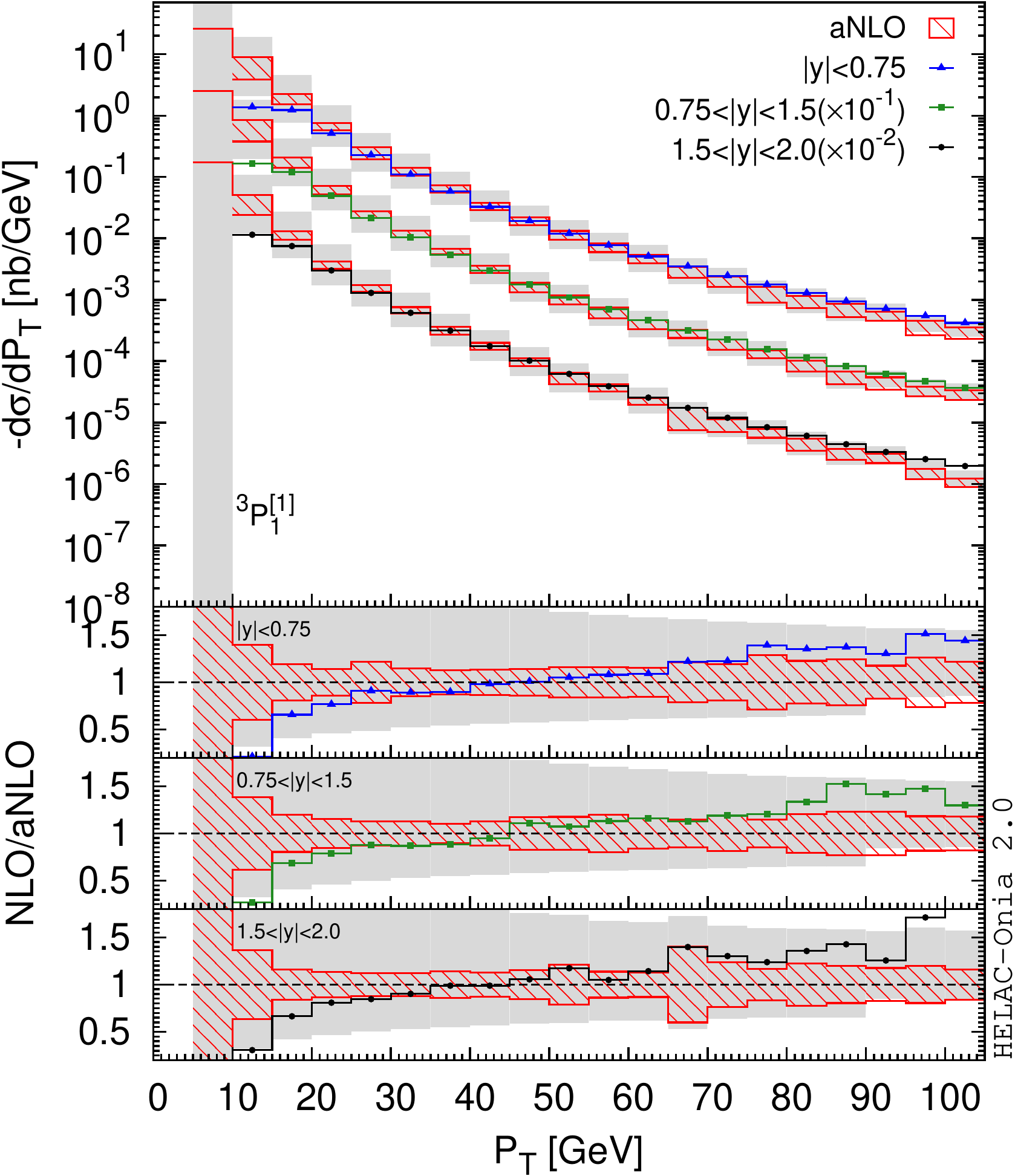}\\
\includegraphics[width=.45\textwidth,draft=false]{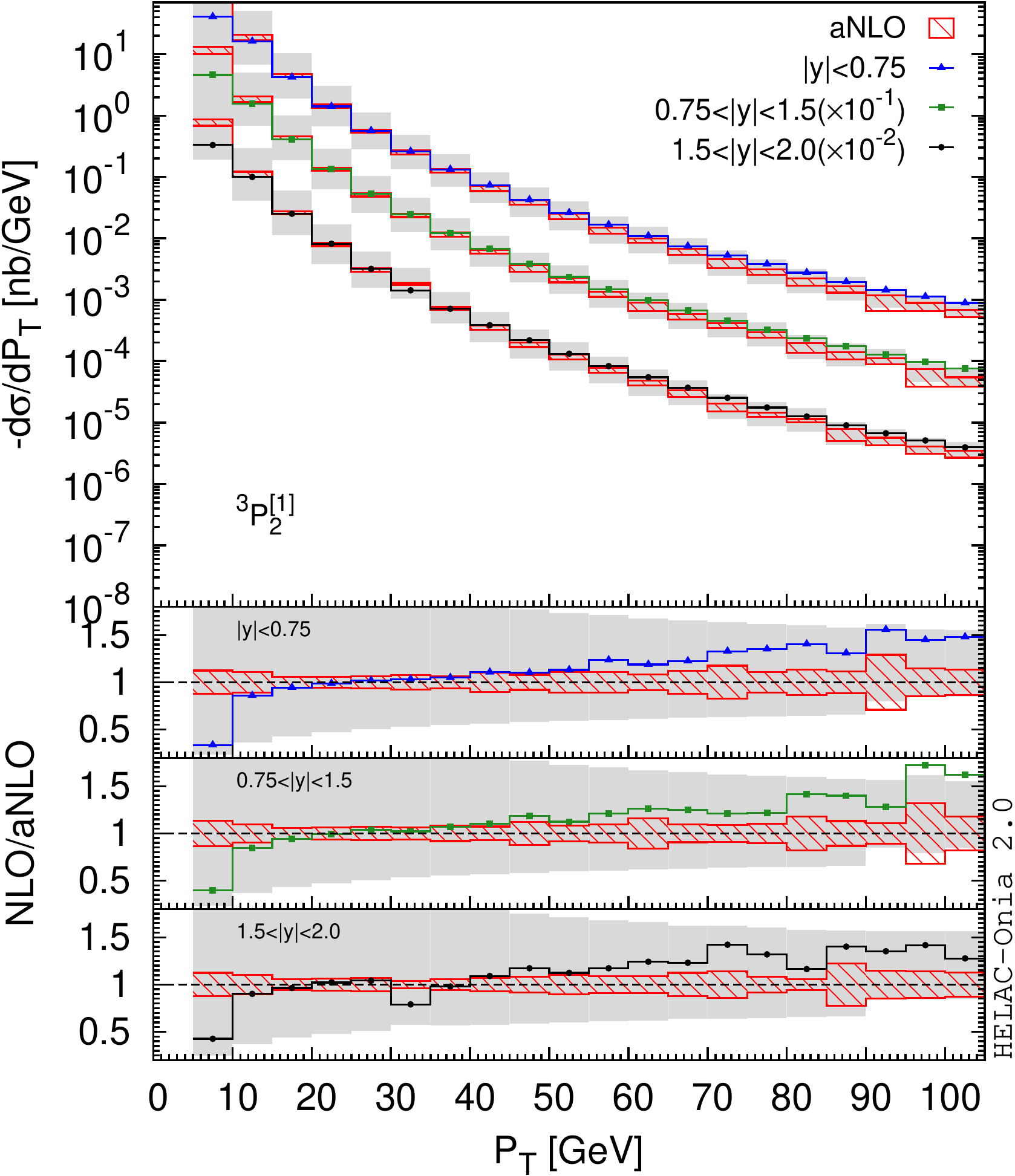}
\caption{Comparisons of spin-summed differential cross sections for the Fock states $\sps,\pj,\tpzs,\tpos,\tpts$ between our aNLO calculations and the complete NLO calculations. They are similar to Fig.~\protect\ref{Fig:aNLOvsNLO}. \label{Fig:aNLOvsNLO2}}
\end{figure}

\begin{figure}[ht!]
\centering
\includegraphics[width=.45\textwidth,draft=false]{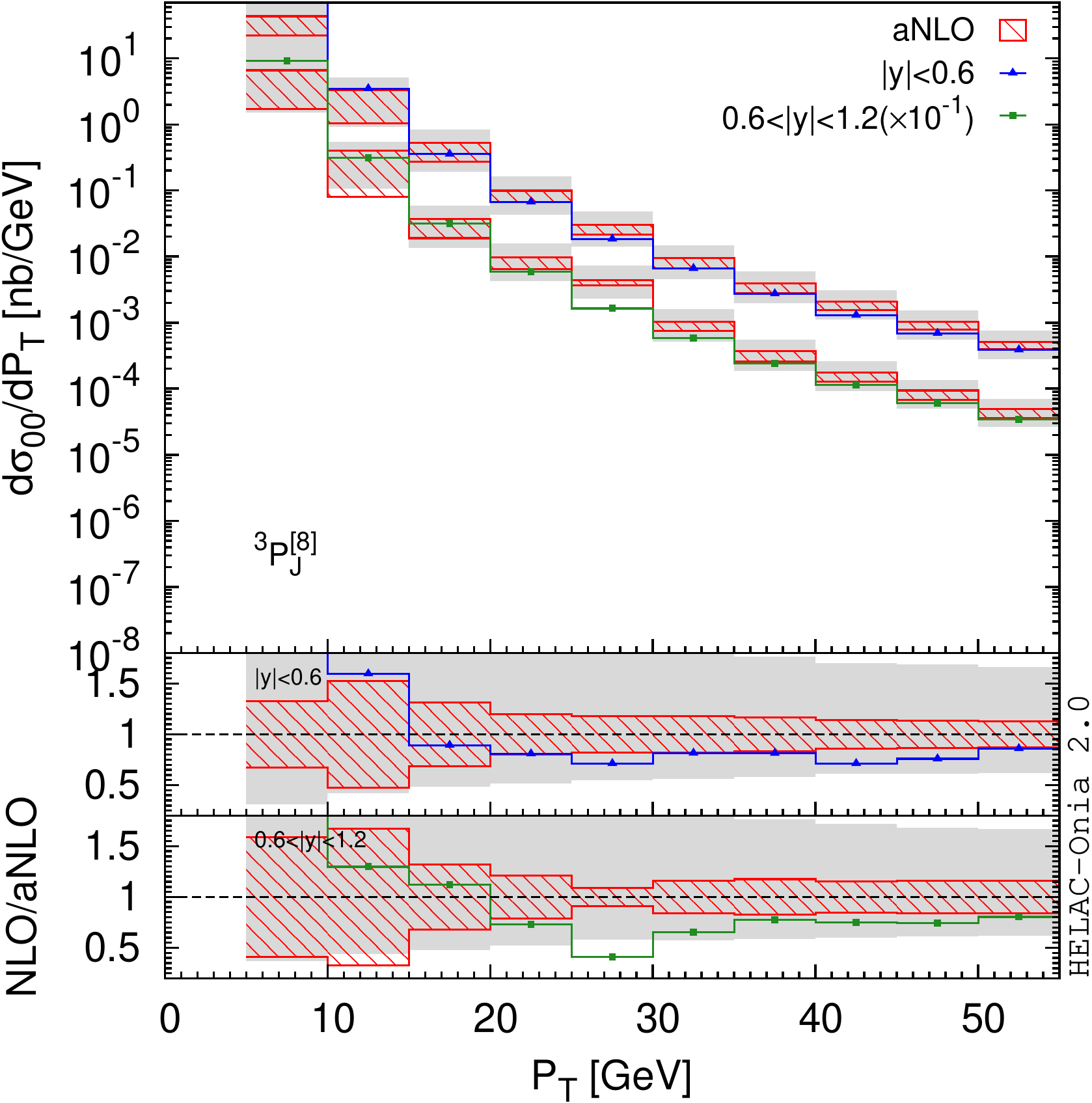}
\includegraphics[width=.45\textwidth,draft=false]{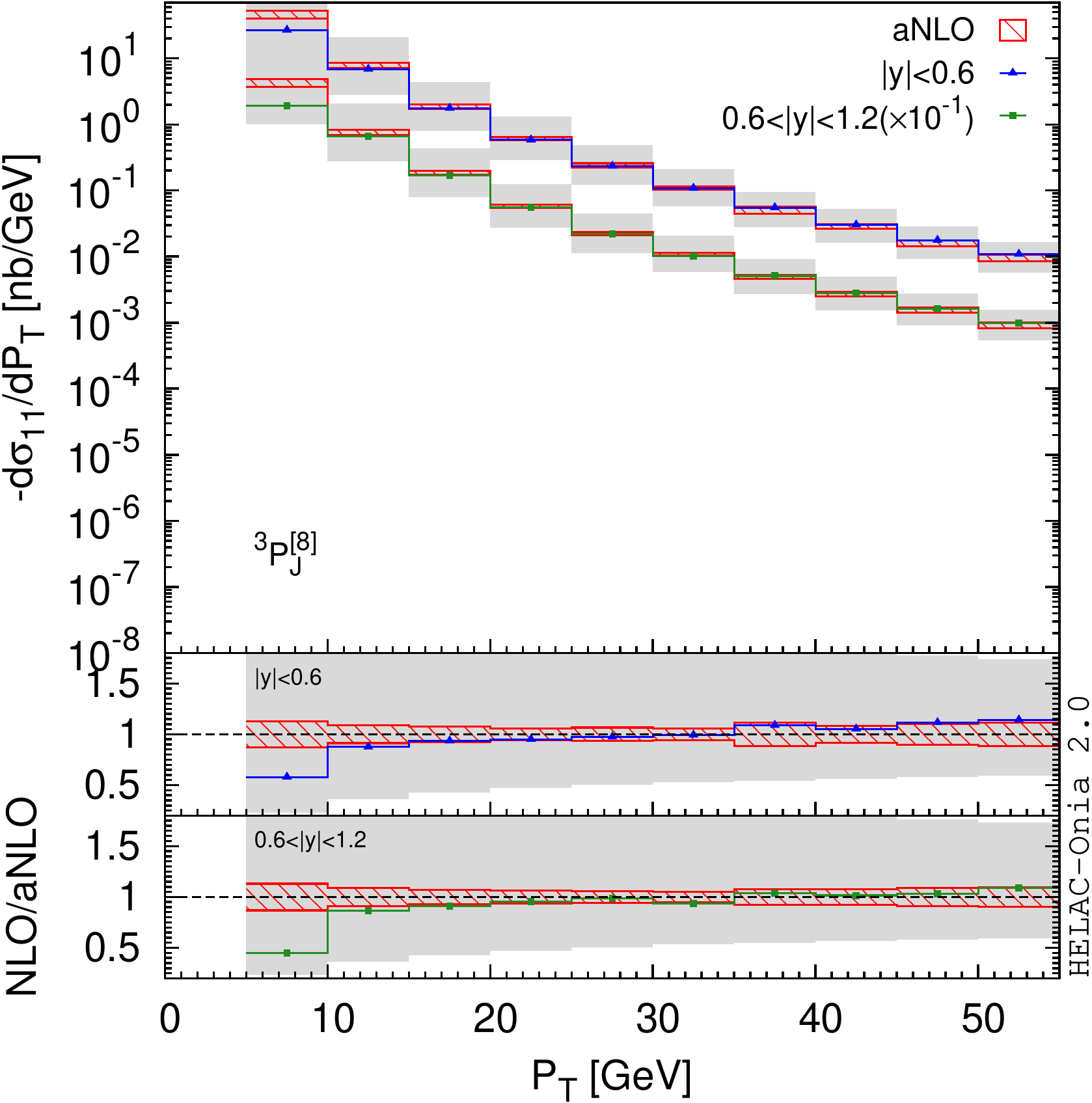}\\
\includegraphics[width=.45\textwidth,draft=false]{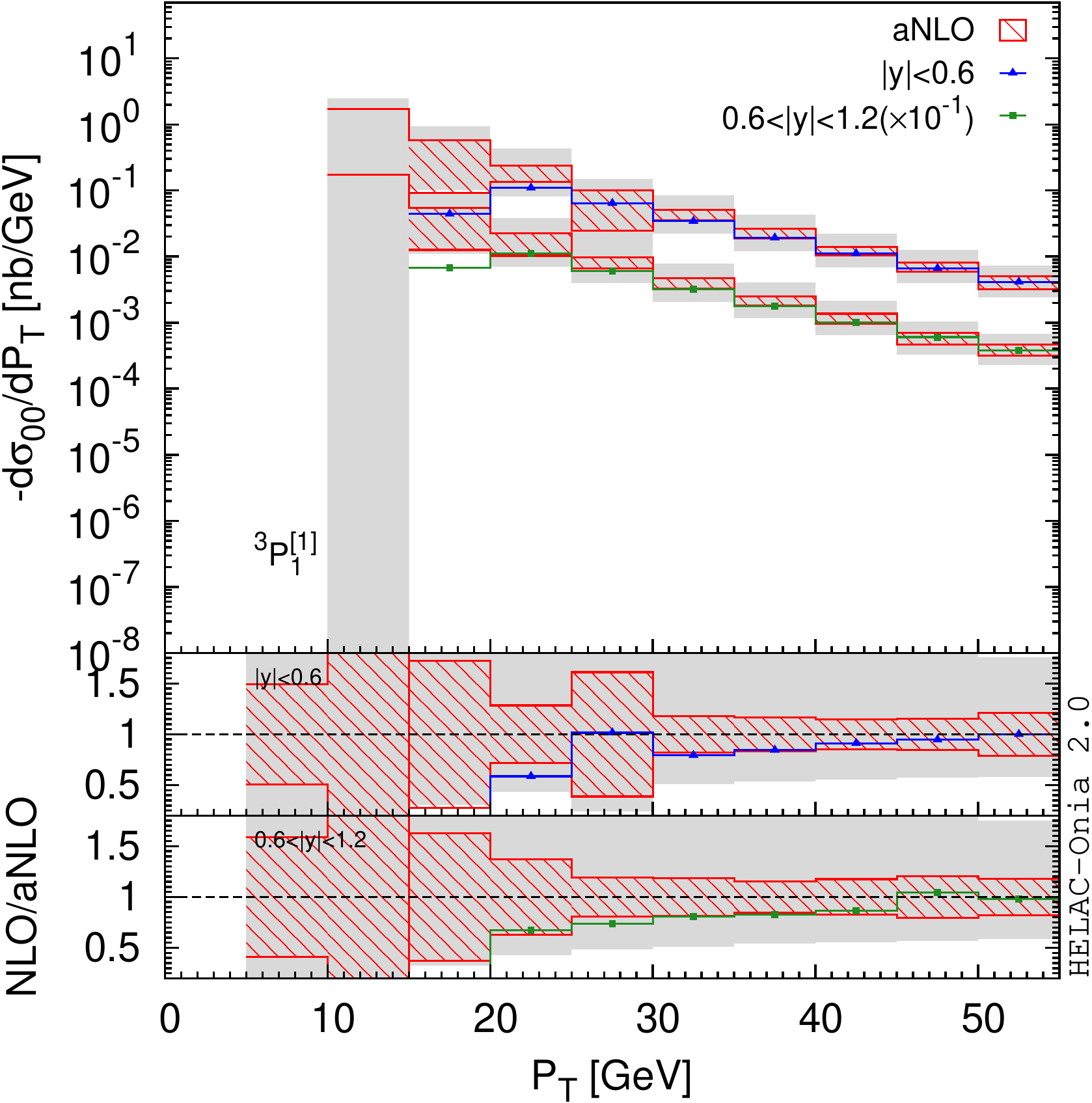}
\includegraphics[width=.45\textwidth,draft=false]{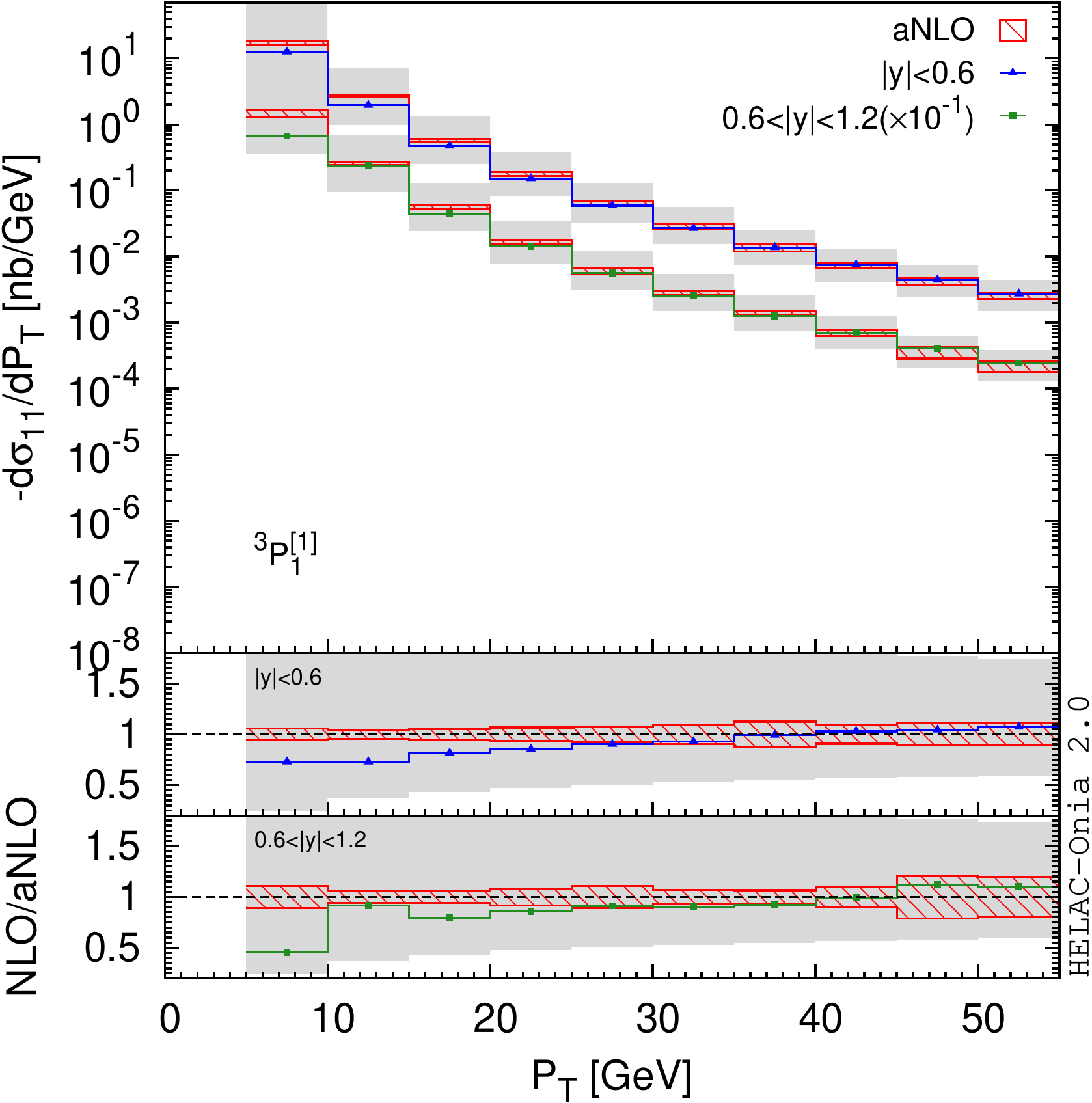}
\caption{Comparisons of spin-dependent differential cross sections for the 2 Fock states $\pj,\tpos$ between our aNLO calculations and the complete NLO calculations. They are similar to Fig.~\protect\ref{Fig:aNLOvsNLOSpin}. \label{Fig:aNLOvsNLOSpin2}}
\end{figure}

\begin{figure}[ht!]
\centering
\includegraphics[width=.45\textwidth,draft=false]{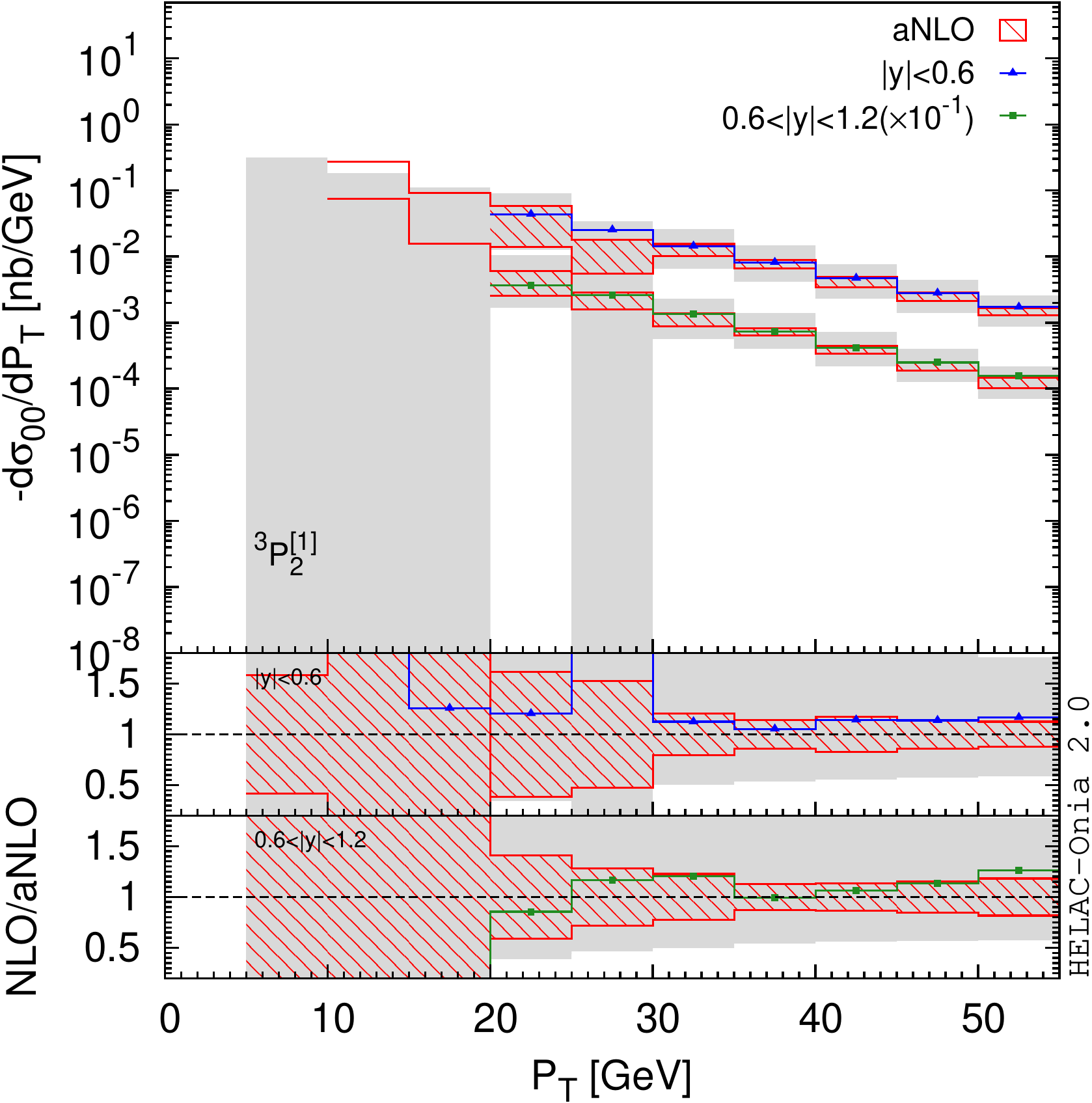}
\includegraphics[width=.45\textwidth,draft=false]{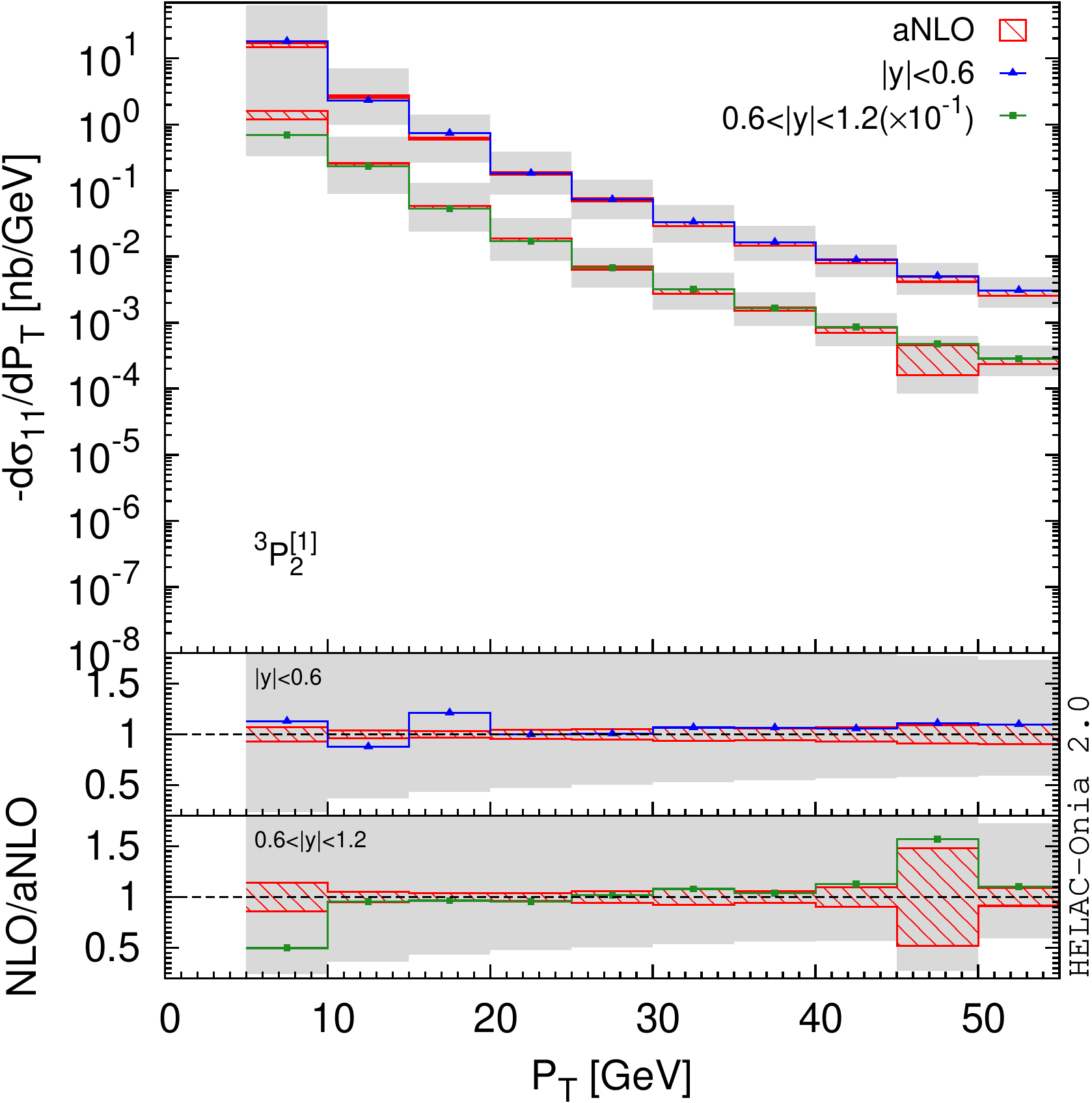}\\
\includegraphics[width=.45\textwidth,draft=false]{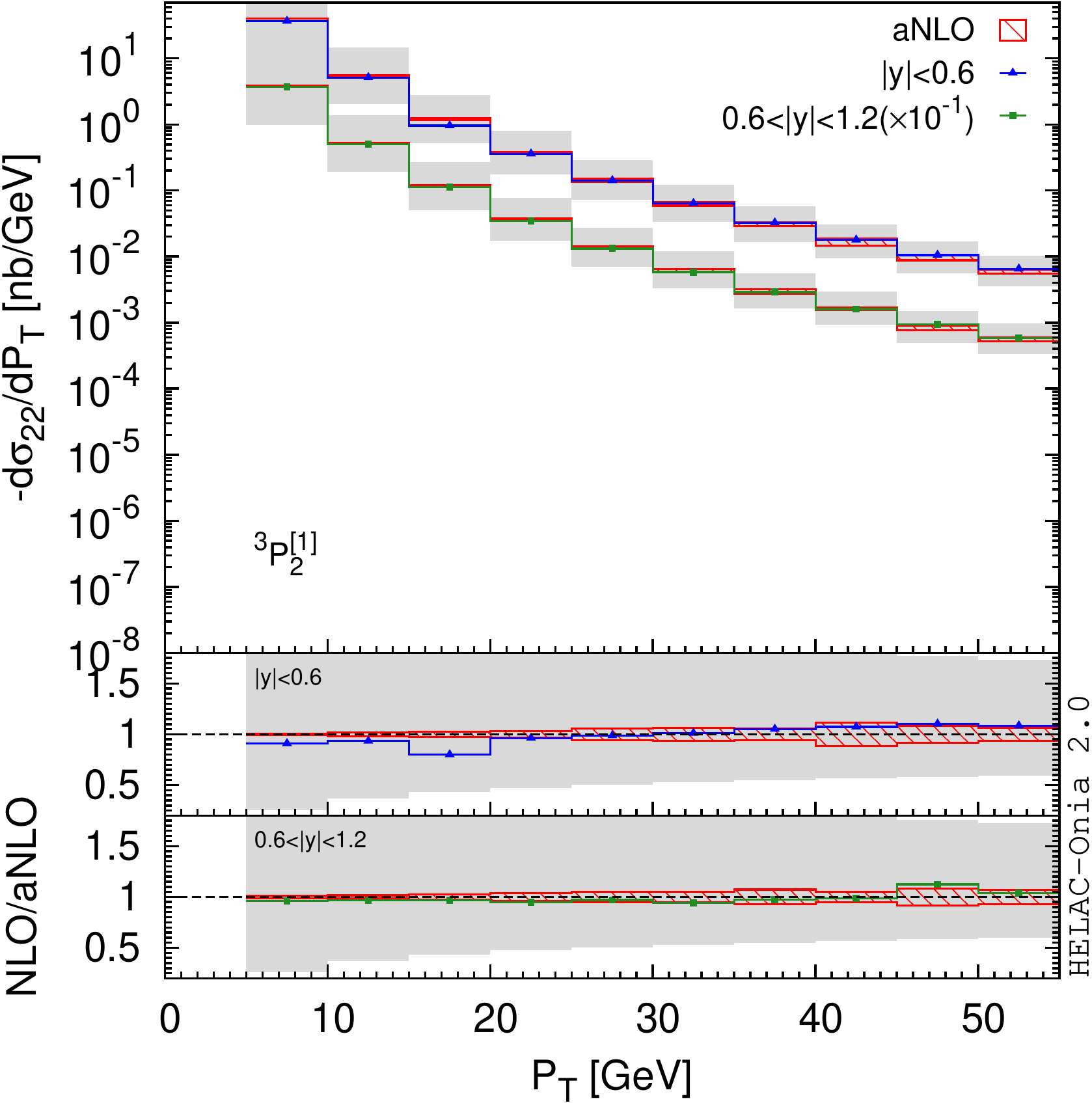}
\caption{Comparisons of spin-dependent differential cross sections for the Fock state $\tpts$ between our aNLO calculations and the complete NLO calculations. They are similar to Fig.~\protect\ref{Fig:aNLOvsNLOSpin}. \label{Fig:aNLOvsNLOSpin3}}
\end{figure}

\begin{figure}[ht!]
\vspace{-1cm}
\centering
\includegraphics[width=.45\textwidth,draft=false]{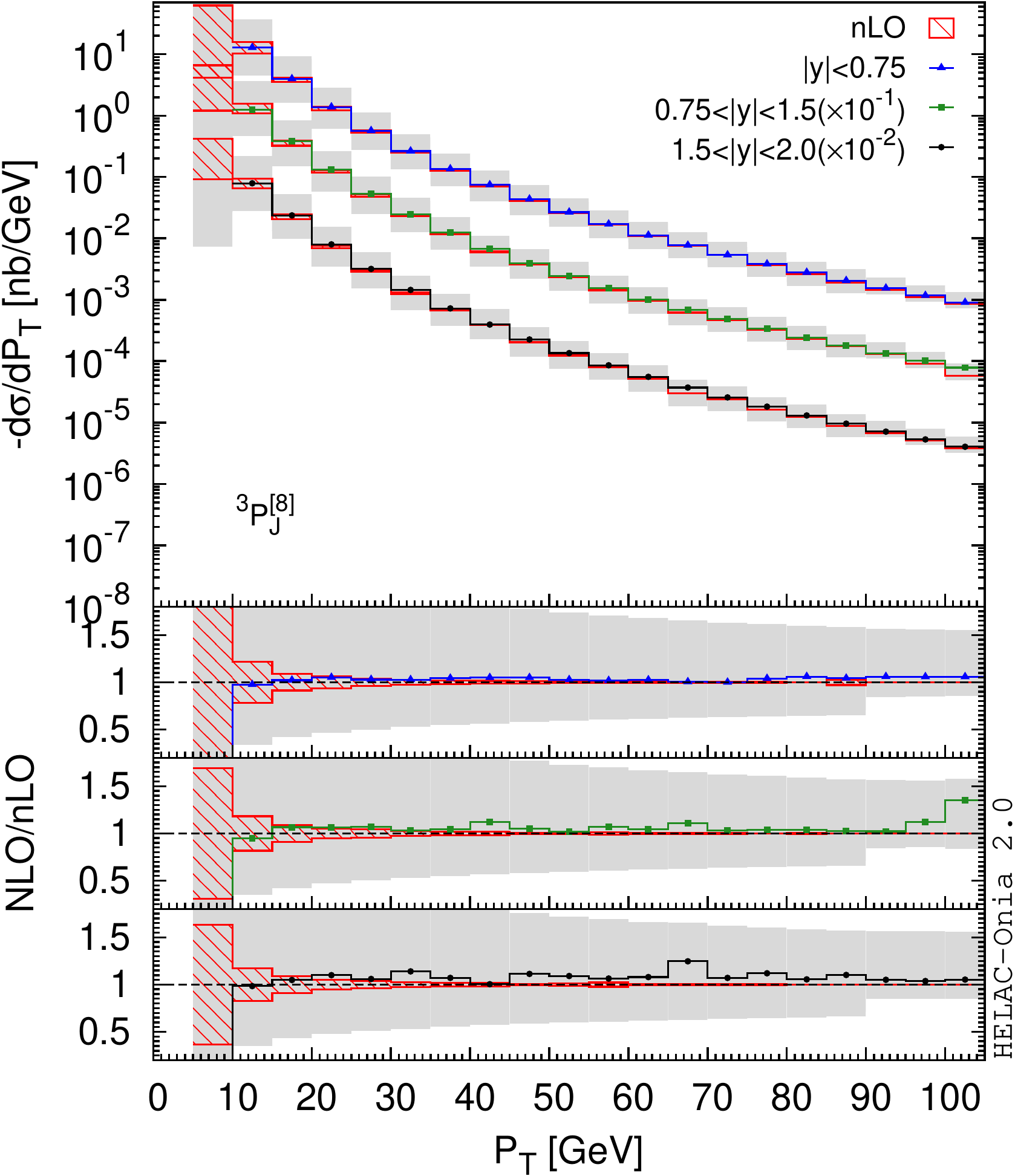}
\includegraphics[width=.45\textwidth,draft=false]{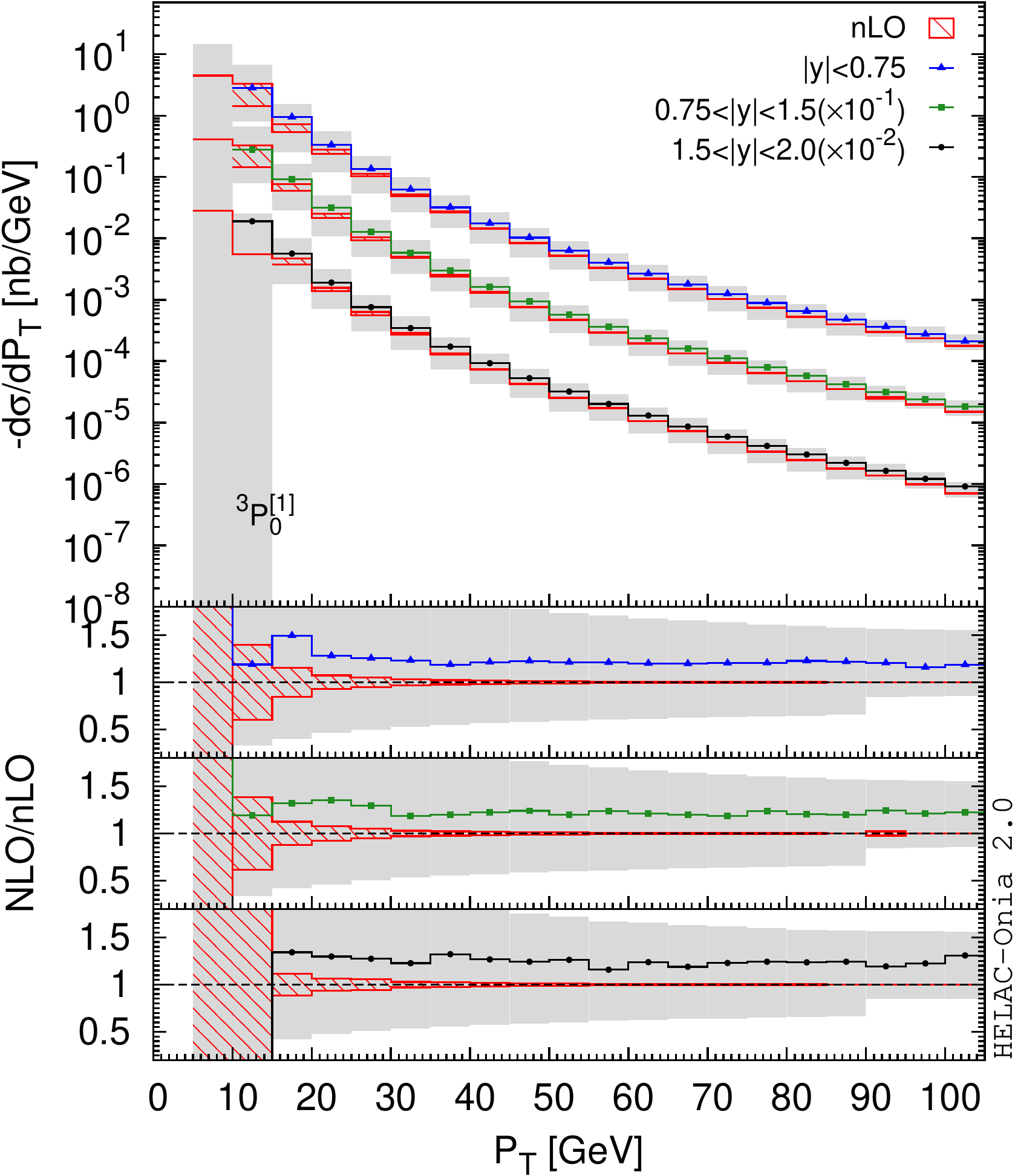}\\
\includegraphics[width=.45\textwidth,draft=false]{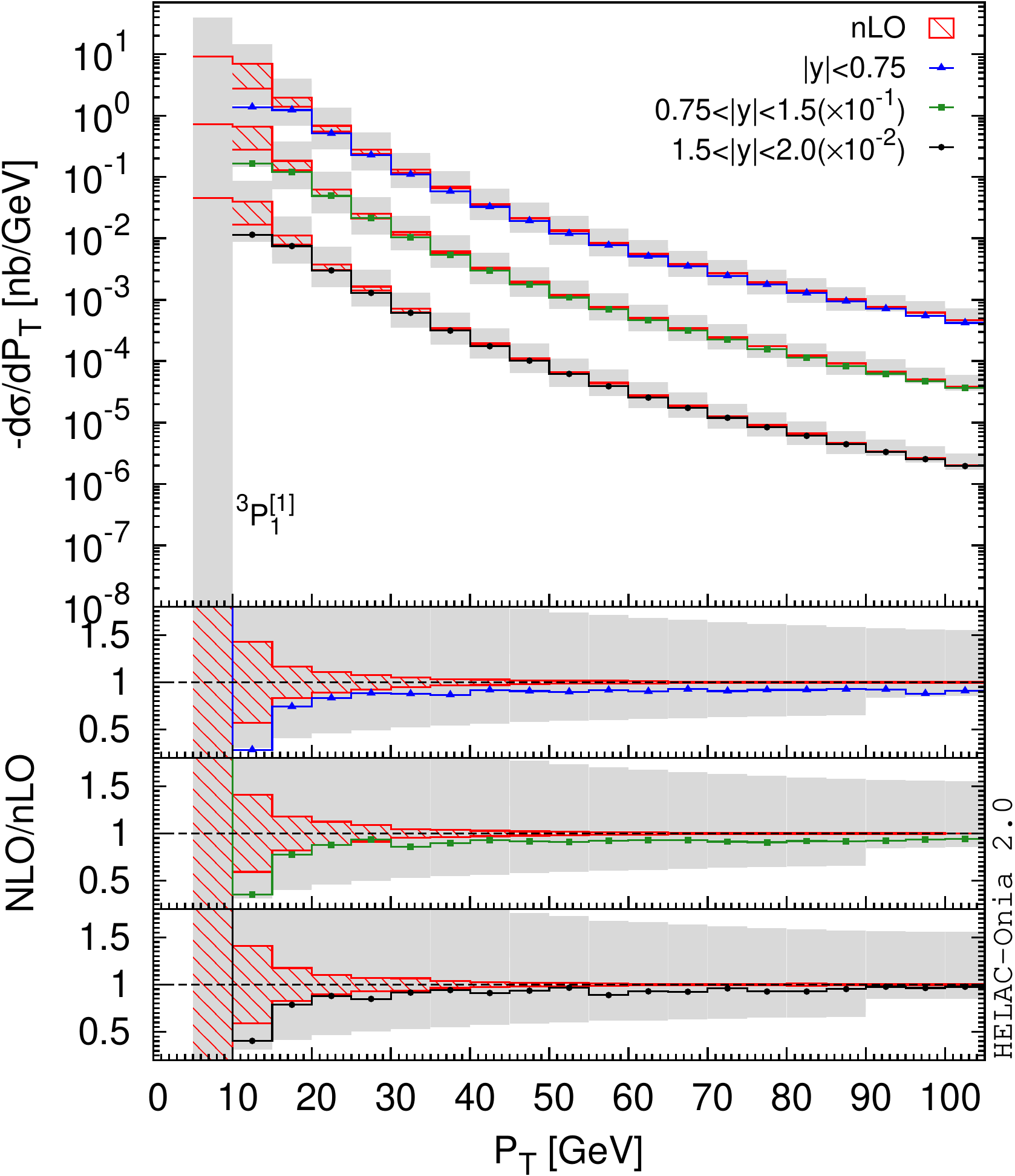}
\includegraphics[width=.45\textwidth,draft=false]{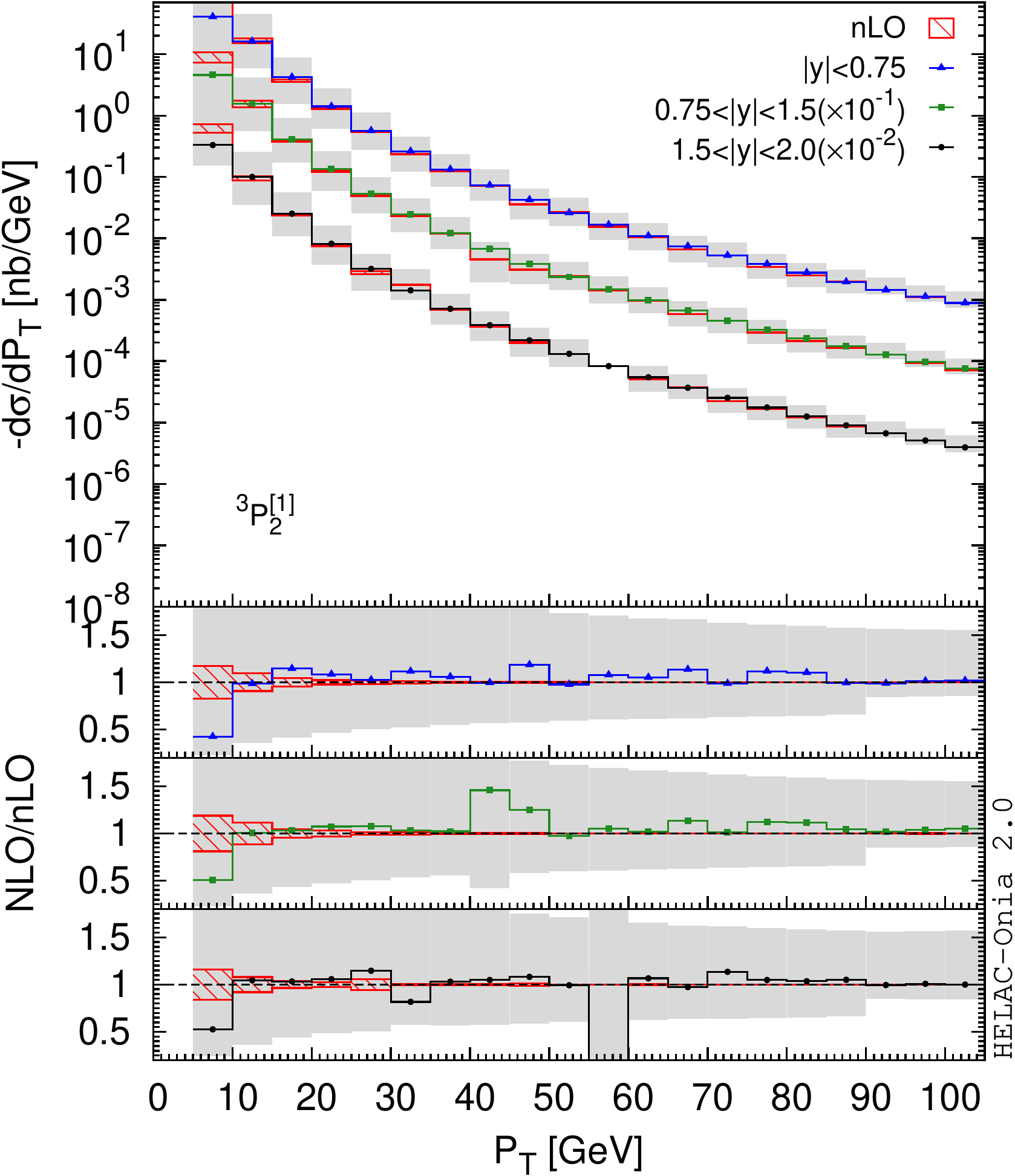}
\caption{Comparisons of spin-summed differential cross sections for the Fock states $\pj,\tpzs,\tpos,\tpts$ between our nLO calculations and the complete NLO calculations. They are similar to Fig.~\protect\ref{Fig:nLOvsNLO}. \label{Fig:nLOvsNLO2}}
\end{figure}

\begin{figure}[ht!]
\centering
\includegraphics[width=.45\textwidth,draft=false]{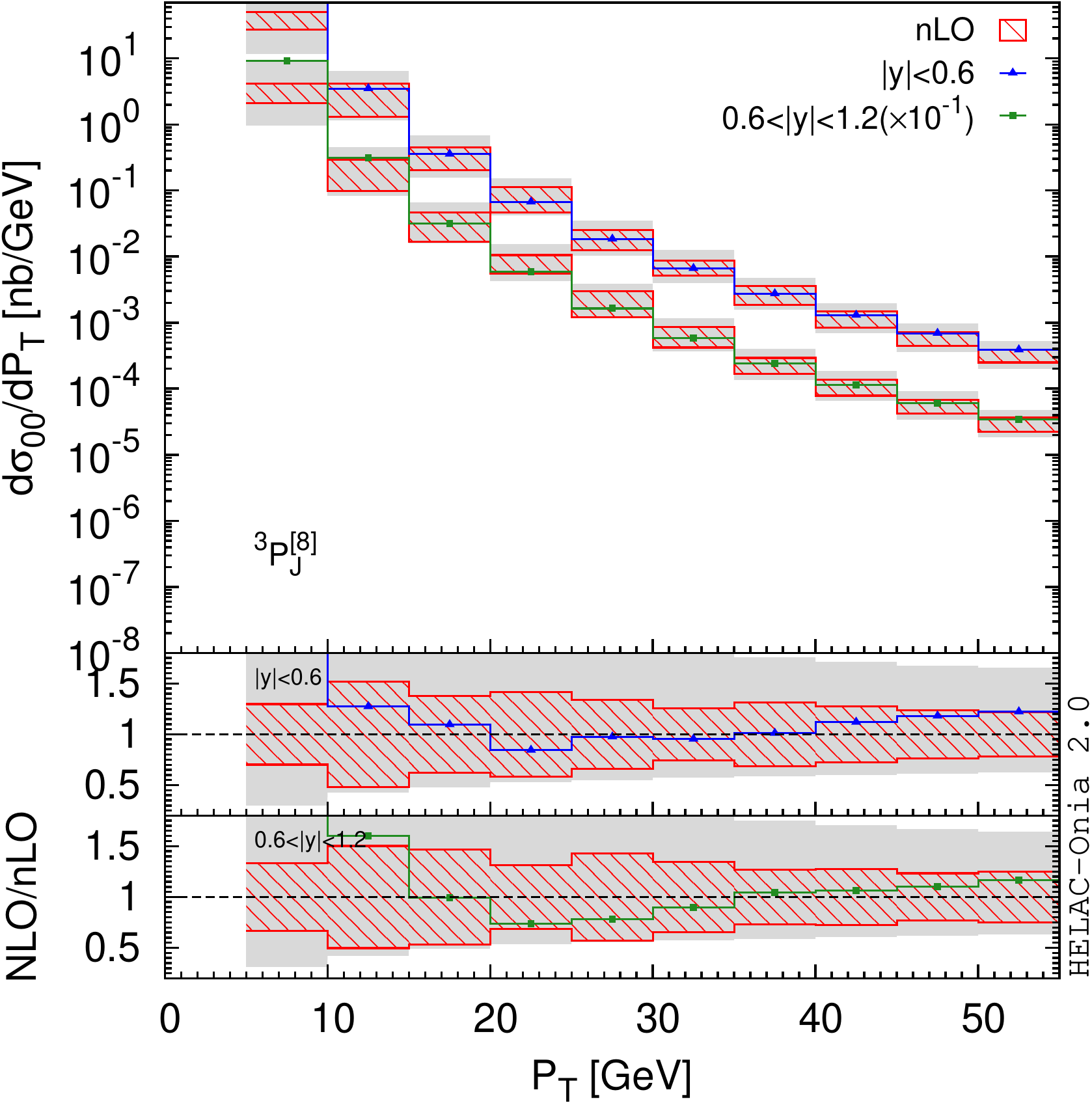}
\includegraphics[width=.45\textwidth,draft=false]{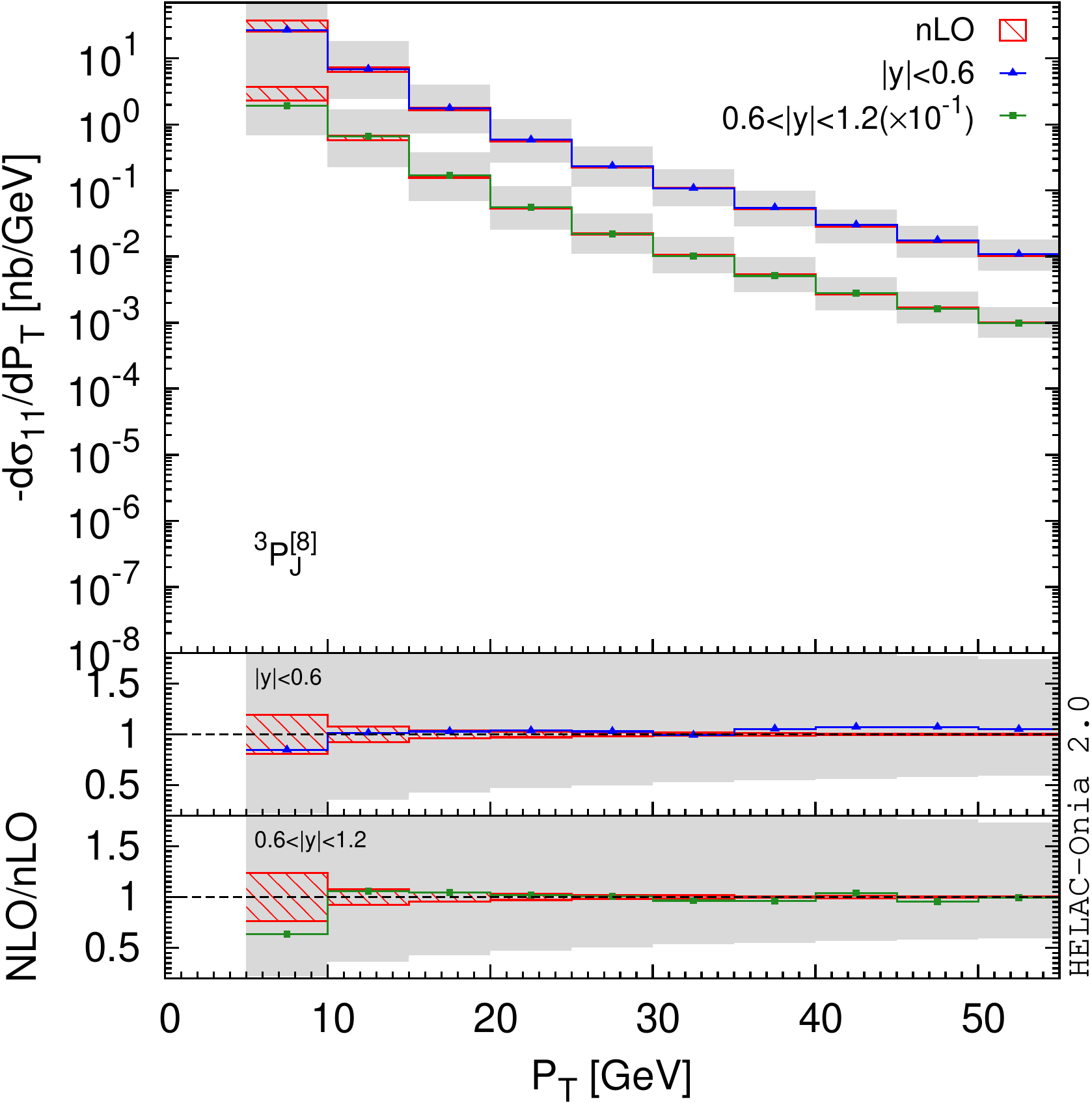}\\
\includegraphics[width=.45\textwidth,draft=false]{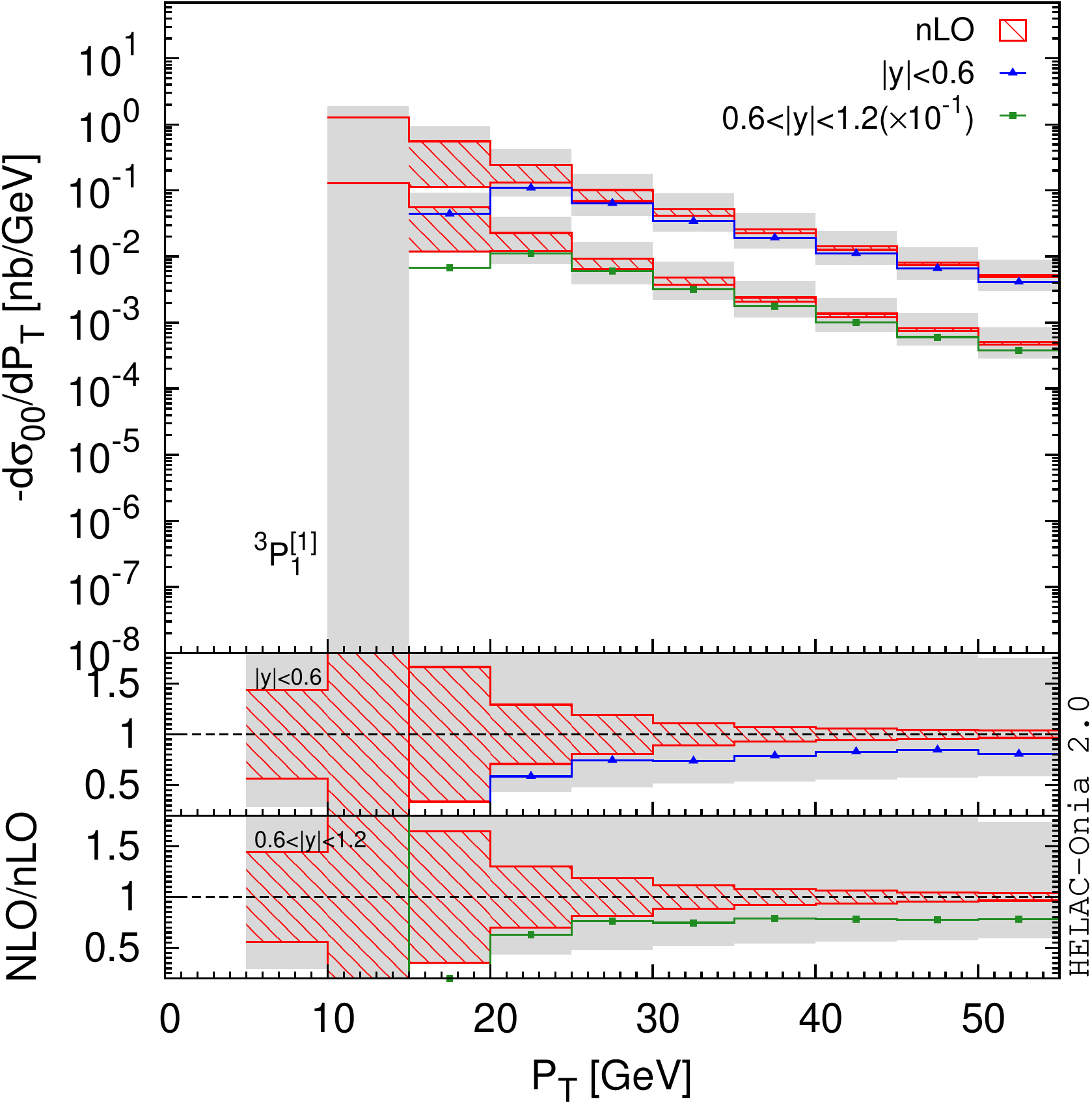}
\includegraphics[width=.45\textwidth,draft=false]{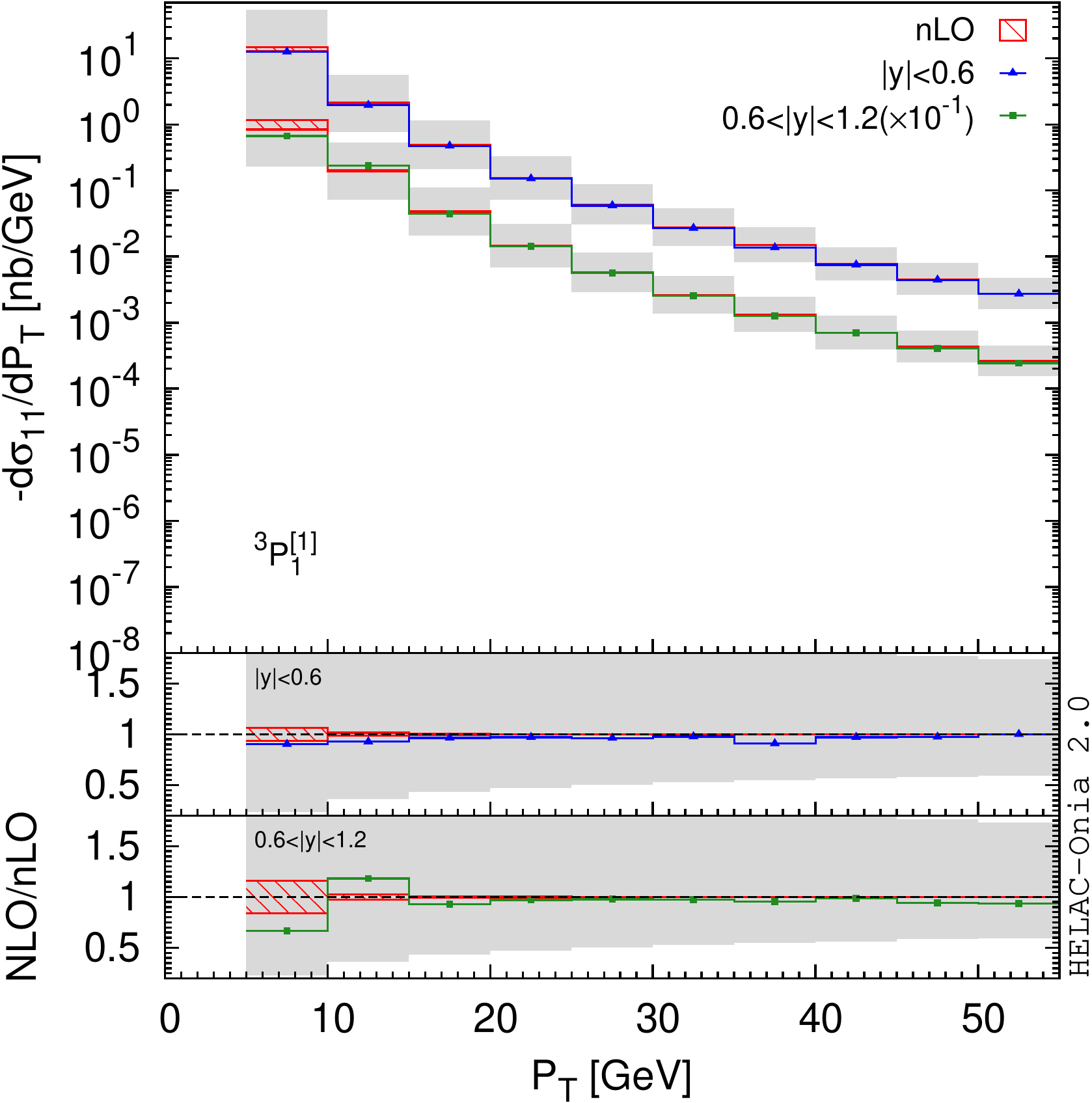}
\caption{Comparisons of spin-dependent differential cross sections for the 3 Fock states $\pj,\tpos$ between our nLO calculations and the complete NLO calculations. They are similar to Fig.~\protect\ref{Fig:nLOvsNLOSpin}. \label{Fig:nLOvsNLOSpin2}}
\end{figure}

\begin{figure}[ht!]
\centering
\includegraphics[width=.45\textwidth,draft=false]{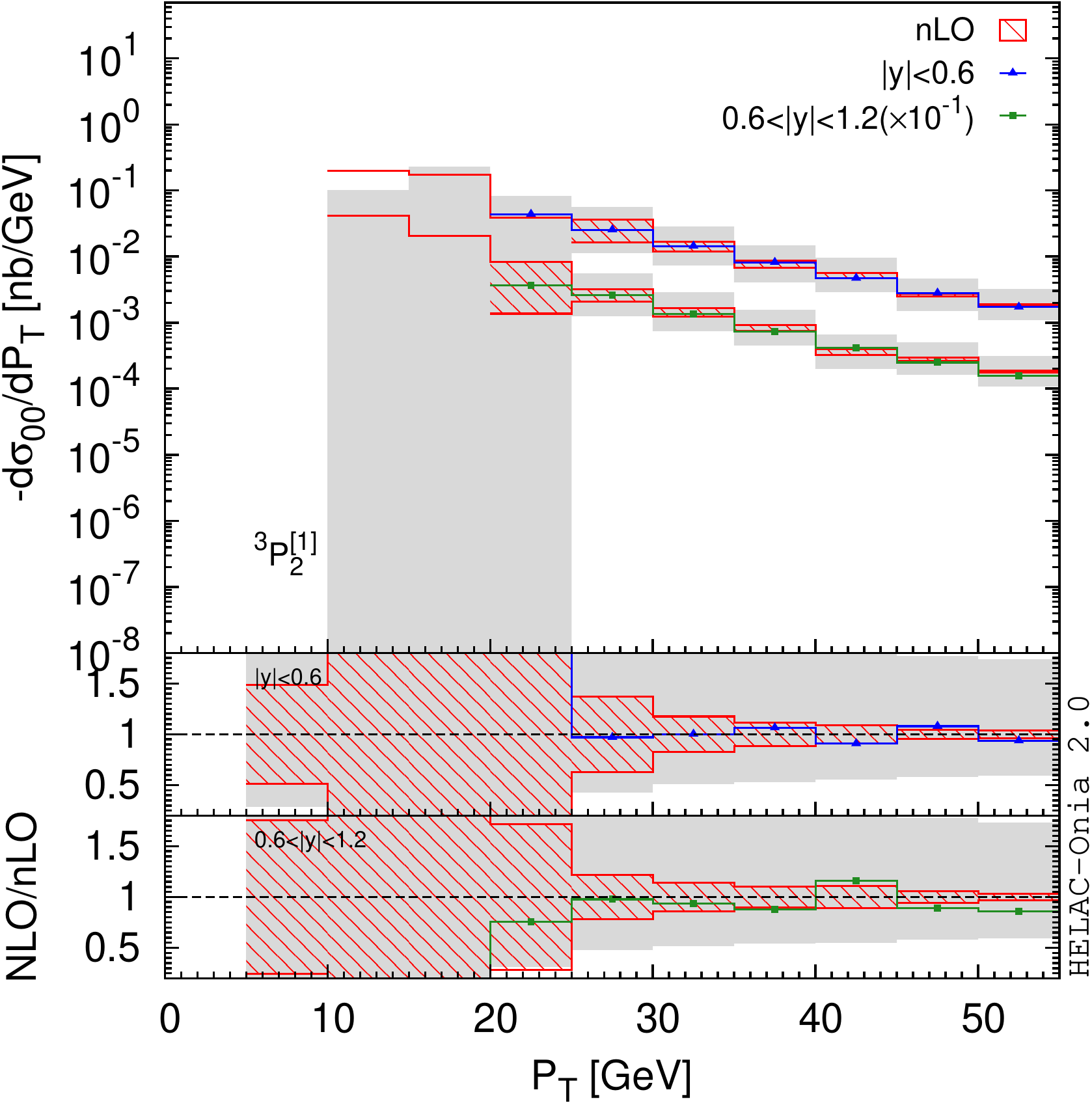}
\includegraphics[width=.45\textwidth,draft=false]{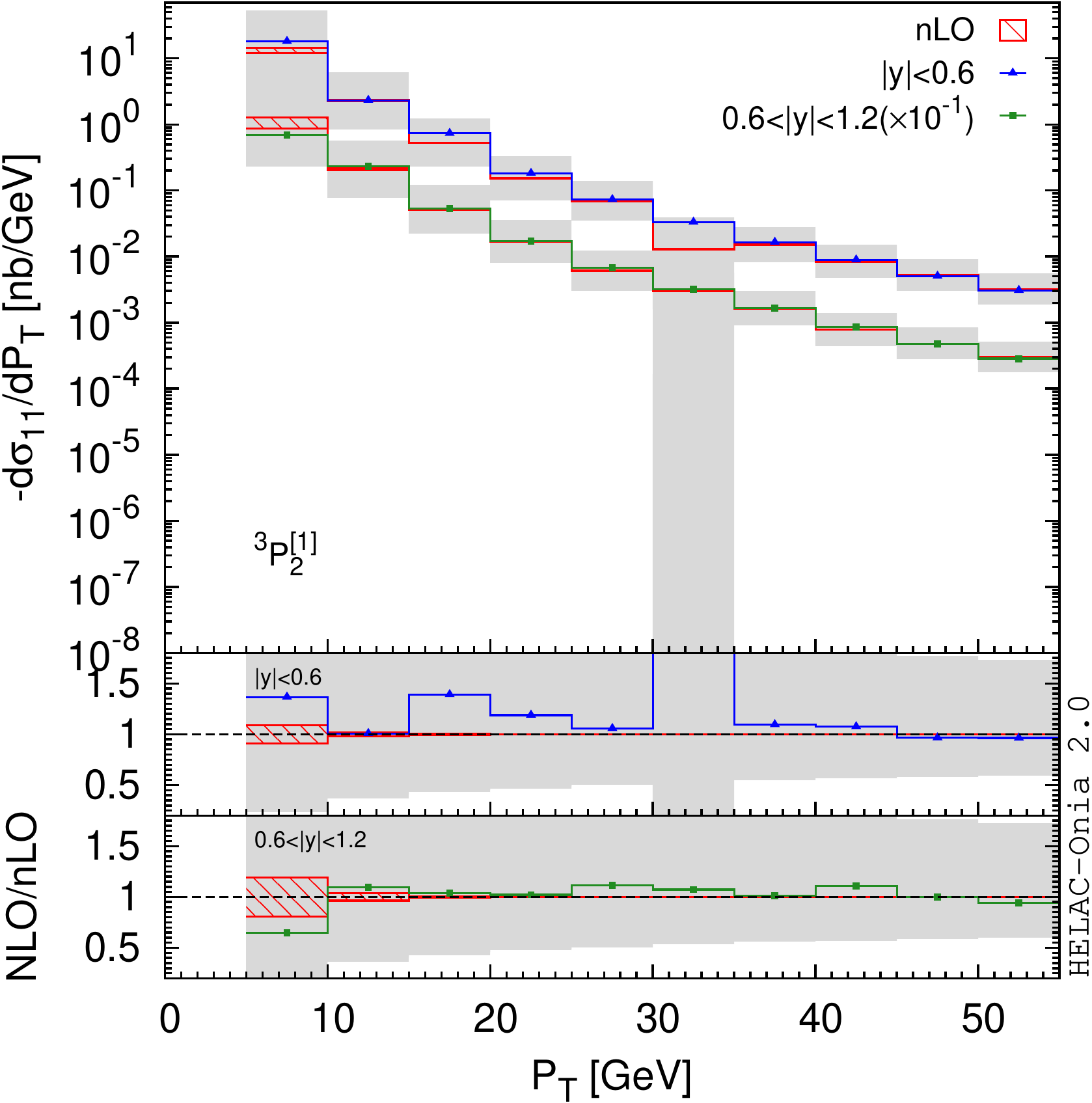}\\
\includegraphics[width=.45\textwidth,draft=false]{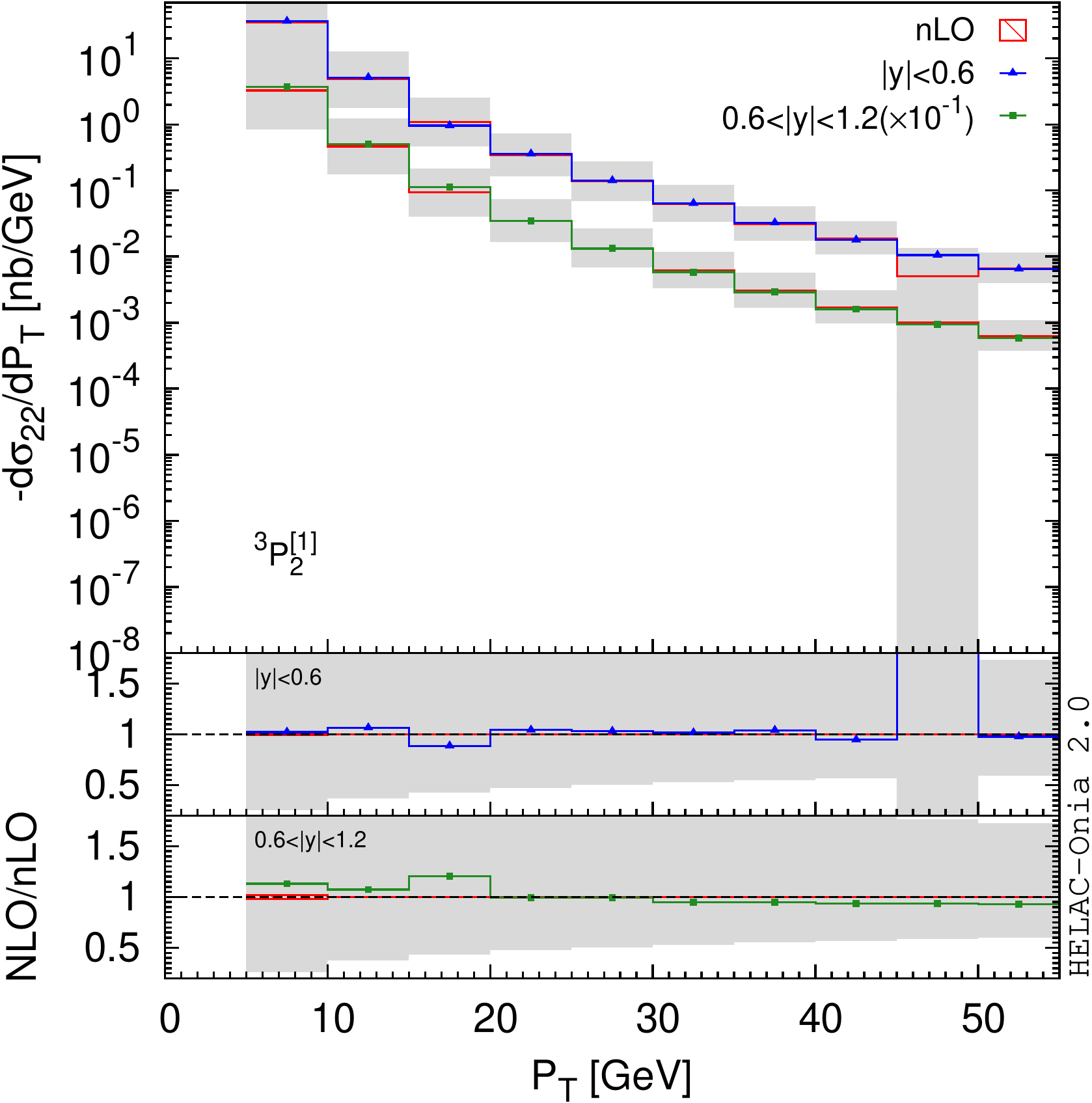}
\caption{Comparisons of spin-dependent differential cross sections for the Fock state $\tpts$ between our nLO calculations and the complete NLO calculations. They are similar to Fig.~\protect\ref{Fig:nLOvsNLOSpin}. \label{Fig:nLOvsNLOSpin3}}
\end{figure}

\begin{figure}[ht!]
\vspace{-1cm}
\centering
\includegraphics[width=.45\textwidth,draft=false]{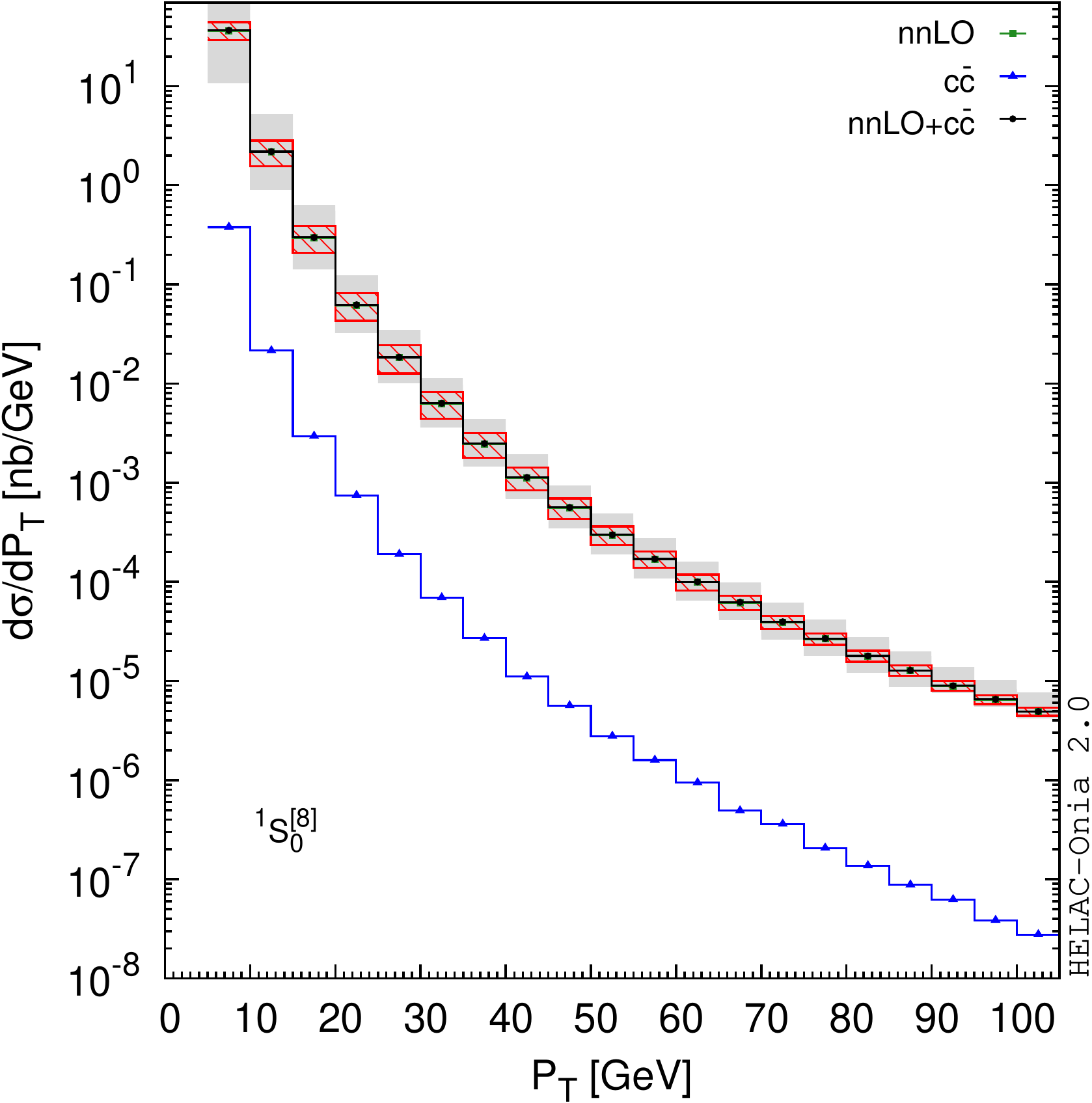}
\includegraphics[width=.45\textwidth,draft=false]{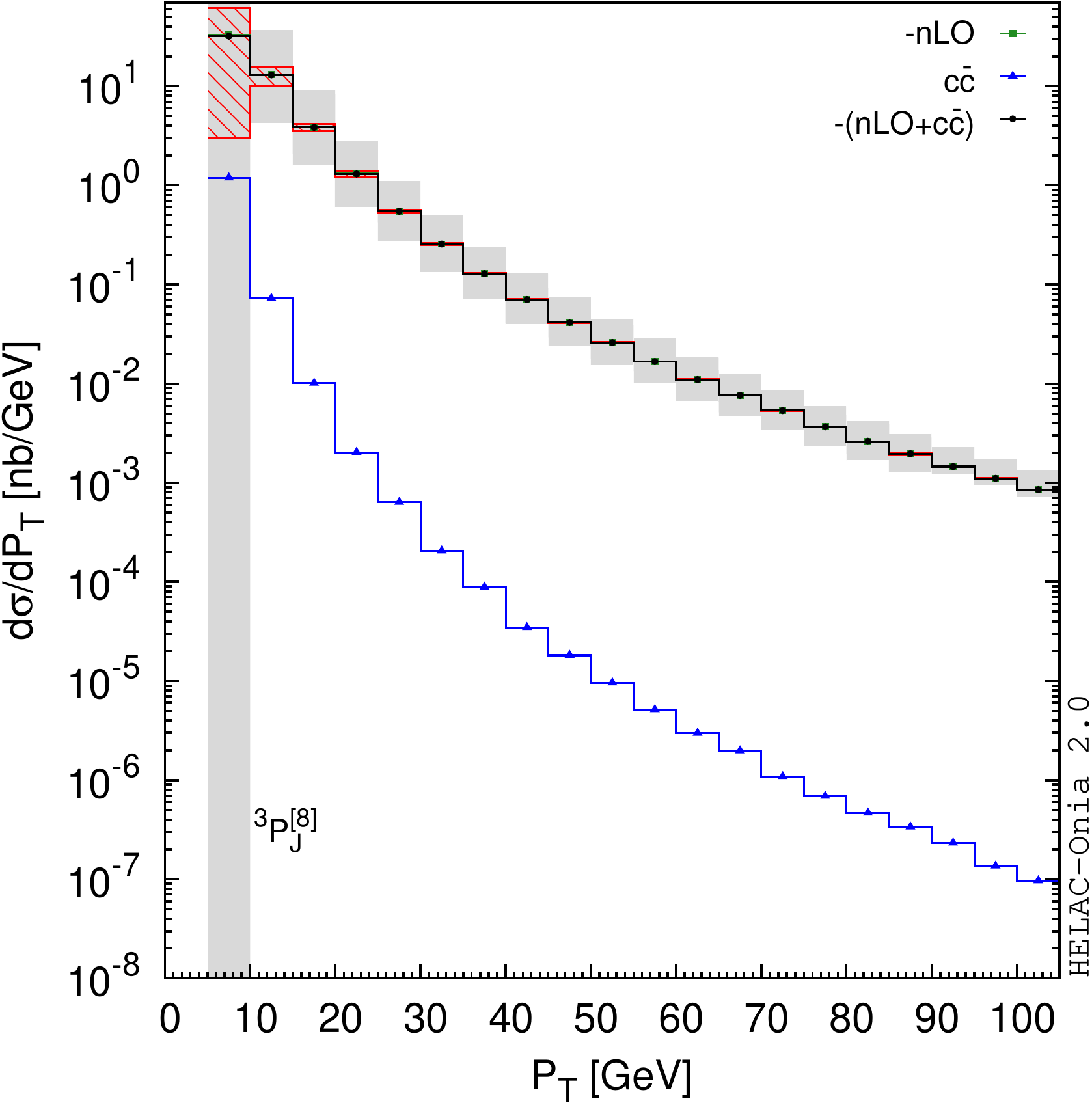}\\
\includegraphics[width=.45\textwidth,draft=false]{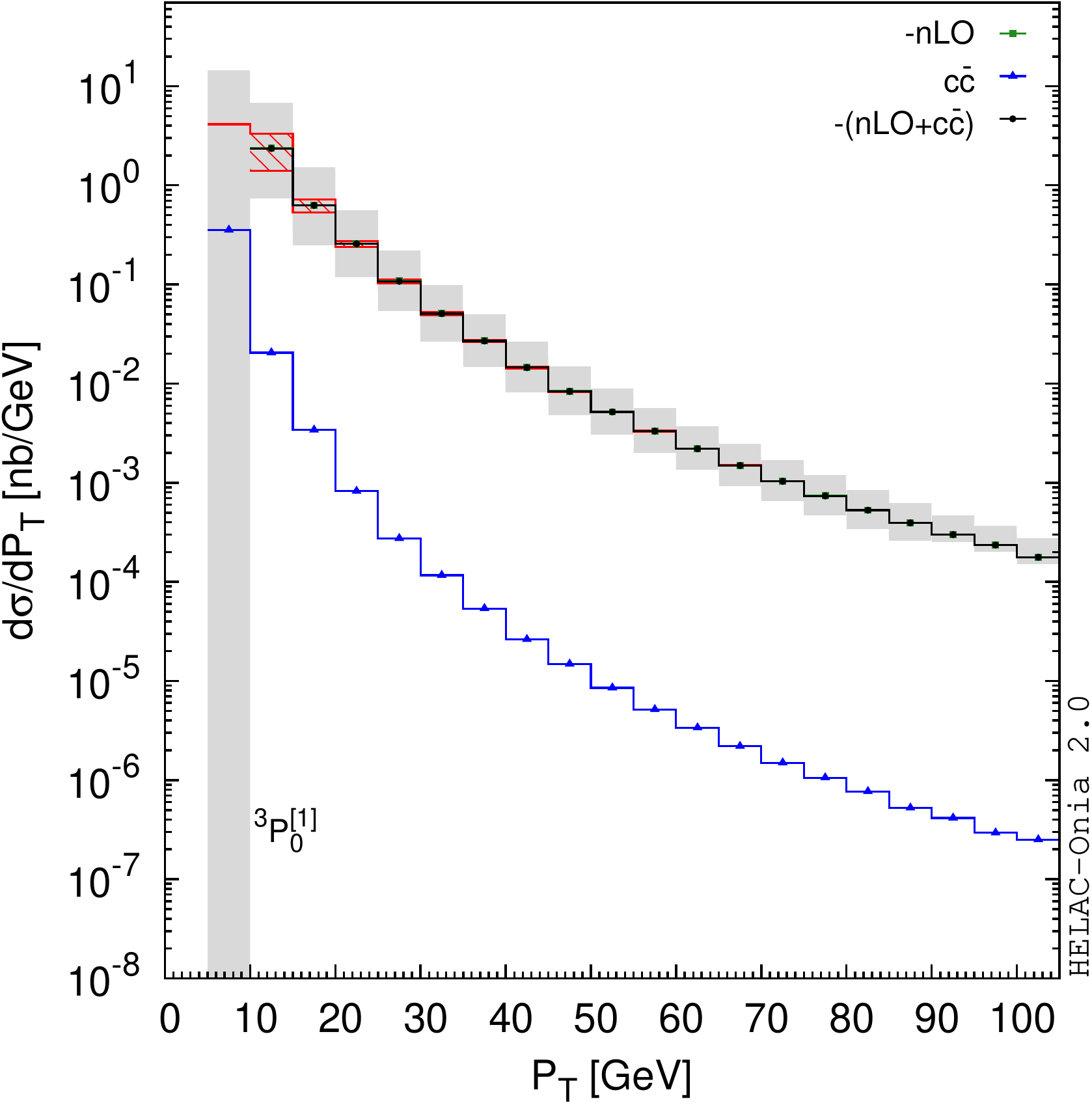}
\includegraphics[width=.45\textwidth,draft=false]{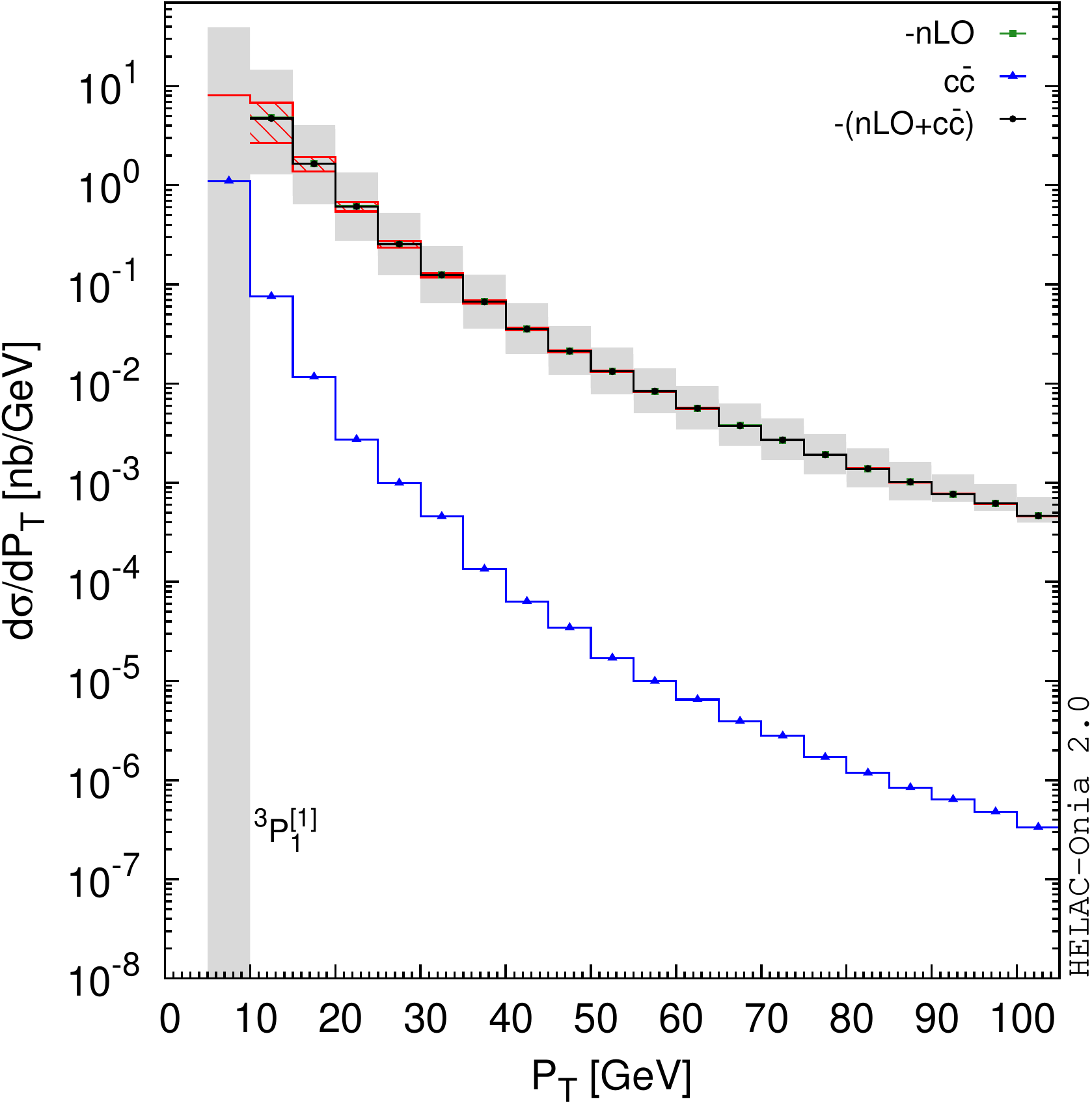}\\
\includegraphics[width=.45\textwidth,draft=false]{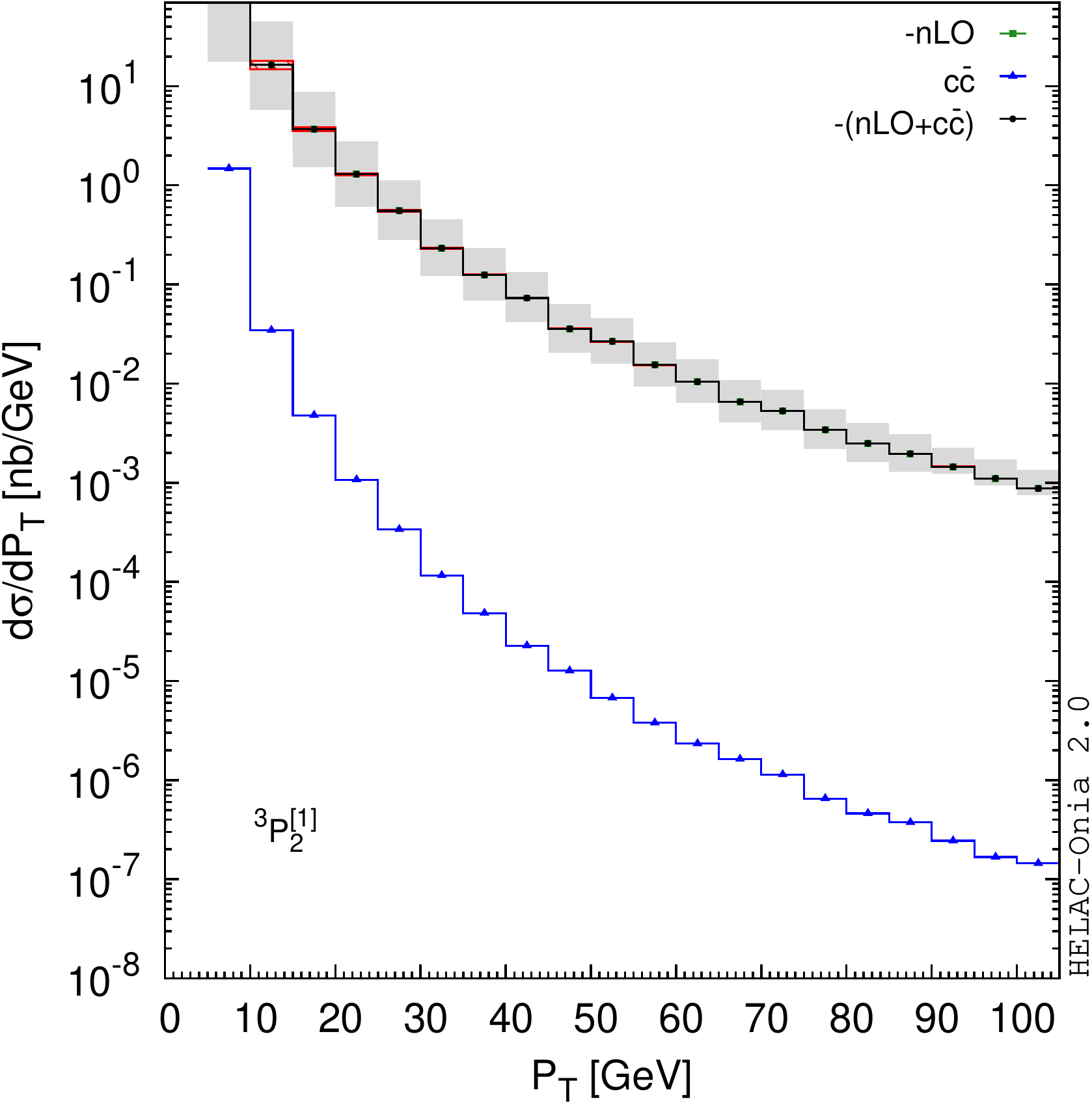}
\caption{Comparisons of spin-summed differential cross sections $\frac{d\sigma}{dP_T}$ for the 5 Fock states $\sps,\pj,\tpzs,\tpos,\tpts$ between our nLO calculations and the LO charmonium plus charm quark pair calculations. They are similar to Fig.~\protect\ref{Fig:nLOvsccx4CO}. \label{Fig:nLOvsccx4CO2}}
\end{figure}

\begin{figure}[ht!]
\centering
\includegraphics[width=.45\textwidth,draft=false]{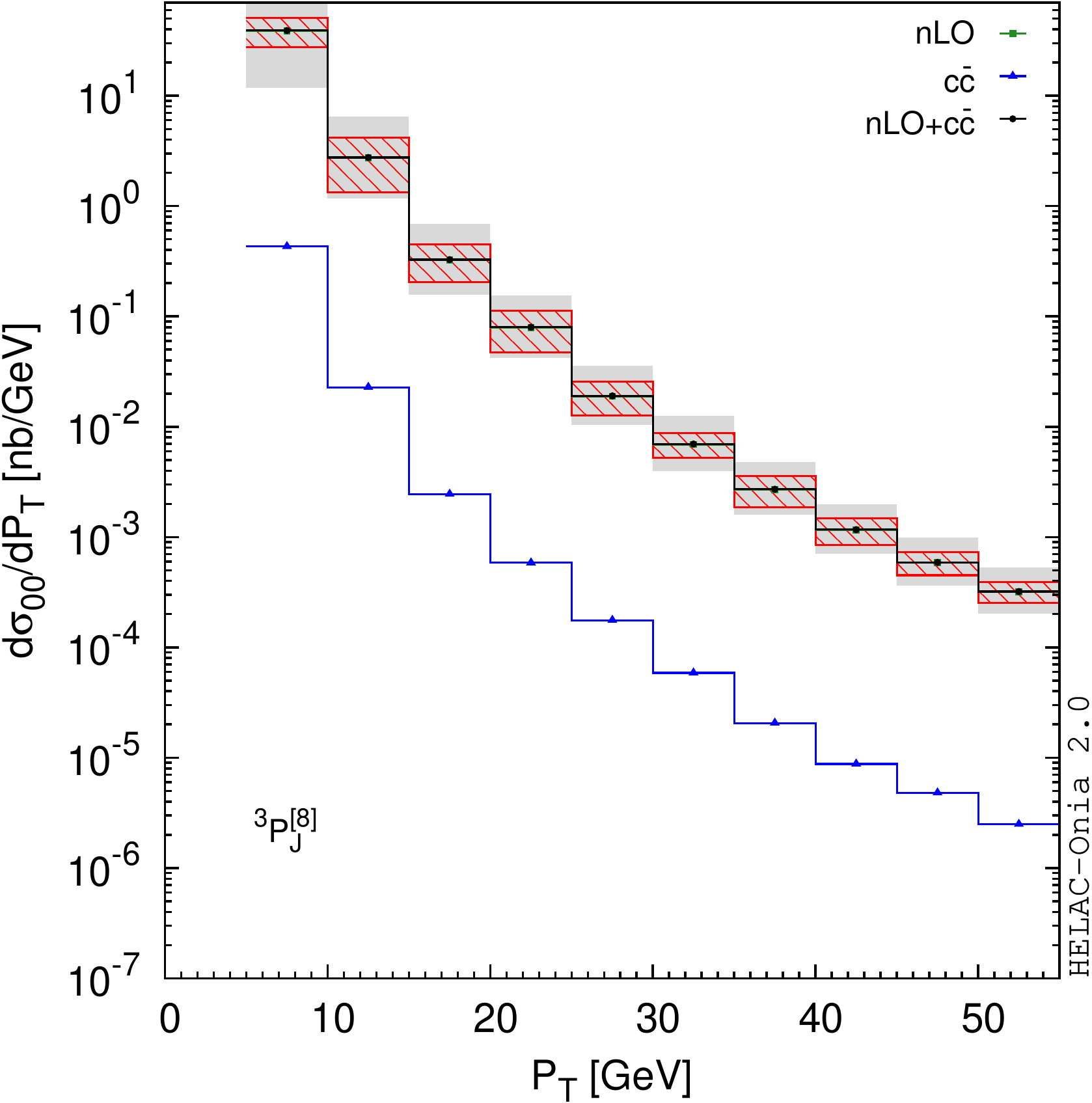}
\includegraphics[width=.45\textwidth,draft=false]{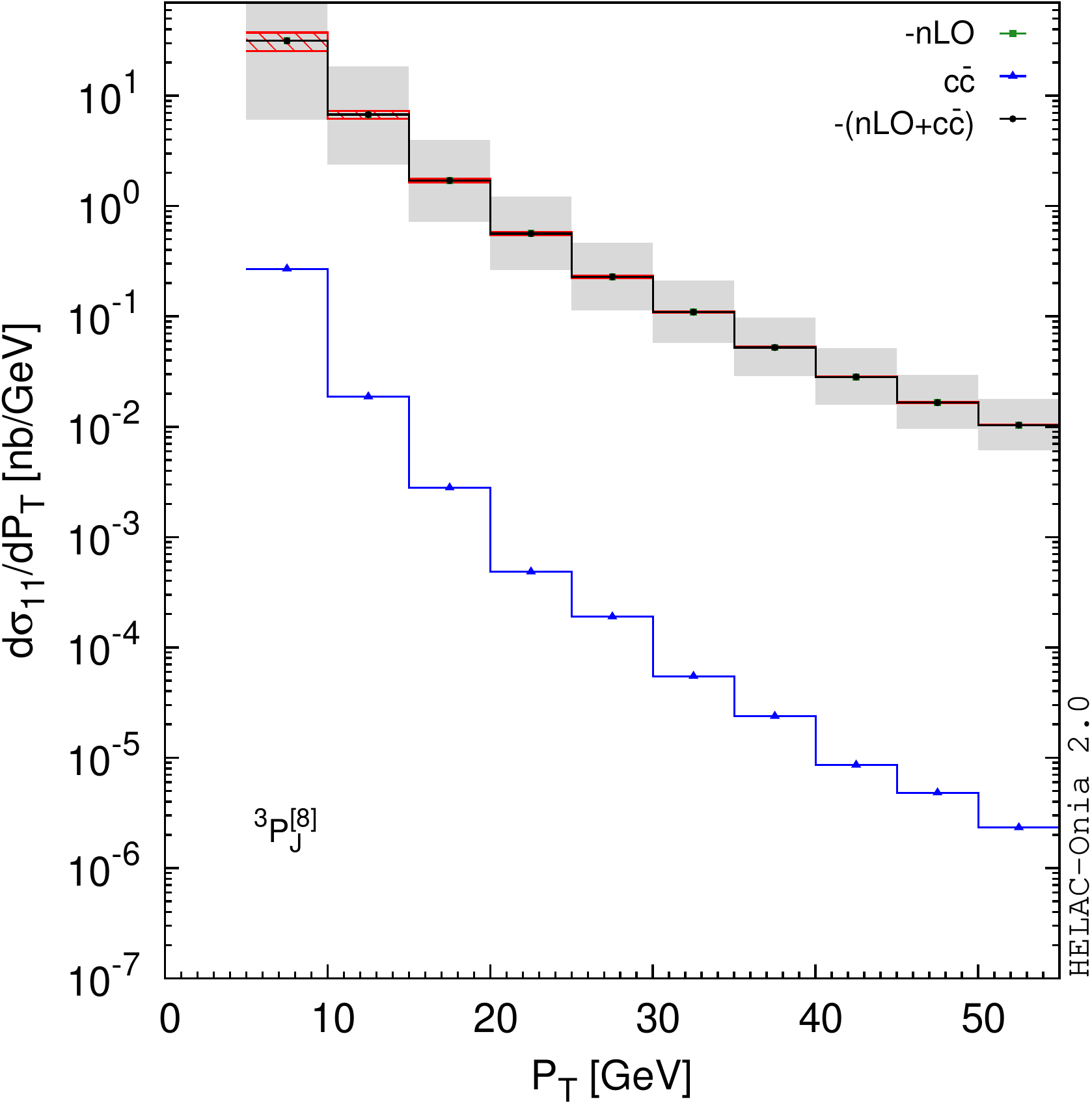}\\
\includegraphics[width=.45\textwidth,draft=false]{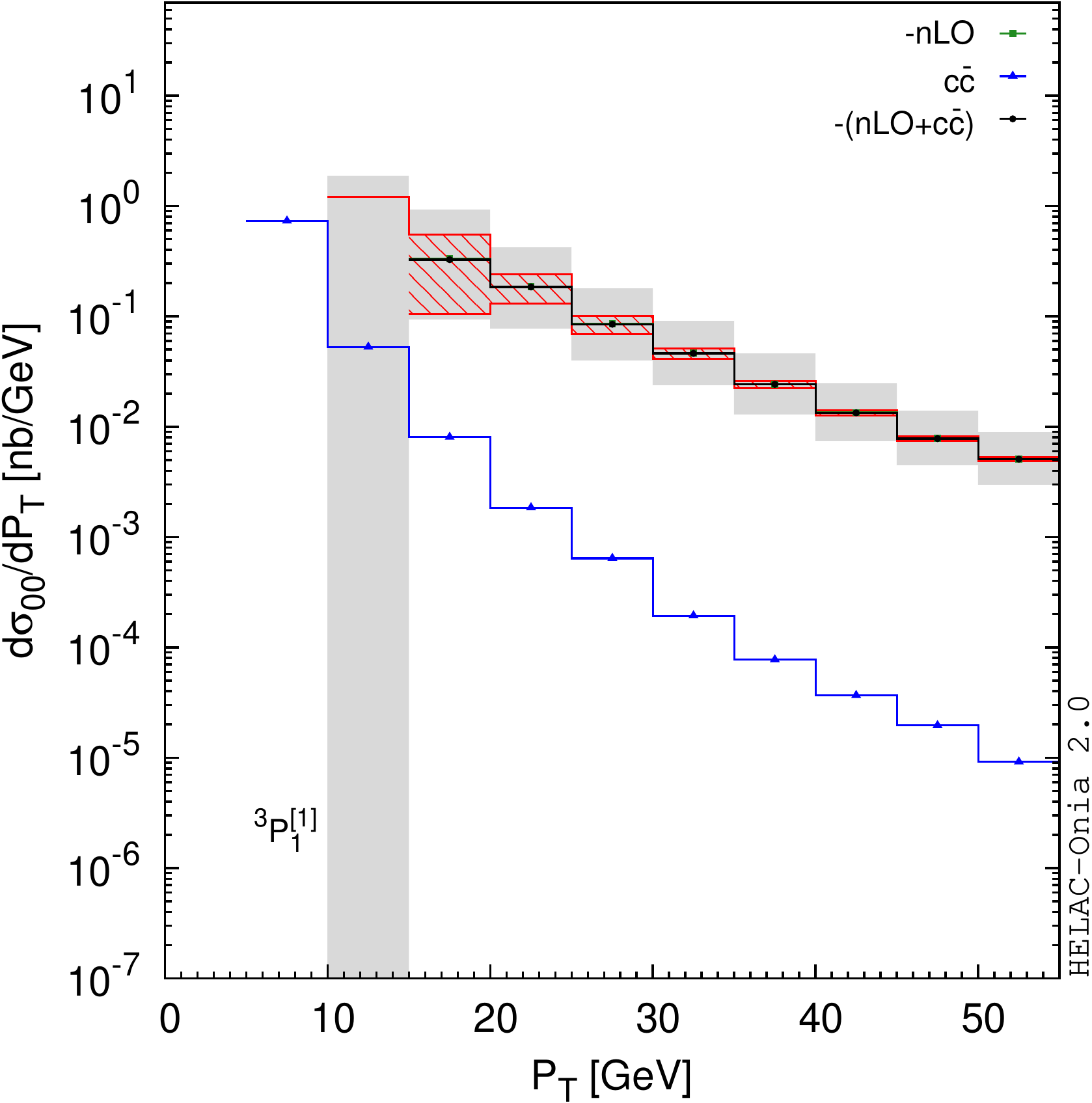}
\includegraphics[width=.45\textwidth,draft=false]{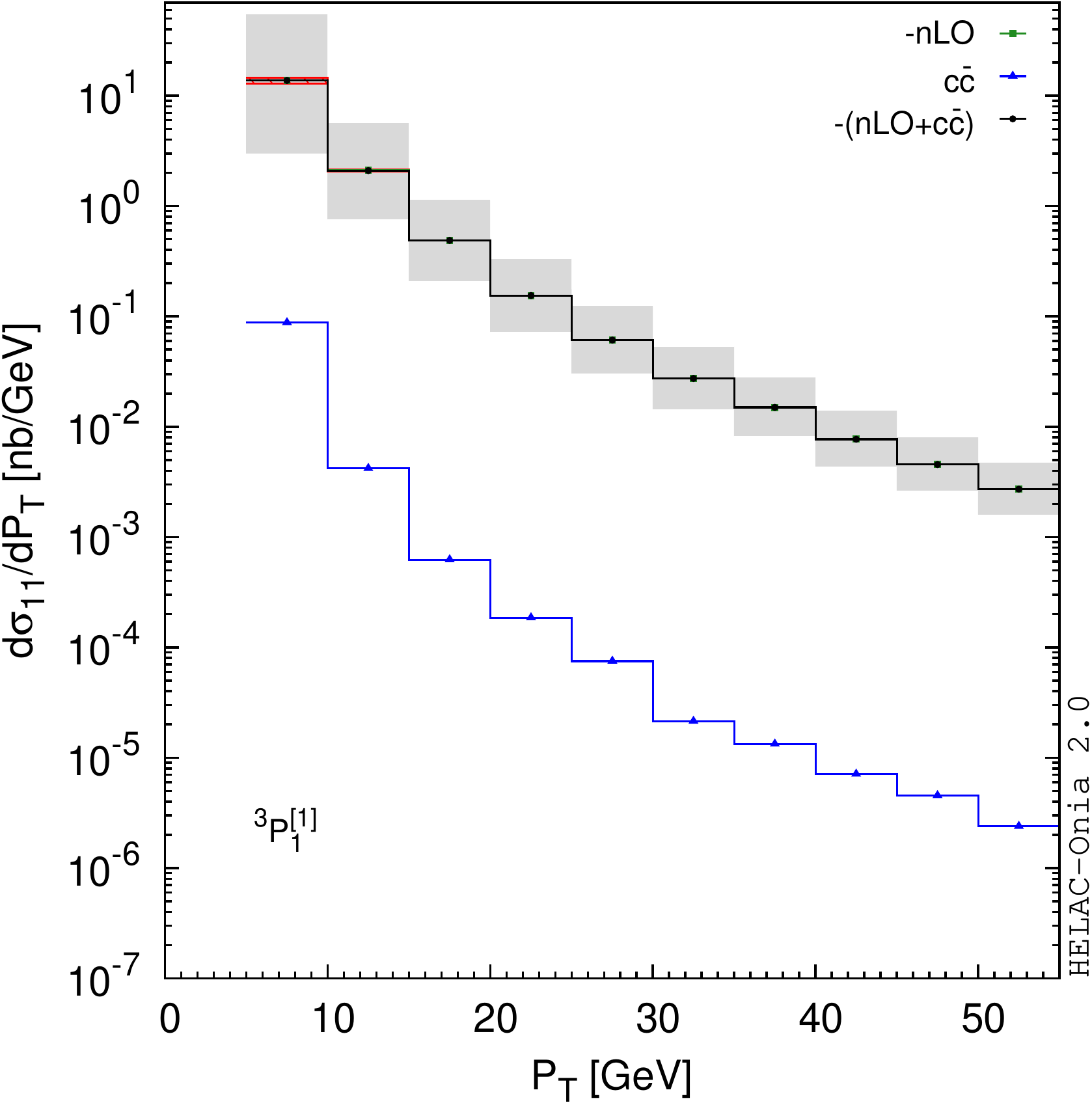}
\caption{Comparisons of spin-dependent differential cross sections $\frac{d\sigma}{dP_T}$ for the Fock states $\pj,\tpos$ between our nLO calculations and the LO charmonium plus charm quark pair calculations. They are similar to Fig.~\protect\ref{Fig:nLOvsccxSpin4CO}. \label{Fig:nLOvsccxSpin4CO2}}
\end{figure}

\begin{figure}[ht!]
\centering
\includegraphics[width=.45\textwidth,draft=false]{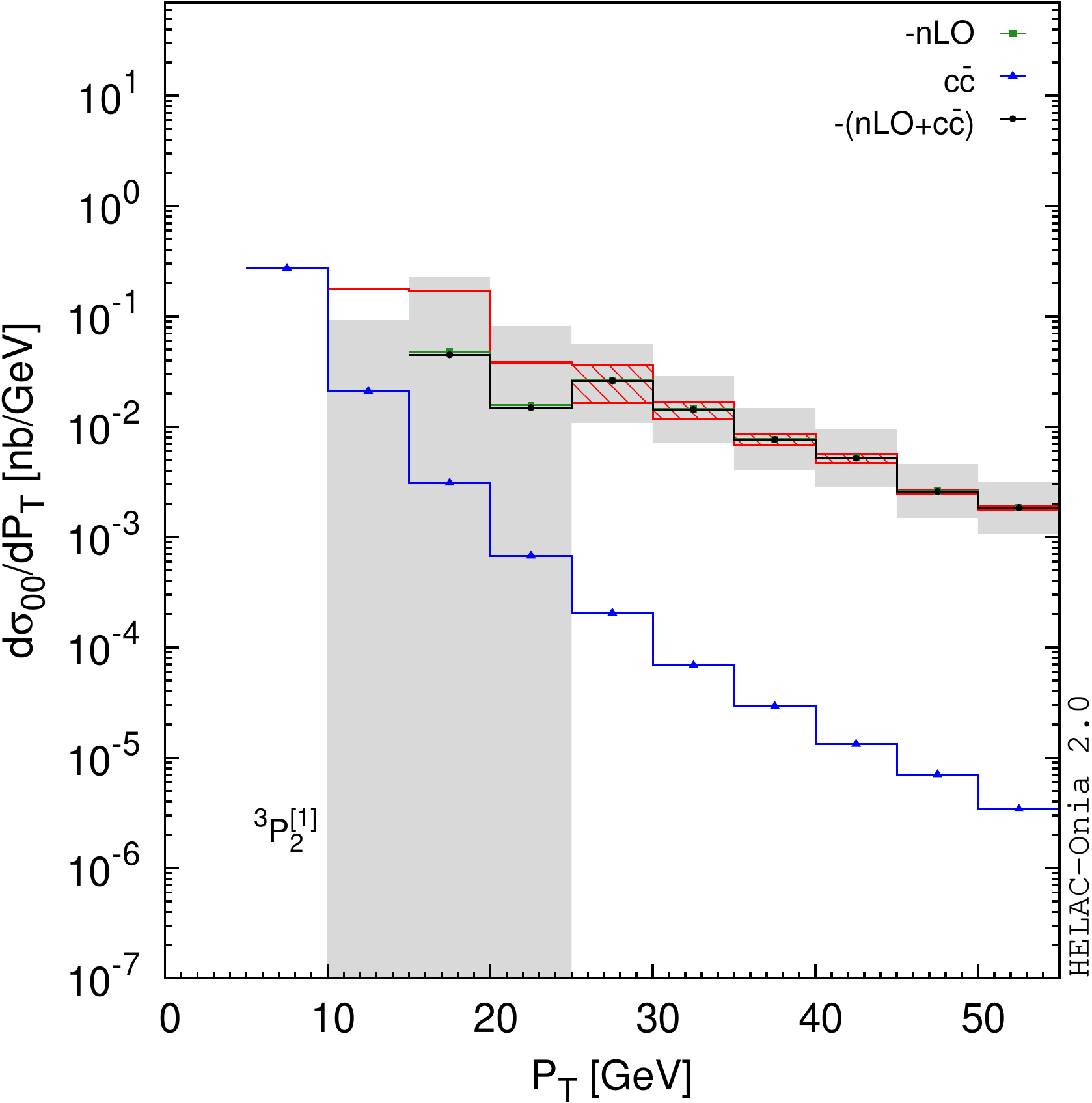}
\includegraphics[width=.45\textwidth,draft=false]{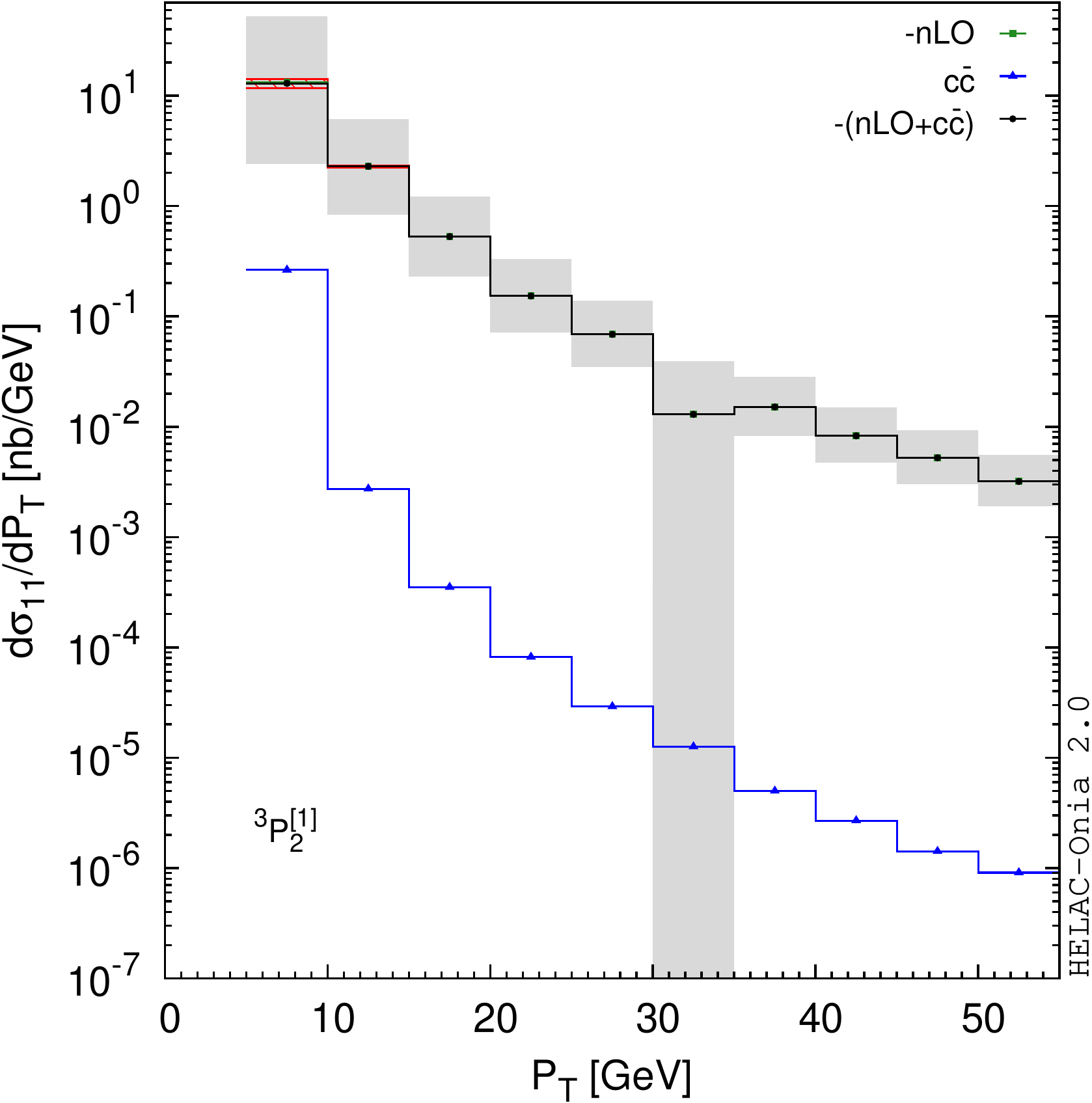}\\
\includegraphics[width=.45\textwidth,draft=false]{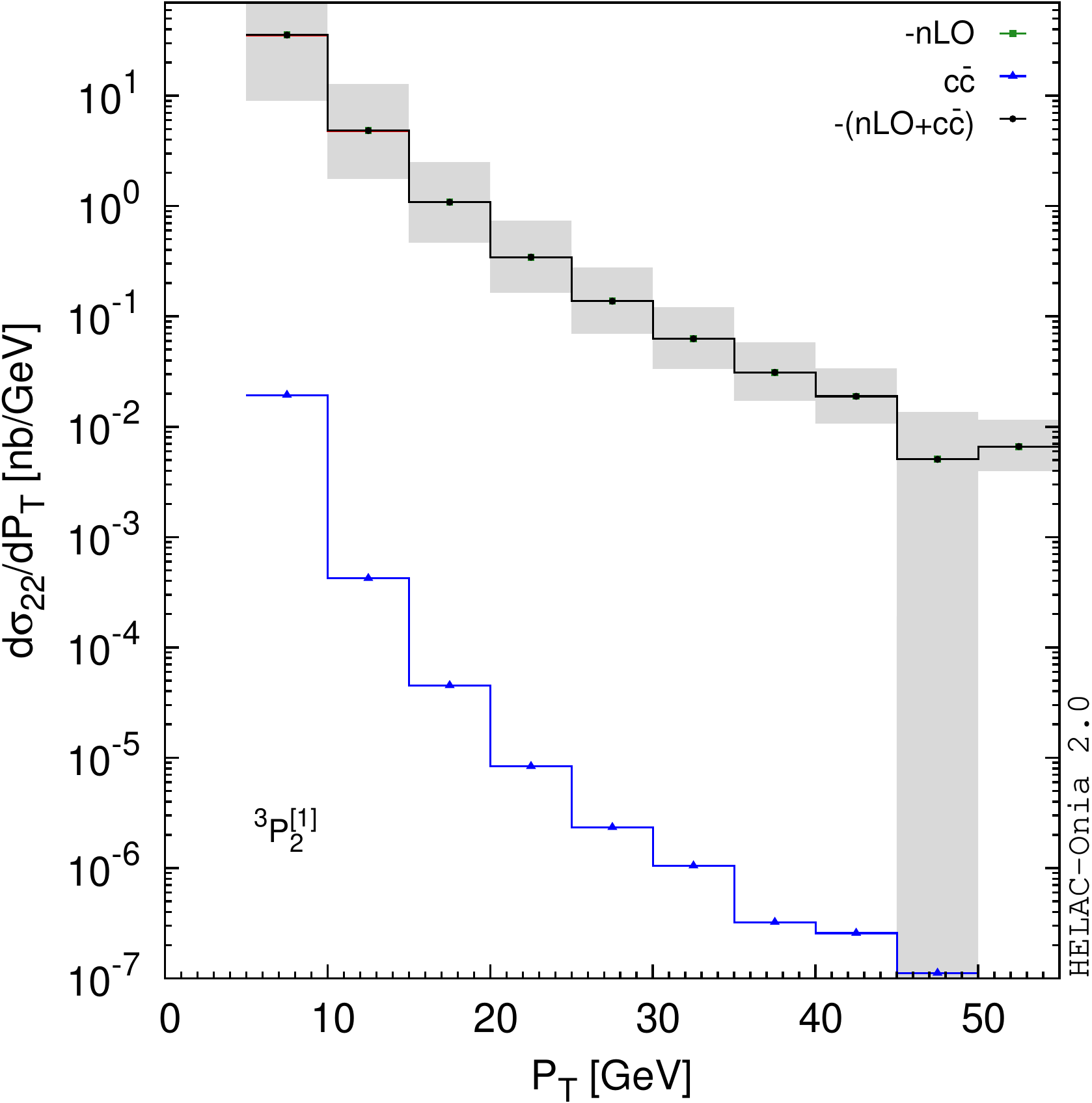}
\caption{Comparisons of spin-dependent differential cross sections $\frac{d\sigma}{dP_T}$ for the Fock state $\tpts$ between our nLO calculations and the LO charmonium plus charm quark pair calculations. They are similar to Fig.~\protect\ref{Fig:nLOvsccxSpin4CO}.\label{Fig:nLOvsccxSpin4CO3}}
\end{figure}



\bibliographystyle{JHEP}        
\bibliography{paper.bib}



\end{document}